\documentclass[]{article}
\usepackage{graphicx}
\usepackage{xcolor}
\usepackage{amsfonts}
\usepackage{amssymb}
\usepackage{pxfonts}
\def\ligne#1{\hbox to\hsize{#1}}
\def\leurre{\noindent\leftskip0pt\small\baselineskip 10pt}
\newtheorem{thm}{\textrm{\sc Theorem}}
\newtheorem{cor}{\textrm{\sc Corollary}}
\newtheorem{lemm}{\textrm{\sc Lemma}}
\newtheorem{fig}{\textrm{Figure}}
\newtheorem{tab}{\textrm{Table}}
\newtheorem{algo}{\textrm{Algorithm}}

\newcounter{laform}
\author{Maurice {\sc Margenstern}}
\title{About Fibonacci trees. 
\begin{center} $-$ III $-$ :\\ multiple Fibonacci trees\end{center}}

\begin{document}
\maketitle
\begin{abstract}
In this third paper, we revisit the question to which extent the properties of the 
trees associated to the tilings $\{p,4\}$ of the hyperbolic plane are still true if 
we consider a finitely generated tree by the same rules but rooted at a black node? 
What happens if, considering the same distinction between black and white nodes
but changing the place of the black son in the rules. What happens if we change the 
representation of the numbers by another set of digits?
	We tackle all of these questions in the paper. Section~\ref{intro} indicates
which Section or Sub-section is devoted to which problem. Section~\ref{conclude}
concludes the paper. The present paper is an extension of the previous 
papers~\cite{mmarXiv1,mmarXiv2}.
\end{abstract}

\section{Introduction}\label{intro}
\def\zz#1{{\footnotesize\tt #1}}

Paper~\cite{mmarXiv1} investigated the question what happens if the rules generating 
the standard Fibonacci tree are applied to a tree whose root is a black node. The question
was investigated with what is called in this paper the leftmost assignment: in the
generating rules, the black son is always the first one. In that paper too, the
question was raised of what happens if instead the standard Fibonacci sequence we
consider what was called the golden sequence which is associated with the square of the
golden number while the standard Fibonacci sequence is associated with the golden 
number itself. The question was considered for both the white and the black Fibonacci
trees. 

Paper~\cite{mmarXiv2} generalizes the context of the same questions. Instead of
considering trees more or less connected with the tilings~$\{5,4\}$ and~$\{7,3\}$
of the hyperbolic plane, that paper considers the trees which span the tilings 
$\{p,4\}$ and $\{p$+$2,3\}$ of the same plane. Those trees are finitely generated 
by rules which generalize the rules of the case \hbox{$p=5$} to which 
paper~\cite{mmarXiv1} limited its study. Also the new trees entail the consideration
of new families of numbers, we called metallic numbers in~\cite{mmarXiv2}, which 
allow representations of the natural numbers which look like the golden family of 
the case \hbox{$p=5$}. In the present paper, we perform two new steps
in the generalization. On one hand, we consider the definition of the generating rules
themselves. Instead of considering that the black son is the first one in each rule,
we consider various possibilities if the position of that son is changed, whether the
change is always the same or if the change is also submitted to variations.
On another hand, we consider the representation of the numbers. 
In paper~\cite{mmarXiv2}, the metallic sequences are defined with digits in 
\hbox{$\{0..p$$-$$3\}$}. What if we impose the digits to be in \hbox{$\{1..p$$-$$2\}$},
considering only positive integers which is the case for the numbers we attach to the 
tiles? 

Section~\ref{trees} recalls definitions about trees and about the trees we
consider in the present paper. Section~\ref{metal} recalls the results about
the metallic numbers and the standard representation of positive numbers we can infer 
from them. The section also considers the representation where the set of digits
is \hbox{$\{1..p$$-$$2\}$}. Section~\ref{assign} defines what we call an assignment,
a way of applying rules in the construction process of a tree and the section studies
the properties of the representations studies in Section~\ref{metal} with respect
to various assignments. Section~\ref{conclude} investigates the contribution of
the paper and the problems which remain open.

\def\bbzz{\hbox{\bf b0}}
\def\bbuu{\hbox{\bf b1}}
\def\wwzz{\hbox{\bf w0}}
\def\wwaa{\hbox{\bf wa}}
\def\bbb{\hbox{\bf b}}
\def\aa{\hbox{\bf a}}
\def\zzz{\hbox{\bf 0}}
\def\xxx{\hbox{\bf x}}
\def\uu{\hbox{\bf 1}}
\def\ddd{\hbox{\bf d}}
\def\ccc{\hbox{\bf c}}
\def\eee{\hbox{\bf e}}
\def\fff{\hbox{\bf f}}
\def\kkk{\hbox{\bf k}}
\def\www{\hbox{\bf w}}
\def\ppp{\hbox{\bf p}}
\def\numq{\hbox{\bf 4}}
\def\numd{\hbox{\bf 2}}
\def\numt{\hbox{\bf 3}}
\def\sympl{\hbox{\bf +}}
\section{Metallic trees}\label{trees}

   In this section, Sub-section~\ref{strees} recalls the vocabulary we shall use
in the considerations of the trees which appear in the paper and in the properties
connected with them, considering, in particular, the numbering we may attach to the nodes
of a finitely generated tree. In Sub-section~\ref{smrules}, we consider the metallic 
trees which we shall study in this paper.

\subsection{Preliminary definitions and properties of trees}\label{strees}

Consider an infinite tree~$\cal T$ with finite branching at each node. Number the nodes
from the root which receives~1, then, level by level and, on each level, from left to 
right with the conditions that for each node, the numbers of its sons are consecutive
numbers. We then say that $\cal T$ is {\bf numbered} or that it is endowed with
its {\bf natural numbering}. In what follows, we shall consider
numbered trees only. Clearly, a sub-tree $\cal S$ of $\cal T$ can also be numbered
in the just above described way but it can also be numbered by the numbers of its
nodes in $\cal T$. In that case, a node~$\nu$ may receive two numbers: $n_{\cal S}$,
the number defined in~$\cal S$ as a numbered tree and $n_{\cal T}$, its number as a
node of~$\cal T$. A node may have no son, it is then called a {\bf leaf}.
A {\bf path} from~$\mu$ to~$\nu$ is a finite sequence of nodes 
\hbox{$\{\lambda_i\}_{i\in[0..k]}$}, if it exists, such that $\lambda_0=\mu$, 
$\lambda_k=\nu$ and, for all $i$ with \hbox{$i\in[0..k$$-$$1]$}, $\lambda_{i+1}$ is a 
son of~$\lambda_i$.  
A {\bf branch} of $\cal T$ is a maximal finite or infinite sequence of paths~$\{\pi_i\}$ 
from the root of~$\cal T$ to nodes of that tree such that for all~$i$, $j$, 
$\pi_i\subseteq\pi_j$ or $\pi_j\subseteq\pi_i$. Accordingly, a branch connects the root
to a leaf or it is infinite. It is clear that for any node, they are connected to 
the root by a unique path. The {\bf length} of the path from a node to one of its son 
is always~1. If the length of a path from~$\mu$ to~$\nu$ is~$k$, the length of the
path from~$\mu$ to any son of~$\nu$, assuming that $\nu$ is not a leaf, is $k$+1.
The length of the path leading from the root to a node~$\nu$ of~$\cal T$ is
called the {\bf distance} of~$\nu$ to the root~$\rho$ and it is denoted by
dist$\{\rho,\nu\}$. We also define \hbox{dist$\{\rho,\rho\}=0$}. 
The {\bf level}~$k$ of~$\cal T$ is the set of its nodes which are at the distance~$k$
from its root. Denote it by \hbox{${\cal L}_{k,\cal T}$}. 
Define~${\cal T}_n$ as the set of levels~$k$ of~$\cal T$ with
\hbox{$k\leq n$}. Say that the {\bf height} of ${\cal T}_n$ is~$n$.
By definition, ${\cal T}_n$ is a sub-tree of~$\cal T$. 
For each node~$\nu$ of~$\cal T$, $\lambda_{\cal T}(\nu)$ is its level 
in~$\cal T$, {\it i.e.} its distance from the root, and $\sigma_{\cal T}(\nu)$ is the 
number of its sons. Clearly, if $\nu \in {\cal T}_n$ and if $\lambda_{\cal T}(\nu)=n$, 
then $\sigma(\nu)=0$. If $\cal S$ is a sub-tree of $\cal T$, denote it
by ${\cal S}\lhd\cal T$, and if $\nu\in\cal S$,
then $\lambda_{\cal S}(\nu)\leq\lambda_{\cal T}(\nu)$ and the numbers may be not equal.

Consider two infinite numbered trees $T_1$ and $T_2$. 
Say that $T_1$ and $T_2$ are {\bf isomorphic} if 
there is a bijection~$\beta$ from~$T_1$ onto~$T_2$ such that:
\vskip 5pt
\ligne{\hfill
$\vcenter{\vtop{\leftskip 0pt\hsize=150pt
\ligne{$f(n_{T_1})=n_{T_2}$ for any $n\in\mathbb N$.\hfill}
\ligne{$\lambda_{T_2}(f(n_{T_1}))=\lambda_{T_1}(n)$.\hfill}
\ligne{$\sigma_{T_2}(f(n_{T_1}))=\sigma_{T_1}(n)$.\hfill}
}}$
\hfill$(0)$\hskip 10pt}
\vskip 5pt
Clearly, if $T_1$ and $T_2$ are two infinite numbered trees, they are isomorphic
if and only if there is a bijection from the nodes of~$T_1$ into those of~$T_2$ such that
a node of~$T_1$ and its image in~$T_2$ have the same number, they are on the same level 
of their respective trees and they have the same number of sons.

\subsection{White and black metallic trees}\label{smrules}

    We call {\bf metallic tree} an infinite tree constituted by two kind of nodes,
\bbb- and \www-ones called 
{\bf black} and {\bf white} respectively, finitely generated by
the following rules:
\vskip 5pt
\ligne{\hfill$\bbb{}\rightarrow \bbb\www^{p-4}$ and $\www{}\rightarrow \bbb\www^{p-3}$.
\hfill (1)\hskip 10pt}
\vskip 5pt
\noindent
with \hbox{$p\geq5$}.

The property for a node to be white or black is called its {\bf literal status}.
We also associate to the node its {\bf numerical status}: 0 or~1 depending on whether
the node is white or black respectively. If it is not specified, {\bf status} will 
refer to the literal one.

We shall mainly investigate two kinds of infinite metallic trees.
When the root of the tree is a white, black node, we call such a metallic tree a 
{\bf white}, {\bf black metallic tree} respectively. We denote the
infinite white metallic tree by $\cal W$ and we endow it with its natural numbering.
We do the same with the infinite black metallic tree $\cal B$. Note
that we can construct a bijective morphism between $\cal B$ and a part $\mathbb B$ 
of~$\cal W$ as follows. The morphism is the identity on~$\mathbb B$
and we fix the following conditions: 
\vskip 5pt
\ligne{\hfill
\vtop{\leftskip 0pt\hsize=200pt
\ligne{\hfill$\sigma_{\mathbb B}(1) = \sigma_{\cal W}(1)-1$,\hfill}
\ligne{\hfill$\sigma_{\mathbb B}(n) = \sigma_{\cal W}(n)$, for all positive integer~$n$.
\hfill}
}
\hfill}
\vskip 5pt
\noindent
Moreover, the nodes numbered by~$n\in [1..p$$-$$2]$ in $\cal W$ also belong
to $\mathbb B$ and receive the same numbers in the natural numbering of $\mathbb B$.
This morphism allows us to identify $\mathbb B$ with $\cal B$, so that in our
sequel, we shall speak of $\cal B$ only. From what we just said, it is plain that
for a node $\nu\in \cal B$, if $\nu_{\cal B} > p$$-$2, then
$\nu_{\cal B} < \nu_{\cal W}$. We shall look closer to the connection between
$\nu_{\cal B}$ and $\nu_{\cal W}$ in Section~\ref{compnumwb}.
Later on, we shall use $\rightleftharpoons$ to introduce a notation for an expression.

We may wonder whether the simplicity of the rules~$(1)$ allow us to give a precise 
connection between the number of a node and those of its sons, whether in~$\cal W$ or
in~$\cal B$. These questions were partially studied in~\cite{mmarXiv2}.
We shall turn to them in Section~\ref{assign}. But before, we need to recall the
introduction of an appropriate representation of the numbers used to number the nodes
of a metallic tree. It is the goal of Section~\ref{metal} to which we new turn.

\section{Metallic numbers}\label{metal}

These numbers are introduced by the computation of the number of nodes which lie on
a given level of a metallic tree. We consider that point in 
Sub-section~\ref{smetalsuites}. The sequence allow us to represent the numbers. We
consider some basic properties of the standard representation in 
Sub-section~\ref{smetalcodes}. We study the same properties in the representation
where the digits are restricted to \hbox{$\{1..p$$-$$2\}$} in
Sub-section~\ref{snzmcodes}.

\subsection{The metallic sequences}\label{smetalsuites}

Let $m_n$, $b_n$ be the number of nodes on ${\cal L}_{n,\cal W}$ 
and ${\cal L}_{n,\cal B}$ respectively. We also define $M_n$, and $B_n$ as the number
of nodes of ${\cal W}_n$ and ${\cal B}_n$ respectively. It appears that these
numbers are defined by a simple induction equation as stated in the following statement:

\begin{thm}{\rm\cite{mmJUCStools,mmarXiv2}}\label{tmetallevelw}
Consider the numbers $m_n$ defined as the number of nodes on ${\cal L}_{n,\cal W}$,
where $\cal W$ is the white metallic tree. The numbers $m_n$ satisfy the
following induction equation:
\vskip 5pt
\ligne{\hfill$m_{n+2} = (p$$-$$2)m_{n+1}-m_n$ with $m_0=1$ and $m_{-1}=0$.\hfill
$(2)$\hskip 10pt}
\vskip 5pt
\noindent
We call {\bf white metallic sequence} the sequence $\{m_n\}_{n\in\mathbb N}$.
\end{thm}

\noindent
See the proof in~\cite{mmarXiv2} for instance.

As the black metallic tree is defined by the same rules, we may conclude that
the same equation rules the sequence $\{b_n\}_{n\in\mathbb N}$:

\begin{thm}\label{tmetallevelb}
The sequence $\{b_n\}_{n\in\mathbb N}$ of the number of nodes on ${\cal L}_{n,\cal B}$
satisfies the equation:
\vskip 5pt
\ligne{\hfill$b_{n+2} = (p$$-$$2)b_{n+1}-b_n$ with $b_1=p$$-$$3$ and $b_0=1$.\hfill
$(3)$\hskip 10pt}
\vskip 5pt
We call {\bf black metallic sequence} the sequence $\{b_n\}_{n\in\mathbb N}$. 
\end{thm}

Note that we could define the white metallic sequence by the initial conditions
$m_1=p$$-$2 and $m_0=1$. In our sequel we shall say {\bf metallic sequence} instead
of {\bf white metallic sequence} for a reason which will be made more clear in a while.

Before turning to the properties of the integers with respect to the metallic numbers,
we have to consider the numbers $M_n$ and $B_n$ already introduced with respect 
to the finite trees ${\cal W}_n$ and ${\cal B}_n$.

\begin{thm}{\rm (see \cite{mmbook1})}\label{theadlevelwb}
On the level~$k$ of $\cal W$, with non-negative~$k$, the rightmost node has the number
$M_k$, so that the leftmost node on the level~$k$$+$$1$
has the number $M_k$$+$$1$.

On the level~$k$ of $\cal B$ with non-negative~$k$, the rightmost node has the
number~$m_k$, so that the leftmost node on the level~$k$$+$$1$ has the
number~$m_k$$+$$1$.

The sequence $\{M_n\}_{n\in\mathbb N}$ satisfies the following induction equation:
\vskip 5pt
\ligne{\hfill$M_{n+2}=(p$$-$$2$$)M_{n+1}-M_n+1$,\hfill 
$(4)$\hskip 10pt}
\vskip 5pt
\noindent
with the initial conditions $M_0=1$ and $M_{-1}=0$, while the sequence 
$\{B_n\}_{n\in\mathbb N}$ satisfy the equation~$(2)$ with the same
initial conditions, which means that \hbox{$B_n=m_n$} for any non-negative~$n$.
We also have:
\vskip 5pt
\ligne{\hfill $M_{n+1} = B_{n+1}+M_n$\hskip 10pt and
\hskip 10pt $m_{n+1} = b_{n+1}+m_n$\hfill $(5)$\hskip 10pt}
\end{thm}

\noindent
See the proof in~\cite{mmarXiv2}.

\subsection{Metallic codes
for the nodes of the metallic trees}~\label{smetalcodes}.

Let us go back to the sequence $\{m_n\}_{n\in\mathbb N}$ of metallic numbers.
It is clear that the sequence defined by~$(2)$ is increasing starting from~$m_1$:
from~$(2)$, we get that \hbox{$m_{n+2}> (p$$-$3$)m_{n+1}$} if we assume that
\hbox{$m_n<m_{n+1}$}. As \hbox{$p\geq5$}, we get that the sequence is increasing starting
from~$m_1$. Now, as the sequence is increasing, it
is known that any positive integer~$n$ can be written as a sum of distinct metallic 
numbers whose terms are defined by Theorem~\ref{tmetallevelw}:
\vskip 5pt
\ligne{\hfill$n=\displaystyle{\sum\limits_{i=0}^k a_im_i}$ with $a_i\in\{0..p$$-$$3\}$.
\hfill
(6)\hskip 10pt}
\vskip 5pt
\noindent
The sum of $a_im_i$'s in~$(6)$ is called the {\bf metallic representation} of~$n$
and the $m_i$'s in~$(6)$ are the {\bf metallic components} of~$n$.

From now on, we use bold characters for the digits of a metallic representation of
a number. In particular, we define \ddd{} to represent $p$$-$3, 
\ccc{} to represent $p$$-$4 and \eee{} to represent $p$$-$5 when \hbox{$p>5$}. Of 
course, \zzz, \uu, \numd{} and \numt{} represent 0, 1, 2 and 3 respectively.

   First, note that the representation~$(6)$ is not unique.

\begin{lemm}{\rm \cite{mmgsJUCS,mmbook1,mmarXiv2}}\label{ltechforbid}
For any integers $n$ and $h$ with \hbox{$0\leq h\leq n$}, we have:
\vskip 5pt
\ligne{\hskip 20pt $(p$$-$$3)m_{n+1}+
	\displaystyle{\sum\limits_{k=h+1}^n\hbox{$(p$$-$$4)m_k$}}+(p$$-$$3)m_h$\hfill}
\ligne{\hskip 40pt
$=(p$$-$$3)m_{n+1}+
	\displaystyle{\sum\limits_{k=h+2}^n\hbox{$(p$$-$$4)m_k$}}+(p$$-$$3)m_{h+1}
-m_h+m_{h-1}$\hfill $(7)$\hskip 10pt}
\end{lemm}

\begin{cor}{\rm \cite{mmgsJUCS,mmbook1}}\label{ctechforbid}
For any positive integer~$n$, we have:
\vskip 5pt
\ligne{\hfill $(p$$-$$3)m_{n+1}+
	\displaystyle{\sum\limits_{k=1}^n\hbox{$(p$$-$$4)m_k$}}+(p$$-$$3)m_0
=m_{n+2}$\hfill $(8)$\hskip 10pt}
\end{cor}

\noindent
See the proofs in~\cite{mmarXiv2} for instance.

Let us write the $a_i$'s of~(5) as a word \hbox{\bf a$_k$..a$_1$a$_0$} which we call
a {\bf metallic word} for~$n$ as the digits $a_i$ which occur in~(5) 
are not necessarily unique for a given~$n$. 
They can be made unique by adding the following condition on the corresponding metallic
word for~$n$: the pattern \hbox{\bf dc$^\ast$d} is ruled out from
that word. It is called the {\bf forbidden pattern}. 
Lemma~\ref{ltechforbid} proves that property which is also proved 
in~\cite{mmgsJUCS,mmbook1}. We reproduced it here for the reader's convenience.

\def\sgn#1{\hbox{\bf sg}}
When a metallic representation for~$n$ does not contain the forbidden
pattern it is called the {\bf metallic code} of~$n$ which we denote by $[n]$. We shall 
write \hbox{$\nu = ([\nu])$} when we wish to restore the number from its metallic code. 
Let us call {\bf signature} of $\nu$ the rightmost digit of 
\hbox{\bf $[\nu]$ $=$ a$_k$$..$a$_1$a$_0$} and denote
it by \sgn{$(\nu)$}.
Let $\sigma_1$, $\sigma_2$, ..., $\sigma_k$ with $k=p$$-$2 or $k=p$$-$3 be the sons 
of~$\nu$. We call {\bf sons signature} of~$\nu$ the word \hbox{\bf s$_1$...s$_k$},
where \hbox{\bf s$_i$ $=$ \sgn{}$(\sigma_i)$}. We shall denote the literal status
of~$\nu$ by \hbox{$\ell s(\nu)$} and its numerical one by $sn(\nu)$.

\subsection{The non-zero metallic codes}\label{snzmcodes}

It is known that given a basis~$b$ with \hbox{$b\geq3$}
any positive number $n$ can be written 
\vskip 5pt
\ligne{\hfill
$n=\displaystyle{\sum\limits_{i=0}^k a_ib^i}$\hskip 10pt with $a_i\in\{1..b\}$
\hfill$(9)$\hskip 10pt}
\vskip 5pt
Let \hbox{$a^+$ $=$ $a$+1} and \hbox{$a^-$ $=$ $a$$-$1} for any positive integer~$a$.
The representation (9) was used by Quine in order to encode any finite sequence
of natural numbers: writing $n$ as in $(9)$, $b$ is used as a separator and the 
other digits which lie in \hbox{$[1..b^-]$} can be interpreted as the representations 
of positive numbers in the base $b^-$ which requires \hbox{$b\geq2$}. Smullyan, 
see~\cite{smullyan} makes use of such a 
representation in order to prove G\"odel's theorem on the incompleteness of
Peano arithmetics.

The question is: taking the metallic numbers $m_n$ as a basis, is it possible to
have a representation which rules out all 0's in the representation of a positive
number? The answer is given by the following proposition:

\begin{thm}\label{tnzmcode}
Let $n$ be a positive natural number. Then it is possible to write~$n$ as:
\vskip 5pt
\ligne{\hfill
$n=\displaystyle{\sum\limits_{i=0}^k a_i\,m_i}$\hskip 10pt with $a_i\in\{1..b\}$,
\hfill$(10)$\hskip 10pt}
\vskip 5pt
\noindent
where $b=p$$-$$2$.
\end{thm}

\noindent
Proof. As all digits of~$n$ in $(10)$ should be not smaller than~\uu, we consider $k$ 
defined by the unique value such that \hbox{$M_k\leq n< M_{k+1}$}.
We then define \hbox{$n_1\rightleftharpoons n-M_k$}. Consider the digits $a_i$ of
$[n_1]$, the metallic code of~$n_1$. We have 
\hbox{$n_1=\displaystyle{\sum\limits_{i=0}^k a_i\,m_i}$} and
\hbox{$M_k=\displaystyle{\sum\limits_{i=0}^k m_i}$}, so that we get
\hbox{$n=\displaystyle{\sum\limits_{i=0}^k s_i\,m_i}$} where \hbox{$s_i=a_i$+$1$}
with \hbox{$i\in\{0..k\}$}. As \hbox{$0\leq a_i<b$} for all~$i$ in \hbox{$\{0..k\}$}
we get \hbox{$1\leq s_i\leq b$} for the same indices.\hfill $\Box$

When we use the representation~$(10)$ to write a positive integer, we use~\xxx{} to
denote the digit whose value is \hbox{$b=p$$-$2}.
We know that the representation with the metallic numbers of a positive integer is
not necessarily unique. This is why we needed to select a criterion in order to
ensure the uniqueness of the metallic code. It is also the case that the representation
$(10)$ is not unique, despite the fact that it satisfies the constraint of no \zzz{}
among the digits. To see that point, we need the following result which enlarges
a lemma from~\cite{mmarXiv2}:

\begin{lemm}\label{lnotuniquenzmcode}
For any integers $n$ and $h$ with \hbox{$0\leq h\leq n$}, we have:
\vskip 5pt
\ligne{\hskip 20pt $(p$$-$$2)m_{n+1}+
	\displaystyle{\sum\limits_{k=h+1}^n\hbox{$(p$$-$$3)m_k$}}+(p$$-$$2)m_h$\hfill}
\ligne{\hskip 40pt
$=(p$$-$$2)m_{n+1}+
	\displaystyle{\sum\limits_{k=h+2}^n\hbox{$(p$$-$$3)m_k$}}+(p$$-$$2)m_{h+1}
+m_{h-1}$\hfill $(11)$\hskip 10pt}
\end{lemm}

\def\nzm{{\bf nzm}}
\def\nnn{{\bf n}}
\def\mmm{{\bf m}}
\def\hhh{{\bf h}}
\def\jjj{{\bf j}}
\def\iii{{\bf i}}
\def\rrr{{\bf r}}
\begin{cor}\label{cnzmcodeforbid}
For any positive integer~$n$, it has a unique representation $(10)$ provided
that the pattern \hbox{\xxx\ddd$^\ast$\xxx} is ruled out and is called a 
{\bf forbidden pattern}. A metallic code where the \zzz-digit is ruled out and where 
the forbidden pattern does
not occur is called a {\bf non-zero metallic code}, \nzm-code for short. The \nzm-code
of~$n$ is denoted by $[\nu]_{nz}$.

In order to get a $[\nu]_{nz}$  from $(10)$, we apply the conversion rules:
\vskip 5pt
\ligne{\hfill \nnn\xxx\ddd$^k$\xxx\mmm\hskip 5pt $=$\hskip 5pt
	\nnn$^+$\uu$^{k+2}$\mmm$^+$
	\hskip 10pt and \hskip 10pt
	\nnn$^+$\zzz$^{k+2}$\mmm$^+$\hskip 5pt $=$\hskip 5pt
\nnn\ddd\ccc$^k$\ddd\mmm\hfill $(12)$\hskip 10pt}
\vskip 5pt
\noindent
where \nnn, \mmm{} are non-zero digits in $\{\uu..\xxx\}$, 
\hbox{\nnn$^+$ = \nnn$\oplus$\uu} and \hbox{\mmm$^+$ = \mmm$\oplus$\uu},
with \hbox{$\aa\oplus\uu = \aa$$+$$1$} if \hbox{$a$$+$$1<p$$-$$2$}
and \hbox{$\aa\oplus\uu = \zzz$} if \hbox{$\aa = p$$-$$3$}.
\end{cor}

We call the forbidden pattern defined in Corollary~\ref{cnzmcodeforbid} the
\nzm-forbidden pattern in order to distinguish it from that of Lemma~\ref{ltechforbid}. 
Note that the forbidden pattern of Lemma~\ref{ltechforbid} is no more forbidden in
an \nzm-code. Of course, the application of~$(12)$ may be repeated in~$(10)$ as long 
as all occurrences of the \nzm-forbidden pattern are replaced by their permitted 
equivalent expression given in~$(12)$.

It is not difficult to adapt the algorithms of~\cite{mmarXiv2} to operations on
\nzm-codes. We just mention the change about the incrementation and decrementation
algorithms.

We take the notations given in the caption of Algorithm~\ref{anzmincr}. The algorithm
first detects whether the \nzm-code of~$\nu$ has a suffix \xxx\ddd$^h$.
If it is the case, appending 1 to \aa$_0$ would create a \nzm-forbidden pattern.
And so, in that case, \xxx\ddd$^h$ is replaced by \uu$^{h+1}$ according to~(12) 
and 1~is added to the digit
which is to the left of~\xxx and which is less than $p$$-$2. If appending~1 would not 
raise a forbidden pattern, the algorithm looks at whether \aa$_0$ is \xxx{} or not. 
If it is~\xxx, \aa$_0$ is replaced by \uu{}
and 1 is added to~\aa$_1$ which is not~\xxx{} as \aa$_0$ was \xxx. If \aa$_0$ is
is not \xxx, 1 is added to it. 

\def\wwhile{{\bf while}}
\def\iff{{\bf if}}
\def\ccase{{\bf case}}
\def\iis{{\bf is}}
\def\wwhen{{\bf when}}
\def\oothers{{\bf others}}
\def\endcase{{\bf end case}}
\def\ffor{{\bf for}}
\def\endloop{{\bf end loop}}
\def\endif{{\bf end if}}
\def\iin{{\bf in}}
\def\faux{{\it false}}
\def\vrai{{\it true}}
\def\nnot{{\bf not}}
\def\lloop{{\bf loop}}
\def\inreverse{{\bf in reverse}}
\def\tthen{{\bf then}}
\def\eelse{{\bf else}}
\def\eet{{\bf and}}
\def\etaussi{{\bf and then}}
\def\oou{{\bf or}}
\def\oudautre{{\bf or else}}
\def\sls{{\bf s}}
\def\cry{{carry}}
\def\pproc{{\bf proc}}
\def\bbegin{{\bf begin}}
\def\endproc{{\bf end proc}}
\vtop{
\begin{algo}\label{anzmincr}
Algorithm giving $[\nu$$+$$1]_{nz}$ from that of~$[\nu]_{nz}$.
Recall that $[\nu]_{nz}$ does not contain a \nzm-forbidden pattern.
We assume that \hbox{$[\nu]_{nz}$ $=$ \aa$_k$$..$\aa$_0$}.
\end{algo}
\vskip-5pt
\ligne{\hfill
\vtop{\leftskip 0pt\hsize=210pt
\hrule height 0.3pt depth 0.3pt width \hsize
\vskip 5pt
\ligne{\hfill
\vtop{\leftskip 0pt\hsize=190pt
\ligne{$i$ := 0;\hfill}
\ligne{\wwhile{} \aa$_i$ = \ddd{} \lloop{} $i$ := $i$+1; \endloop;\hfill}
\ligne{\iff{} $i$ $>$ 0 \hfill}
\ligne{\hskip 10pt\tthen{} \iff{} \aa$_i$ = \xxx\hfill}
\ligne{\hskip 10pt\hskip 23pt \hskip 10pt\tthen{} \ffor{} $j$ \iin{} $[0..i$]\hfill}
\ligne{\hskip 10pt\hskip 23pt \hskip 10pt\hskip 23pt\lloop{} \aa$_j$ :=  \uu; \endloop;
\hfill}
\ligne{\hskip 10pt\hskip 23pt \hskip 10pt\hskip 23pt\aa$_{i+1}$ := \aa$_{i+1}$ + \uu;
\hfill}
\ligne{\hskip 10pt\hskip 23pt \hskip 10pt\eelse{} \aa$_0$ := \xxx;\hfill}
\ligne{\hskip 10pt\hskip 23pt \endif;\hfill}
\ligne{\hskip 10pt \eelse{} \iff{} \aa$_0$ $=$ \xxx \hfill}
\ligne{\hskip 10pt\hskip 23pt \hskip 10pt\tthen{} \aa$_0$ := \uu; 
\aa$_1$ := \aa$_1$+1;\hfill}
\ligne{\hskip 10pt\hskip 23pt \hskip 10pt\eelse{} \aa$_0$ := \aa$_0$ + \uu;\hfill}
\ligne{\hskip 10pt \hskip 23pt\endif;\hfill}
\ligne{\endif;\hfill}
}
\hfill}
\vskip 5pt
\hrule height 0.3pt depth 0.3pt width \hsize
}
\hfill}
}
\vskip 10pt
The decrementation algorithm works exactly in the opposite way: if \aa$_0$ is not \uu,
we can replace it by \aa$_0$$-$\uu. Otherwise, we look at the position~$h$ of the
leftmost item of consecutive~\uu's. We replace \aa$_{h+1}$ by \aa$_{h+1}$$-$\uu,
we replace \aa$_h$ by~\xxx{} and then we replace all \aa$_i$ from~0 up to $h$$-$1
by~\ddd: see Algorithm~\ref{anzmdecr}. Note that when \aa$_i$ is found different from
\uu{} in the \wwhile-loop, \hbox{$i>0$}, so that \xxx{} is  always written in place of
the digit at $i$$-$1. If \hbox{$i=1$} when the execution leaves the \wwhile-loop
we write \xxx{} instead of \aa$_0$ and the range of the \ffor-loop is empty. Accordingly,
Algorithm~\ref{anzmdecr} works in all cases for a positive integer~$\nu$.

\vtop{
\begin{algo}\label{anzmdecr}
Algorithm giving the $[\nu$$-$$1]_{nz}$ from~$[\nu]_{nz}$,provided that
$\nu$ is positive. Recall that $[\nu]_{nz}$ does not contain an \nzm-forbidden pattern.
Assume that \hbox{$[\nu]_{nz}$ $=$  \aa$_k$$..$\aa$_0$}.
\end{algo}
\vskip-5pt
\ligne{\hfill
\vtop{\leftskip 0pt\hsize=210pt
\hrule height 0.3pt depth 0.3pt width \hsize
\vskip 5pt
\ligne{\hfill
\vtop{\leftskip 0pt\hsize=190pt
\ligne{\iff{} \aa$_0$ $\not=$ \uu\hfill}
\ligne{\hskip 10pt \tthen{} \aa$_0$ := \aa$_0$ $-$ \uu;\hfill}
\ligne{\hskip 10pt \eelse{}  $i$ := 0;\hfill}
\ligne{\hskip 10pt\hskip 23pt \wwhile{} \aa$_i$ = \uu{} 
\lloop{} $i$ := $i$+1; \endloop;\hfill}
\ligne{\hskip 10pt\hskip 23pt \aa$_i$ := \aa$_i$ $-$ \uu; $i$ := $i$$-$1;
\aa$_i$ := \xxx;\hfill}
\ligne{\hskip 10pt\hskip 23pt \ffor{} $j$ \iin{} $[0..i$$-$$1]$ \inreverse\hfill}
\ligne{\hskip 10pt\hskip 23pt \lloop{} \aa$_j$ := \ddd; \endloop;\hfill}
\ligne{\endif;\hfill}
}
\hfill}
\vskip 5pt
\hrule height 0.3pt depth 0.3pt width \hsize
}
\hfill}
}
\vskip 10pt
It can be noted that if we provide Algorithms~\ref{anzmincr} and~\ref{anzmdecr} with
\nzm-codes, the result is again an \nzm-code in both cases.

\section{Metallic trees and their assignments}\label{assign}

In this section, we consider the notion of assignment which we define in
Sub-section~\ref{sassign} where we consider a tool to compare the assignments.
In Sub-section~\ref{spenultimate}, we focus our attention on a particular assignment,
which we call the penultimate one. In Sub-section~\ref{spreferred}, we characterise 
a property shared by the assignments, a property which we shall discover with
the penultimate one. In Sub-section~\ref{smassmid}, we look at another assignment, the
mid-assignment which allows us to have a new look on the particular assignments 
investigated in the previous sections. In Sub-section~\ref{snzmassign}, we shall 
investigate the properties of the assignments in the frame of the \nzm-codes.

\subsection{Assignments in the metallic trees}\label{sassign}

In Sub-section~\ref{strees}, we recalled the definition of the {\bf natural 
numbering} of $\cal W$ and~$\cal B$. Consider those trees.
We can see each of them as an infinite sequence ${\cal L}_n$ of finite sequences of 
numbers defined by \hbox{${\cal L}_{n+1} = \{U_n$+$1..U_{n+1}\}$}, where
$U_n=M_n$ for all~$n$ or $U_n=B_n$ for all~$n$, depending on whether we consider $\cal W$ 
or $\cal B$. In both cases, we can see the application
of the rules~$(1)$ as an application $\alpha$ which, to each node~$\nu$ of the level~$n$
associates three numbers $\ell_{\nu}$, $s_{\nu}$ and $b_{\nu}$ such that $s_{\nu}$ is 
the numeral status of~$\nu$ under~$\alpha$, $\ell_\nu$ is the leftmost node
of an interval~$I_{\nu}$ of~${\cal L}_{n+1}$ with the conditions:
\vskip 5pt
\ligne{\hfill
for all~$\nu$, \hskip 10pt $I_{\nu}\cap I_{n+1}=\emptyset$ \hskip 10pt and
\hskip 10pt
$\displaystyle{\sum\limits_{\nu\in{{\cal L}_n}}\vert I_{\nu}\vert} = m_{n+1}$,
\hfill $(13)$\hskip 10pt}
\vskip 5pt
\noindent
and $b_{\nu}$ is the position of the black node associated to~$\nu$ among the
nodes of the interval~$I_{\nu}$, the leftmost position being~1. The nodes
belonging to~$I_{\nu}$ are called the {\bf sons of~$\nu$ under~$\alpha$}, for short they 
are called sons only when it is clear which assignment is considered. For short they are 
also called {\bf $\alpha$-sons}. The conditions $(13)$ can equivalently be stated as:
\vskip 5pt
\ligne{\hfill
$\vcenter{\vtop{\leftskip 0pt\hsize=275pt
\ligne{for all~$\nu$ with $\nu\in{\cal L}_n$,
$\alpha(\nu) = (\alpha_\ell(\nu),\alpha_b(\nu),\alpha_s(\nu))$,\hfill}
\ligne{with $\alpha_{\ell}(\nu)\in{\cal L}_{n+1}$, $\alpha_s(\nu)\in\{0,1\}$,
$\alpha_b(\nu)\in\{1..p$$-$$2$$-$$s_{\nu}\}$,\hfill}
\ligne{for all $\nu$, 
$\displaystyle{\sum\limits_{k=\alpha_{\ell}(\nu)}^{\alpha_{\ell}(\nu+1)-1}%
\alpha_s(\nu)=1}$\hskip 5pt
and, for any $\nu\in[M_{n-1}$+$1..M_n]$,\hfill}
\ligne{for any positive $\nu$, $\alpha_s(\alpha_b(\nu))=1$, 
\hfill}
\ligne{$\alpha_{\ell}(\nu+1)=\alpha_{\ell}(\nu)+p$$-$2$-$$\alpha_s(\nu)$,
and $\alpha_{\ell}(M_n)=M_{n+1}-p$+$3$+$\alpha_s(M_n)$.\hfill}
}}$
\hfill $(14)$\hskip 10pt}
\vskip 10pt
We call {\bf assignment} an application~$\alpha$ which satisfies~$(14)$. 
We denote by ${\cal W}_{\alpha}$ the white metallic tree $\cal W$ dotted with the
assignment~$\alpha$: it means that, starting from the root, the status of each node~$\nu$
under~$\alpha$ is defined by $\alpha_s(\nu)$ and that the position of the black son
of~$\nu$ among its $\alpha$-sons is defined by $\alpha_b(\nu)$.
When $\alpha$ is associated with the rules~$(1)$, we additionally have that
\hbox{$\alpha_b(\nu) = 1$} for all node~$\nu$. That assignation is called
the {\bf leftmost assignment}. It was called the {\bf standard assignment} 
in~\cite{mmJUCStools} which considers white metallic trees only in the case
when \hbox{$p=5$}. In~\cite{mmgsJUCS}, another assignment 
was considered, defined by:
\vskip 5pt
\ligne{\hfill $\alpha(\nu) = (\ell_{\nu},p$$-$3$-$$sn(\nu),sn(\nu))$ for all $\nu$. 
\hfill $(15)$\hskip 10pt}
\vskip 5pt
It is not difficult to see that whether $\ell s(\nu)$ is~\bbb{} or \www, the black
son is the penultimate son of~$\nu$. For this reason, we call $(15)$ 
the {\bf penultimate assignment}. Similarly, we define the {\bf rightmost assignment}
by
\vskip 5pt
\ligne{\hfill $\alpha(\nu) = (\ell_{\nu},p$$-$2$-$$sn(\nu),sn(\nu))$ for all $\nu$. 
\hfill $(16)$\hskip 10pt}
\vskip 5pt
   
   Say that an assignment~$\alpha$ is an {\bf \aa-assignment} if and only if for
any node~$\nu$, one of its sons exactly has \aa{} as its signature. It means that
for one son of~$\nu$ and for one of them only, its code has \aa{} among its suffixes. 
We say that an assignment $\alpha$ has the {\bf preferred son property} if,
for any node~$\nu$, exactly one of its sons has the code {\bf [$\nu$]\zzz}. 
Note that an assignment which possesses the preferred son property is also
a \zzz-assignment. Accordingly, the preferred son property assumes that we consider
the representation of the numbers by their metallic codes.

    Say that an assignment~$\alpha$ is a {\bf \bbb-\aa-assignment} if and only if
all black nodes of the tree and only them have \aa{} as their signature. We 
can note that the notion of \bbb-\aa-assignment is meaningful also in the case
of the representation of the numbers by their \nzm-code when it exists. We shall
see a bit later that there are many values of~\aa{} for which the \bbb-\aa-assignment
exists.

    In order to establish the property characterised in Sub-section~\ref{spreferred},
we consider the following tool which measures the distance between two assignments
as follows. Let $\alpha$ and~$\beta$ be two assignments of the white metallic tree.
Call {\bf apartness between $\alpha$ and~$\beta$} denoted by 
\hbox{$\delta_{\alpha\beta}$} the function defined by
\hbox{$\delta_{\alpha\beta}(\nu)=\beta_{\ell}(\nu)-\alpha_{\ell}(\nu)$} for any
node~$\nu$ of~$\cal W$. We have the easy property:

\begin{lemm}\label{lapart}
Let $\alpha$, $\beta$ and~$\gamma$ be three assignments on the white metallic tree.
For any node~$\nu$ of~$\cal W$ we have: 
\vskip 5pt
\ligne{\hfill $\delta_{\alpha\beta}(\nu)=
\delta_{\gamma\beta}(\nu)-\delta_{\gamma\alpha}(\nu)$. \hfill$(17)$\hskip 10pt}
\end{lemm}

   Consider the metallic codes which are associated to the numbers by~$(6)$, see 
Sub-section~\ref{smetalcodes}. Consider the metallic codes of the nodes which lie
on~${\cal L}_1$. One of them only has the signature~\zzz: it is the node numbered by
$p$$-$2 whose metallic code is \uu\zzz. Consider the nodes on ${\cal L}_2$.
Their numbers grow from $M_1$+1 up to~$M_2$ and the metallic codes go from
\uu\numd{} to \uu\uu\uu. The nodes whose signature is~\zzz{} are:
\numd\zzz, \numt\zzz, ..., \ccc\zzz, \ddd\zzz, \uu\zzz\zzz, \uu\uu\zzz. We can see that
the distance between two consecutive such nodes is $p$$-$2 except for \ddd\zzz{}
with \uu\zzz\zzz{} whose distance is $p$$-$3. More generally, call {\bf \zzz-node}
any node of~$\cal W$ whose signature is~\zzz. On ${\cal L}_{n+1}$, the \zzz-nodes
run from \uu$^{n-1}$\numd\zzz{} up to \uu$^{n+1}$\zzz. Note that if we erase the last 
digit of those codes, we get the codes from \uu$^{n-1}$\numd{} up to \uu$^{n+1}$, 
{\it i.e.} the metallic codes of the nodes on ${\cal L}_n$. Accordingly,
the number of \zzz-nodes on ${\cal L}_{n+1}$ is the number of nodes on~${\cal L}_n$.
Moreover, we can observe that the distance between two consecutive \zzz-nodes on
a level is $p$$-$2 of $p$$-$3. When it is $p$$-$3? On the level ${\cal L}_2$,
the distance $p$$-$3 occurs between \ddd\zzz{} and \uu\zzz\zzz. Indeed, 
\hbox{\uu\zzz\zzz$\ominus$\uu $=$ \ddd\ccc} and the distance between \ddd\zzz{}
and \ddd\ccc{} is $p$$-$4. More generally, we can state:

\setbox110=\hbox{\bf\small +}
\setbox120=\hbox{\bf\small -}
\begin{lemm}\label{lzzdist}
On the level~${\cal L}_n$ let $\mu$ and~$\nu$ be two consecutive \zzz-nodes,
with \hbox{$\mu < \nu$}. Then \hbox{$\nu-\mu = p$$-$$3$} if and only if
\hbox{\bf [$\nu$] $=$ [$\omega$]\zzz$^k$} with \hbox{$k\geq2$}. When it is not the
case, \hbox{$\nu-\mu = p$$-$$2$}.
\end{lemm}

\noindent
Proof. The decrementation algorithm tells us that 
\hbox{\bf [[$\omega$]\zzz$^k$$\ominus$\uu] $=$ [[$\omega$]$\ominus$\uu]\ddd\ccc$^{k-1}$} 
and the distance between that latter metallic code and 
\hbox{\bf [[$\omega$]$\ominus$\uu]\ddd\ccc$^{k-2}\zzz$}, the metallic code of the previous
\zzz-node
is $p$$-$4 so that the distance on ${\cal L}_n$ between \hbox{\bf [$\omega$]\zzz$^k$}
and the previous \zzz-node is $p$$-$3. If \hbox{\bf [$\nu$] = [$\omega$]\zzz}, assume
that \hbox{[$\omega$] = [$\omega_1$]\ddd\ccc$^k$} with \hbox{$k>0$}. Then,
\hbox{[$\omega_1$\ddd\ccc$^k$\zzz]$\ominus$\uu = 
[$\omega_1$]\ddd\ccc$^{k-1}$\ccc$^{\copy120}$\ddd}, 
where \hbox{\ccc$^{\copy120}$ $=$ \ccc$$-$$\uu}. That latter metallic code
does not contain the forbidden pattern. And so, the distance from that
node to the previous \zzz-node is $p$$-$3, so that \hbox{$\nu-\mu = p$$-$2}.
Now, if \hbox{\bf [$\omega$]} does not contain the suffix \ddd\ccc$^+$,
the last digit of $\omega$ which is greater than \zzz{} can be reduced by~\uu{}
so that in that case too \hbox{$\nu - \mu = p$$-$2}.\hfill $\Box$

Is there a connection between this two values
between two consecutive \zzz-nodes and the smaller occurrence of the smaller distance
with the distinction between white and black nodes which have $p$$-$2 and $p$$-$3
nodes respectively?

   That issue is addressed by the next sub-section.

\subsection{The penultimate assignment}\label{spenultimate}

   Say that an assignment $\alpha$ possesses the {\bf preferred son} property
if and only if for any node~$\nu$ of~$\cal W$, the signature of one of its sons
under~$\alpha$ and one of them only is \zzz{} and if the metallic code of that son
is \hbox{\bf [$\nu$]\zzz}. When an assignment $\alpha$ possesses the preferred son
property, for any node~$\nu$, the node whose signature is [$\nu$]\zzz{} is called
its {\bf preferred son} under~$\alpha$. Note that if $\alpha$ and~$\beta$ are
two assignments which possess the preferred son property, for each node, the
preferred son under~$\alpha$ and that under~$\beta$ coincide.

 We can state:

\begin{thm}\label{tprefpenult}
The penultimate assignment possesses the preferred son property and it is the
\bbb-\zzz-assignment.
\end{thm}

\noindent
Proof. It is based on the following property on the signatures and on the metallic codes
of the sons with respect to those of the node.
\def\llaligne #1 #2 #3 #4 #5 {%
\hsize=260pt
\ligne{\hfill
\hbox to 40pt{\hfill #1\hfill}
\hbox to 50pt{\hfill #2\hfill}
\hbox to 30pt{\hfill #3\hfill}
\hbox to 100pt{\hfill #4\hfill}
\hbox to 20pt{\hfill {\footnotesize\bf #5}\hfill}
\hfill}
}

\def\fnb #1{\hbox{\footnotesize\bf #1}}
\begin{lemm}\label{lpenultsgnmcodes}
Let $\nu$ be a node of~$\cal W$ equipped with the penultimate assignment~$\pi$. 
The signatures of its sons under~$\pi$ is defined by the following rules:
\vskip 5pt
\ligne{\hfill
$\vcenter{\vtop{\leftskip 0pt\hsize=280pt
\ligne{\hfill \bbb\zzz{} $\rightarrow$ 
\www\numd$($\www\aa$)^{p-7}$\www\ccc.\bbb\zzz.\www\uu,
\www\aa{} $\rightarrow$ \www\numd$($\www\aa$)^{p-6}$\www\ddd.\bbb\zzz.\www\uu, \hfill}
}}$
\hfill$(18)$\hskip 10pt}
\vskip 5pt
The metallic codes of the sons of~$\nu$ under~$\pi$ are given by the following table,
where \hbox{\aa$_k$$..$\aa$_1$\aa$_0$ $\rightleftharpoons$ $[\nu]$} and 
\hbox{\bbb$_k$$..$\bbb$_0$ $\rightleftharpoons$	
$[$$[$\aa$_k$$..$\aa$_0]$$\ominus$\uu$]$} and \aa{} is in
\hbox{$\{$\uu$..$\ddd$\}$} in lines~\fnb 4 to~\fnb 6.
\vskip 5pt
\ligne{\hfill
$\vcenter{\vtop{\leftskip 0pt\hsize=280pt
\llaligne {$\nu$} {range} {son} {metallic code} {ref.}
\vskip 3pt
\hrule height 0.3pt depth 0.3pt width \hsize 
\vskip 3pt
\llaligne {\bbb\zzz} {$1..p$$-$$5$} {$h$} {\bbb$_k$$..$\bbb$_0$\hhh$^{\copy110}$} {1}
\llaligne {} {$p$$-$$4$} {} {\aa$_k$$..$\aa$_0$\zzz} {2}
\llaligne {} {$p$$-$$3$} {} {\aa$_k$$..$\aa$_0$\uu} {3}
\llaligne {\www\aa} {$1..p$$-$$4$} {$h$} {\bbb$_k$$..$\bbb$_0$\hhh$^{\copy110}$} {4}
\llaligne {} {$p$$-$$3$} {} {\aa$_k$$..$\aa$_0$\zzz} {5}
\llaligne {} {$p$$-$$2$} {} {\aa$_k$$..$\aa$_0$\uu} {6}
\vskip 3pt
\hrule height 0.3pt depth 0.3pt width \hsize 
}}$
\hfill$(19)$\hskip 10pt}
\vskip 5pt
\end{lemm}

Clearly, Theorem~\ref{tprefpenult} follows from Lemma~\ref{lpenultsgnmcodes}.

\noindent
Proof of Lemma~\ref{lpenultsgnmcodes}. We perform it by complete induction on~$\nu$. 
The lemma is clearly true for the root. It is applied the rule \www\uu{} of~$(18)$
and its sons satisfies the lines \fnb {4}, \fnb {5} and~\fnb {6} of~$(19)$ as 
here \hbox{\bbb $=$ \zzz}.
Assume that the lemma is proved up to~$\nu$ which is on the level~$n$+1.
Let $\lambda_{k+1}$ be the leftmost node of~${\cal L}_{k+1}$, so that
\hbox{$\lambda_{k+1}=M_k+1$}. Clearly, \hbox{\bf [$\lambda_{k+1}$] $=$ \uu$^k$\numd},
so that recursively applying the incrementation algorithm to the code \uu$^{n+1}\numd$, 
which is \hbox{\bf [$\lambda_{n+2}$]}, the leftmost son of $\lambda_{n+1}$,
the sons of~$\lambda_{n+1}$ have the metallic codes given by the lines~\fnb {4}, \fnb {5}
and~\fnb {6} of~$(19)$. 

Consider that $\nu$ is a white node. If its signature is~\aa{} with
\hbox{\aa{} $<$ \ccc}, then the induction hypothesis entails that the metallic
code of the leftmost son of~$\nu$+1 is [$\nu$]\numd. Accordingly, the incrementation 
algorithm ensures that the metallic codes of the sons of~$\nu$+1 satisfy the lines
\fnb 4, \fnb 5 and~\fnb 6 of~$(19)$. If the signature of~$\nu$ is \ccc{} and if
that of~$\nu$+1 is \ddd, we can apply the same argument as when \hbox{\aa{} $<$ \ccc}.
Assume that the signature of~$\nu$+1 is~\zzz. We know from Lemma~\ref{lzzdist} that 
\hbox{[$\nu$]= [$\omega$]\ddd\ccc$^{k+1}$}, so that its rightmost son~$\rho$
satisfies \hbox{[$\rho$] = [$\nu$]\uu}. Accordingly, the leftmost son~$\lambda$ 
of~$\nu$+1 is $\rho$+1 whose metallic code is \hbox{[$\nu$\numd]}. Applying
consecutively $p$$-$6 times the incrementation algorithm, we have
that \hbox{[$\lambda$+$p$$-$6] = $[\nu]$\ccc = [$\omega$]\ddd\ccc$^{k+2}$}.
Accordingly, the metallic code of the next two sons of~$\nu$+1 are
\hbox{[$\nu$+1]\zzz} and \hbox{[$\nu$+1]\uu}. Now, due to the signature of~$\nu$, 
$\nu$ and~$\nu$+1 have the same father which belongs to ${\cal L}_n$ so that the
rules of~$(18)$ apply which means that $\nu$+1 is a black node. Counting the metallic 
nodes we computed for the sons of~$\nu$+1, we get that we have $p$$-$3 of them so that
we have all the sons of~$\nu$+1 and their signatures satisfy~$(18)$ and their metallic
codes are conformal to the lines~\fnb 4, \fnb 5 and~\fnb 6 of~$(19)$.

We remain with the case when $\nu$ is a black node. We have that 
\hbox{[$\nu$] = [$\omega$]\zzz}, so that the signature of~$\nu$+1 is \uu{} as
required. From the induction hypothesis, the metallic code of the rightmost son of~$\nu$
is \hbox{[$\omega$]\zzz\uu}, so that the metallic code of the leftmost son of~$\nu$+1
is \hbox{[$\omega$]\zzz\numd}. Iterating $p$$-$3 times the incrementing algorithm, we
get that the metallic codes of the $p$$-$2 sons of~$\nu$+1 satisfy~$(19)$ as the
metallic code of its penultimate son will be \hbox{[$\omega$]\uu\zzz{} = [$\nu$+1]\zzz}.
\hfill $\Box$.

   We can now establish another property. Consider the metallic tree $\cal W$
and a tree~$T$. Assume that both trees are isomorphic. If we equip $\cal W$ with
the assignment $\alpha$, we can endow $\cal T$ with a way of defining the branchings
of the tree in accordance with the rules entailed by~$\alpha$ on~$\cal W$ by
putting the same branchings on~$T$ by the isomorphism. 
Let $\alpha$ be an assignment which possesses the preferred son property.
Consider the set~$\cal T$ of the \zzz-nodes of~$\cal W$. We define a tree structure on
$\cal T$ as follows. Take as root the node whose metallic code is \uu\zzz. Assuming
that we defined the level~$n$ of $\cal T$, we define the level~$n$+1 as follows.
Let $\nu$ be a node on the level~$n$ of $\cal T$. Let $\mu$ be the father of~$\nu$
in~$\cal W$. We say that the status of~$\nu$ is that of~$\mu$. Moreover,
each son of~$\mu$ has a preferred son~$\pi$ on the level~$n$+1: it is a \zzz-node
which, by definition belongs to~$\cal T$. We define those nodes $\pi$ as the sons 
of~$\nu$. Note that the nodes~$\pi$ and $\nu$ belong to the subtree of~$\cal W$
rooted at~$\mu$. It is not difficult to see that in this way we construct an isomorphism
from~$\cal W$ onto~$\cal T$ and that isomorphism transports $\alpha$ as an assignment 
on~$\cal T$ which coincide with the definition of the status of the nodes of~$\cal T$
which we above indicated. We proved:

\begin{thm}\label{tpenzzisom}
Consider the application $\varphi$ such that 
\hbox{$\varphi($$[\nu]$\zzz$) = ([$$\nu$$])$}, defining a bijection of~$\cal W$ on the
set $\cal T$ of the \zzz-nodes of $\cal W$. 
Define the sons of a node~$\nu$ of~$\cal T$ 
as the \zzz-nodes of~$\cal W$ which are in the subtree~$\cal S$ of $\cal W$ rooted
at $($$[\nu]$\zzz$)$ and which are on the level~$2$ of~$\cal S$. Define the status 
of~$\nu$ in $\cal T$ as the status of~$($$[\nu]$\zzz$)$. Then,
the bijection~$\varphi$ defines an isomorphism of $\cal W$ onto~$\cal T$ which 
transports the assignment~$\alpha$ onto~$\cal T$ whatever $\alpha$.
\end{thm}

\subsection{Assignments and the preferred son property}\label{spreferred}

   In this subsection, we shall see that all assignments on the white metallic
tree possess the preferred son property.

To that purpose, we compare the assignments to the same one: the leftmost 
assignment which we denote by~$\lambda$.
The reason of this choice lies in the following property:

\begin{lemm}\label{lcompassign}
For any node $\nu$ of~$\cal W$, we have:
\vskip 5pt
\ligne{\hfill $0\leq \delta_{\lambda\alpha}(\nu)\leq 1$ \hfill$(20)$\hskip 10pt}
\end{lemm}

\def\statutdiff #1 {%
\hbox{$st_{\lambda}(#1) \not= st_{\alpha}(#1)$}}
\def\statuteq #1 {%
\hbox{$st_{\lambda}(#1) = st_{\alpha}(#1)$}}
\def\statutsous #1 en #2 is #3 {%
\hbox{$st_{#1}(#2) = #3$}}
We recall that $st_{\alpha}(\nu)$ is the status of the node~$\nu$ under~$\alpha$.
It will be easier to prove the lemma, once we have proved the following one.

\begin{lemm}\label{lecartassign}
Let $\lambda$ be the leftmost assignment on $\cal W$ and let $\alpha$ be an assignment 
on $\cal W$. If \statuteq {\nu} {} then,
\hbox{$\delta_{\lambda\alpha}(\nu$$+$$1)=\delta_{\lambda\alpha}(\nu)$}. Otherwise, if 
\statutsous {\lambda} en {\nu} is {\www} {} 
then \hbox{$\delta_{\lambda\alpha}(\nu$$+$$1)=\delta_{\lambda\alpha}(\nu)$$-$$1$} and if
\statutsous {\lambda} en {\nu} is {\bbb} {},
then \hbox{$\delta_{\lambda\alpha}(\nu$$+$$1)=\delta_{\lambda\alpha}(\nu)$$+$$1$}. 
\end{lemm}

\noindent
Proof of Lemma~\ref{lecartassign}. We set \hbox{$\ell s_{\alpha}(\nu)=1$} if
\hbox{$st_{\alpha}(\nu)=\bbb$} and \hbox{$\ell s_{\alpha}(\nu)=0$} if
\hbox{$st_{\alpha}(\nu)=\www$}. Then 
\hbox{$\alpha_{\ell}(\nu$+$1)=\alpha_{\ell}(\nu)$+$p$$-$2$-$$\ell{}s_{\alpha}(\nu)$}.
Considering the four cases raised by the different distributions between $\lambda$ 
and~$\alpha$ for defining the leftmost son of~$\nu$ and of~$\nu$+1 we get the conclusion 
of the lemma.\hfill $\Box$

\noindent
Proof of Lemma~\ref{lcompassign}. The lemma is of course true for the root. 
However, we shall prove a stronger property. For any node~$\nu$ of~$\cal W$,
we have:
\vskip 5pt
\ligne{\hfill
$\delta_{\lambda\alpha}(\alpha_{\ell})(\nu)=
\delta_{\lambda\alpha}(\alpha_{\ell})(\nu$+$1)=0$
\hfill $(21)$\hskip 10pt}
\vskip 5pt
Note that property $(21)$ says something about the sons of~$\nu$, not about $\nu$
itself. We prove the property by induction on~$\nu$. First, we have to prove
the property for the root, which means that we have to compute 
$\delta_{\lambda\alpha}$ for~\hbox{$\nu$ = \numd} up to \hbox{$\nu$ = \uu\uu}.
Consider the node \numd{} which is $\lambda$-black. Its $\lambda$-sons are 
\uu\numd, \uu\numt, up to \numd\zzz{} included: we check that there are $p$$-$3 nodes 
as \numd{} is $\lambda$-black. If \numd{} is $\alpha$-black too 
\hbox{$\delta_{\lambda\alpha}(\numt) = 0$}, from 
Lemma~\ref{lecartassign}. Now, as a node has a single black node, whatever the assignment
we can see that on $\cal W$, the other sons of the root up to~\uu\uu{} are white nodes 
under both~$\lambda$ and~$\sigma$ so that 
\hbox{$\delta_{\lambda\alpha}(\nu) = 0$} for all $\nu$ up to \uu\uu{} included. 
If \numd{} is an $\alpha$-white node,
\hbox{$\delta_{\lambda\alpha}(\numt) = 1$} and 
\hbox{$\delta_{\lambda\alpha}(\nu) = 1$} as long as $\nu$ is $\alpha$-white 
from \numd{} until we meet the $\alpha$-black son of the root, say \kkk.
Accordingly,
\hbox{$\delta_{\lambda\alpha}(\kkk) = 1$} but, as \kkk{} is $\lambda$-white,
we get that
\hbox{$\delta_{\lambda\alpha}(\kkk$+$1) = 0$} from Lemma~\ref{lecartassign} 
and 
\hbox{$\delta_{\lambda\alpha}(\nu) = 0$} for the further values of~$\nu$ until \uu\uu{}
if any. Note that if \hbox{\kkk{} = \uu\uu}, we already know that
\hbox{$\delta_{\lambda\alpha}(\numd) = 0$} which completes our proof of~$(21)$ for the
root. The same argument holds if we replace $\nu$ by an $\alpha$-black node 
assuming that 
\hbox{$\delta_{\lambda\alpha}(\lambda_{\ell}(\nu)) = 0$}. 

We remain with the case when $\nu$ is $\alpha$-white. From the induction hypothesis
applied to~$\nu$$-$1, we know that
\hbox{$\delta_{\lambda\alpha}(\lambda_{\ell}(\nu)) = 0$}: the leftmost son of~$\nu$
is the same under whatever~$\lambda$ or~$\alpha$. If $\nu$ is $\alpha$-white 
too, the above argument can be repeated: if $\kappa$ is the $\alpha$-black son
of~$\nu$, we necessarily have that 
\hbox{$\kappa < \lambda_{\ell}(\nu$+$1)$}.  If \hbox{$\kappa=\lambda_{\ell}(\nu)$}, there 
is nothing to prove and \hbox{$\delta_{\lambda\alpha}(\sigma) = 0$} for all sons of~$\nu$
under both~$\lambda$ and~$\alpha$. If \hbox{$\kappa\not=\lambda_{\ell}(\nu)$},
as \hbox{$\kappa < \lambda_{\ell}(\nu$+1$)$}, the above argument shows us that
\hbox{$\delta_{\lambda\alpha}(\sigma)$} takes the value~1 on 
\hbox{$\sigma=\lambda_{\ell}(\nu)$+1}, keeping that value until \hbox{$\sigma=\kappa$}
and then, takes back the value~0 from~$\kappa$+1 until the rightmost son of~$\nu$.

If $\nu$ is a $\lambda$-white node and an $\alpha$-black one,
as \hbox{$\kappa\leq \alpha_{\ell}(\nu)<\lambda_{\ell}(\nu)$}, the same argument holds
proving that \hbox{$\delta_{\lambda\alpha}(\alpha_{\ell}(\nu)$+$1)$=0}.
Accordingly, the proof of Lemma~\ref{lecartassign} is completed.\hfill $\Box$

\begin{thm}\label{tpref}
	Any assignment~$\alpha$ on $\cal W$ do possess the preferred son property. In 
${\cal W}_{\alpha}$, whatever~$\alpha$, we have that $m_{n+1}$ is the preferred son 
of~$m_n$. Call the sequence of nodes $\{m_n\}_{n\in\mathbb N}$ the {\bf \zzz-branch}.
\end{thm}

\noindent
Proof. First, we prove that~$\lambda$, the leftmost assignment, possesses the preferred
son property, by proving an analog of the rules~$(18)$ and Table~$(19)$ for~$\lambda$.
We have:

\begin{lemm}\label{lleftmostsgnmcodes}
Let $\nu$ be a node of~$\cal W$ equipped with the leftmost assignment~$\lambda$. 
The signatures of its sons under~$\lambda$ is defined by the following rules:
\vskip 5pt
\ligne{\hfill
$\vcenter{\vtop{\leftskip 0pt\hsize=280pt
\ligne{\hfill \bbb\uu,\bbb\numd{} $\rightarrow$ 
\bbb\numd$($\www\aa$)^{p-6}$\www\ddd.\www\zzz,
\www\aa{} $\rightarrow$ \bbb\uu$($\www\aa$)^{p-5}$\www\ddd.\www\zzz, \hfill}
\ligne{\hfill
\www\zzz{} $\rightarrow$ \bbb\uu$($\www\aa$)^{p-6}$\www\ccc.\www\zzz.\www\uu, 
\www\uu{} $\rightarrow$ \bbb\numd$($\www\aa$)^{p-6}$\www\ddd.\www\zzz.\www\uu. \hfill}
}}$
\hfill$(22)$\hskip 10pt}
\vskip 5pt
The metallic codes of the $\lambda$-sons of~$\nu$ are given by the following table,
where \hbox{\aa$_k$$..$\aa$_1$\aa$_0$ $\rightleftharpoons$ $[\nu]$} and 
\hbox{\bbb$_k$$..$\bbb$_0$ $\rightleftharpoons$	
$[$$[$\aa$_k$$..$\aa$_0]$$\ominus$\uu$]$}, and in lines~\fnb {9} and~\fnb {10},
\aa{} is in \hbox{$\{$\numd{} $..$ \ddd$\}$}.
\vskip 5pt
\ligne{\hfill
$\vcenter{\vtop{\leftskip 0pt\hsize=280pt
\llaligne {$\nu$} {range} {son} {metallic code} {ref.}
\vskip 3pt
\hrule height 0.3pt depth 0.3pt width \hsize 
\vskip 3pt
\llaligne {\bbb\uu$,$\bbb\numd} {$1..p$$-$$4$} {$h$} 
{\bbb$_k$$..$\bbb$_0$\hhh$^{\copy110}$} {1}
\llaligne {} {$p$$-$$3$} {} {\aa$_k$$..$\aa$_0$\zzz} {2}
\llaligne {\www\zzz} {$1..p$$-$$4$} {$h$} {\bbb$_k$$..$\bbb$_0$\hhh} {3}
\llaligne {} {$p$$-$$3$} {} {\aa$_k$$..$\aa$_0$\zzz} {4}
\llaligne {} {$p$$-$$2$} {} {\aa$_k$$..$\aa$_0$\uu} {5}
\llaligne {\www\uu} {$1..p$$-$$4$} {$h$} {\bbb$_k$$..$\bbb$_0$\hhh$^{\copy110}$} {6}
\llaligne {} {$p$$-$$3$} {} {\aa$_k$$..$\aa$_0$\zzz} {7}
\llaligne {} {$p$$-$$2$} {} {\aa$_k$$..$\aa$_0$\uu} {8}
\llaligne {\www\aa} {$1..p$$-$$3$} {$h$} {\bbb$_k$$..$\bbb$_0$\hhh} {9}
\llaligne {} {$p$$-$$2$} {} {\aa$_k$$..$\aa$_0$\zzz} {10}
\vskip 3pt
\hrule height 0.3pt depth 0.3pt width \hsize 
}}$
\hfill$(23)$\hskip 10pt}
\vskip 5pt
\end{lemm}

\noindent
Proof of the lemma. We know from Theorem~\ref{tprefpenult} that the \zzz-nodes coincide
with the black nodes under~$\pi$, the penultimate assignment. Clearly, under~$\lambda$,
the root obeys the rule~\www\uu{} of~$(22)$ and the metallic codes of its $\lambda$-sons
satisfy the lines~\fnb 6, \fnb 7 and~\fnb 8 of~$(23)$. Consider the nodes on the 
level~$n$ of~$\cal W$ under~$\lambda$ and compare the intervals assigned by~$\lambda$
with those assigned by~$\pi$. The lefmost node of the level is a $\lambda$-black node
while it is an $\alpha$-white one. Accordingly, the first \zzz-node of the level~$n$+1 is
the rightmost $\lambda$-son of the first node on the level~$n$.
From our previous studies about the positions of the \zzz-nodes on a level,
we have that the next \zzz-nodes on the level~$n$+1 are the $\lambda$-rightmost sons 
of the first nodes~$\nu$ of the level~$n$ whose signatures run from~\numt{} until \ddd.
The signature of the next node is \zzz{} so that the rule~\bbb\numd{} is observed by
the $\lambda$-sons of~$\nu$. The incrementation algorithm applied to the
$\lambda$-rightmost son of the leftmost node on the level~$n$ so that the rules~\www\aa{}
apply to~$\nu$+1, and the lines~\fnb 9 and~\fnb {10} of~$(23)$ are observed by the
metallic codes of the $\lambda$-sons of~$\nu$+1. Clearly, what the argument can be 
repeated when going from a $\lambda$-white node whose signature is~\aa{} 
with \hbox{\uu $<$ \aa{} $<$ \ddd}, to the new node whose signature is~\aa$\oplus$\uu.
Applying the induction hypothesis on a node whose signature is~\ddd{} and iteratively
applying the incrementation algorithm, we obtain that $\lambda$-white nodes whose
signature is~\zzz{} and~\uu{} are applied the rules~\www\zzz{} and~\www\uu{}
respectively and that the metallic codes of their $\lambda$-sons are those indicated
by the lines~\fnb 3 up to~\fnb 8 included of Table~$(23)$. The proof of
Lemma~\ref{lleftmostsgnmcodes} is completed. \hfill $\Box$

Note that the rules can be identified by their left-hand side part, which we shall do
later on. We also have:

\begin{lemm}\label{lantedigit}
Let $\nu$ be a node of~$\cal W$ and let $\alpha$ be an assignment among the leftmost, 
the penultimate and the rightmost ones. Let $\sigma$ be the leftmost son of~$\nu$
under~$\alpha$. If {$[\omega]$\aa} is the code of~$\nu$, then we have that
\hbox{$[\sigma]$ = $[\omega_1]$\aa$^-${\bf u}}, where
\hbox{$\omega_1 = \omega$} if \hbox{\aa{} $>$ \zzz} and 
\hbox{$[\omega_1] = [\omega]\ominus$\uu} otherwise and 
\hbox{{\bf u} $\in \{\uu,\numd,\numt\}$}.  More exactly,
\hbox{{\bf u} = \numt} when $\alpha$ is the rightmost assignment and $\nu$ is
black or is a \www\uu-node;
\hbox{{\bf u} = \numd} when $\nu$ is the root; when $\alpha$ is the penultimate
assignment;
when $\alpha$ is the leftmost assignment and then $\nu$ is black or is a \www\uu-node;
when $\alpha$ is the rightmost assignment and then $\nu$ is a \www\zzz-node or
a \www\aa-node with \hbox{\aa $\not=$ \uu}.
\end{lemm}

\noindent 
Proof of the lemma. Lemma~\ref{lantedigit} is a reformulation of 
Lemmas~\ref{lpenultsgnmcodes} and~\ref{lleftmostsgnmcodes} for what are
the penultimate and the leftmost assignments respectively. We consider also the
rightmost assignment. For that assignment, we have the following rules and metallic 
codes for the sons of a node, with the same notations as in 
Lemma~\ref{lleftmostsgnmcodes}:

\setbox130=\hbox{\bf\small ++}
\vskip 5pt
\ligne{\hfill
$\vcenter{\vtop{\leftskip 0pt\hsize=280pt
\ligne{\hfill \bbb\uu,\bbb\numd{} $\rightarrow$ 
\www\numt$($\www\aa$)^{p-7}$\www\ddd.\www\zzz.\bbb\uu,
\www\aa{} $\rightarrow$ \www\numd$($\www\aa$)^{p-6}$\www\ddd.\www\zzz.\bbb\uu, \hfill}
\ligne{\hfill
\www\zzz{} $\rightarrow$ \www\numd$($\www\aa$)^{p-7}$\www\ccc.\www\zzz.\www\uu.\bbb\numd, 
\www\uu{} $\rightarrow$ \www\numt$($\www\aa$)^{p-7}$\www\ddd.\www\zzz.\www\uu\bbb\numd. \hfill}
}}$
\hfill$(24)$\hskip 10pt}
\vskip 5pt
\ligne{\hfill
$\vcenter{\vtop{\leftskip 0pt\hsize=280pt
\llaligne {$\nu$} {range} {son} {metallic code} {ref.}
\vskip 3pt
\hrule height 0.3pt depth 0.3pt width \hsize 
\vskip 3pt
\llaligne {\bbb\uu$,$\bbb\numd} {$1..p$$-$$5$} {$h$} 
{\bbb$_k$$..$\bbb$_0$\hhh$^{\copy130}$} {1}
\llaligne {} {$p$$-$$4$} {} {\aa$_k$$..$\aa$_0$\zzz} {2}
\llaligne {} {$p$$-$$3$} {} {\aa$_k$$..$\aa$_0$\uu} {3}
\llaligne {\www\zzz} {$1..p$$-$$5$} {$h$} {\bbb$_k$$..$\bbb$_0$\hhh$^{\copy110}$} {4}
\llaligne {} {$p$$-$$4$} {} {\aa$_k$$..$\aa$_0$\zzz} {5}
\llaligne {} {$p$$-$$3$} {} {\aa$_k$$..$\aa$_0$\uu} {6}
\llaligne {} {$p$$-$$2$} {} {\aa$_k$$..$\aa$_0$\numd} {7}
\llaligne {\www\uu} {$1..p$$-$$5$} {$h$} {\bbb$_k$$..$\bbb$_0$\hhh$^{\copy130}$} {8}
\llaligne {} {$p$$-$$4$} {} {\aa$_k$$..$\aa$_0$\zzz} {9}
\llaligne {} {$p$$-$$3$} {} {\aa$_k$$..$\aa$_0$\uu} {10}
\llaligne {} {$p$$-$$2$} {} {\aa$_k$$..$\aa$_0$\numd} {11}
\llaligne {\www\aa} {$1..p$$-$$3$} {$h$} {\bbb$_k$$..$\bbb$_0$\hhh$^{\copy110}$} {12}
\llaligne {} {$p$$-$$2$} {} {\aa$_k$$..$\aa$_0$\zzz} {13}
\llaligne {} {$p$$-$$2$} {} {\aa$_k$$..$\aa$_0$\uu} {14}
\vskip 3pt
\hrule height 0.3pt depth 0.3pt width \hsize 
}}$
\hfill$(25)$\hskip 10pt}
\vskip 5pt
Note that \aa{} is in \hbox{$\{$\numd$..$\ddd$\}$} in lines~\fnb {12} to~\fnb {14}.

Let $\rho$~denote the rightmost assignment. From the definition, it seems close to the 
penultimate one: the $\rho$-black son is always the last one among the $\rho$-sons of a
node. This explains that the rule \www\aa{} of~$(18)$ is transformed to that
of~$(24)$ by exchanging the status of the nodes with signatures~\zzz{} and~\uu{}
accordingly. The change in the other rules is a bit more complex. However, the proof 
of~$(24)$ and~$(25)$ is very similar to those of Lemmas~\ref{lpenultsgnmcodes} 
and~\ref{lleftmostsgnmcodes}: it is based on the consideration of the penultimate 
assignment, the \zzz-nodes and the iterated applications of the incrementation 
algorithm. This completes the proof of Lemma~\ref{lantedigit} which synthesizes the 
Tables~$(19)$, $(23)$ and~$(25)$.
\hfill $\Box$

\noindent
Proof of Theorem~\ref{tpref}. Consider an assignment~$\alpha$ on~$\cal W$. Let~$\nu$
be a node of~$\cal W$. If 
\hbox{$\delta_{\lambda\alpha}(\nu)$ = 0}, 
then, as the $\lambda$ and~$\alpha$-leftmost 
and rightmost sons of~$\nu$ coincide, and as $\nu$ has a preferred son under~$\lambda$,
its \zzz-son is also its $\alpha$-preferred son. 
If \hbox{$\delta_{\lambda\alpha}(\nu)$ = 1}, clearly 
\hbox{$\varphi$ $\leq$ $\rho_{\lambda}$ $\leq$ $\rho_{\lambda}$} where
$\rho_{\alpha}$, $\rho_{\lambda}$ denote the $\alpha$-, $\lambda$-rightmost son of~$\nu$
respectively and $\varphi$ denotes the $\lambda$-preferred son of~$\nu$. 

We remain with the proof that $m_{n+1}$ is the preferred son of~$m_n$. The proof proceeds
by induction and on the following remark. By definition of $M_n$, we have
that \hbox{$m_n< M_n$}. Note that \hbox{$m_1>M_0=m_0$}. Assume that 
\hbox{$M_n<m_{n+1}$}. Then, 
\hbox{$M_{n+1}=M_n+m_{n+1}<2m_{n+1}<(p$$-$3$)m_{n+1}<m_{n+2}$}. So that, by induction,
we get that \hbox{$M_n<m_{n+1}<M_{n+1}$}. We can write that 
\hbox{$m_{n+1} = M_{n+1}-M_n=M_{n+1}-\displaystyle{\sum\limits_{k=0}^n m_k}$}. This
means that if we assume that $m_{n+1}$ is the penultimate $\lambda$-son of~$m_n$,
we get that $m_{n+2}=M_{n+2}-M_{n+1}=M_{n+2}-\displaystyle{\sum\limits_{k=0}^{n+1}m_k}$.
Now, if we interpret the sum $\displaystyle{\sum\limits_{k=0}^n m_k}$ as the trace
on the level of~$m_{n+1}$ of white trees $T_h$ of heights $h=0$ up to $h=n$,
the sum $\displaystyle{\sum\limits_{k=0}^{n+1}m_k}$ can be interpreted as the trace
on the next level of the same trees, so the height of $T_h$ is now increased by~1,
plus one node which is the rightmost $\lambda$-son of~$m_{n+1}$. Together with the 
application of the rule~\www\zzz{} of~$(22)$ together with the metallic code of the
penultimate son of a \zzz-node, what we just remarked proves that $m_n$ is the
preferred son of~$m_{n+1}$.

Accordingly,
Theorem~\ref{tpref} is proved. \hfill $\Box$
\vskip 10pt
Figures~\ref{fmetalblanc} and~\ref{fmetalblancassder} illustrate the property 
proved in Theorem~\ref{tpref} for the leftmost and the rightmost assignments 
respectively.
In the figures, the red colour is used to mark the black nodes, while the white ones
have a blue and a green colour. The blue and the green colours are used to distinguish
between the different kinds of white nodes which appear in Tables~$(19)$, $(23)$
and~$(25)$. The blue nodes correspond to the nodes marked by \www\aa. The \www\zzz-nodes
are indicated by a green disk with a red circle while the \www\uu-nodes are indicated
by a green disk with a darker green circle. The numbers in red, above the nodes, indicate
the natural numbering of the tree. The metallic code is mentioned vertically, below each
node. In these illustrations, \hbox{$p=9$}. However, in order to indicate the general
form of the properties, in the metallic codes, {\bf 5} and {\bf 6} are replaced by \ccc{}
and \ddd{} respectively. Indeed, it corresponds for \hbox{$p=9$} to the general values
given to \ccc{} and to~\ddd{} respectively. In order to make easier the reading of the 
figures, not all nodes are mentioned. We just mention those which allow us to see the
application of the rules~$(22)$ and to check Table~$(23)$.
\vskip 5pt
\vskip 5pt
\vtop{
\ligne{\hfill
\includegraphics[scale=0.35]{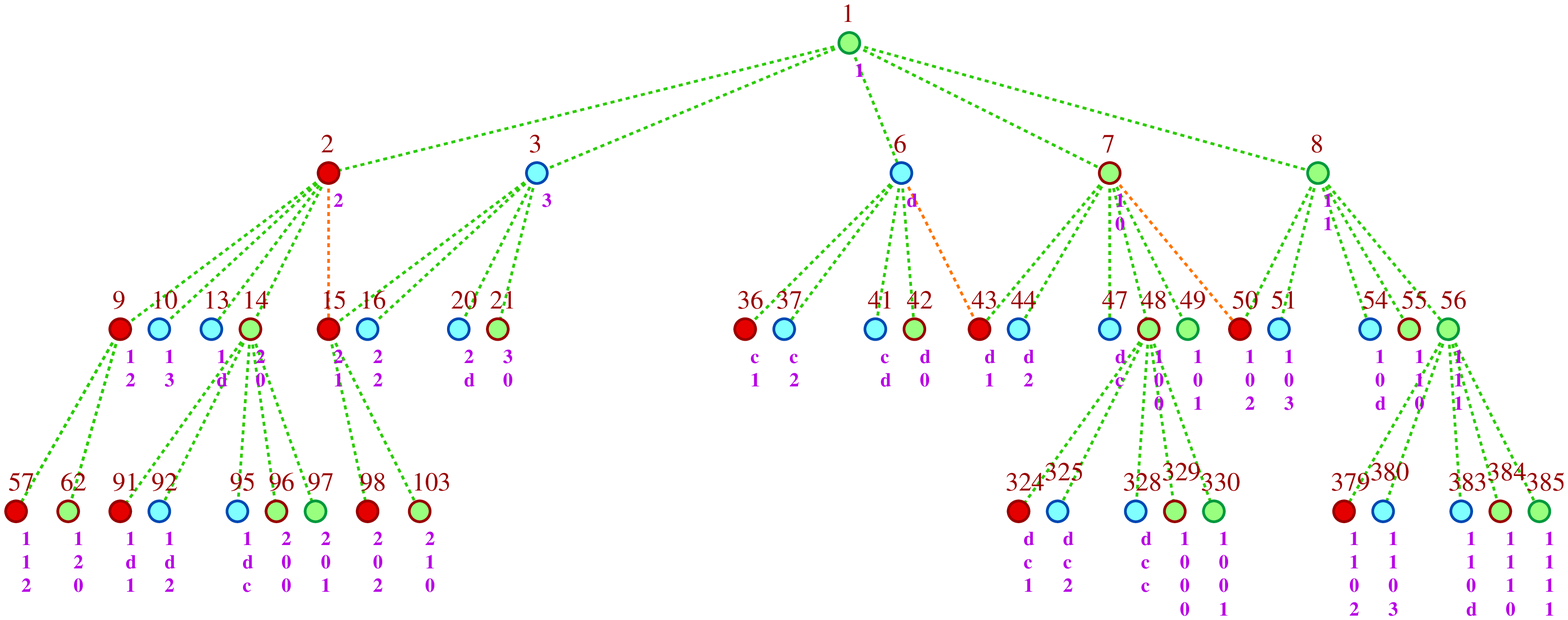}
\hfill}
\vspace{-10pt}
\ligne{\hfill
\vtop{\leftskip 0pt\parindent 0pt\hsize=300pt
\begin{fig}\label{fmetalblanc}
\leurre
The white metallic tree. Partial representation of the first three levels of the tree
when $p=9$
with the conventions mentioned in the text.
\end{fig}
}
\hfill}
}
\vskip 5pt
\vtop{
\ligne{\hfill
\includegraphics[scale=0.35]{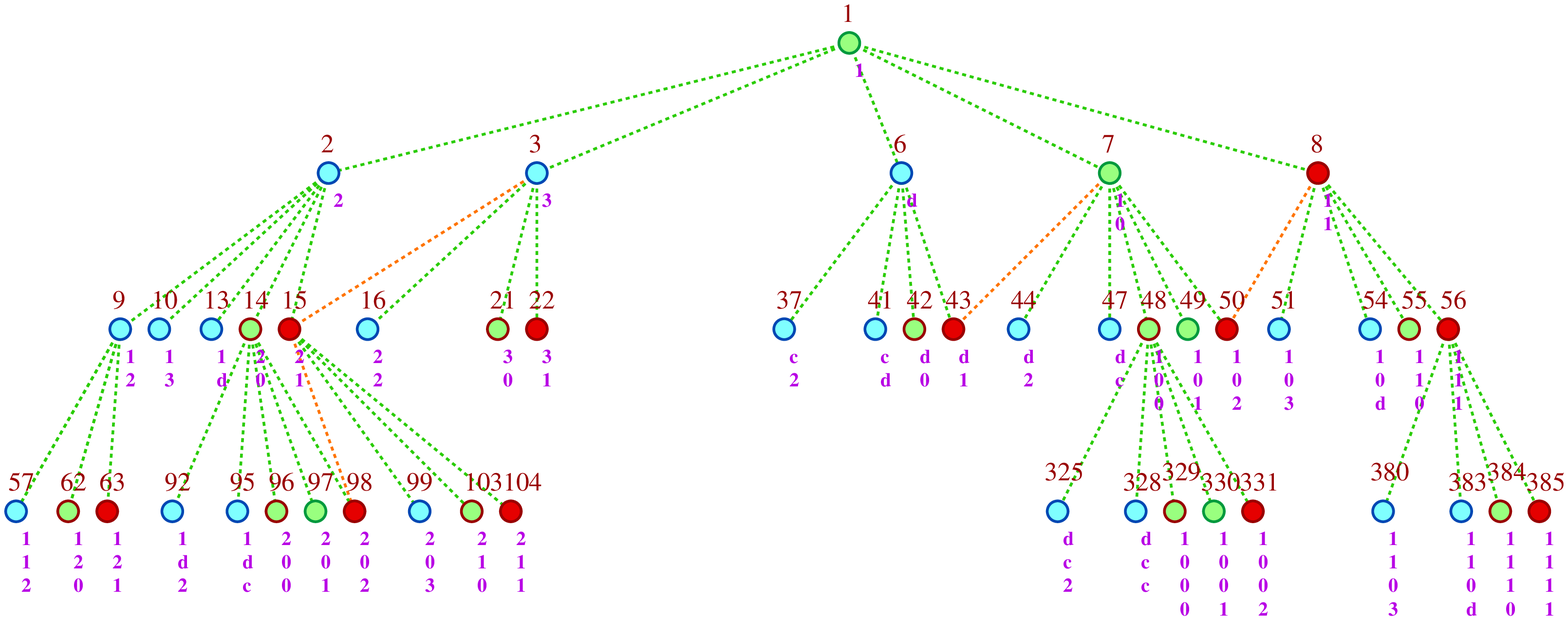}
\hfill}
\vspace{-10pt}
\ligne{\hfill
\vtop{\leftskip 0pt\parindent 0pt\hsize=300pt
\begin{fig}\label{fmetalblancassder}
\leurre
The white metallic tree under the rightmost assignment. Partial representation of the 
first three levels of the tree when $p=9$ with the conventions mentioned in the text.
\end{fig}
}
\hfill}
}

  We defined the \zzz-branch which connects the \zzz-nodes we obtain which are the 
\zzz-son of the previous one except the first one which is the root. We noted that the
\zzz-branch does not depend on the assignment~$\alpha$ with which we equipped $\cal W$.
Now, if we take a node~$\nu$ whose signature is not~\zzz. It has a unique $\alpha$-son
$\sigma$ which is a \zzz-node, and we know that the position of~$\sigma$ in $\cal W$
does not depend on $\alpha$. What depends on~$\alpha$ is the position of~$\sigma$
among the $\alpha$-sons of~$\nu$. As an example, $m_{n+1}$ is the penultimate 
$\lambda$-son of~$m_n$ while it is its ante-penultimate $\rho$-son. From what we just
mentioned, we can construct a sequence \hbox{$\{\varphi_n\}_{n\in\mathbb N}$} of nodes
such that \hbox{$\varphi_0=\nu$} and $\varphi_{n+1}$ is the \zzz-son of~$\varphi_n$
for any~$n$. From Theorem~\ref{tpref}, we know that $\varphi_{n+1}$ is always an 
$\alpha$-son of~$\varphi_n$ and again, its position does not depend on $\alpha$.
Call the sequence \hbox{$\{\varphi_n\}_{n\in\mathbb N}$} the {\bf \zzz-path 
issued from~$\nu$}. From Lemma~\ref{lzzdist},we can state:

\begin{thm}\label{tzzpaths}
For any assignment $\alpha$, the \zzz-paths indicate the nodes in 
${\cal W}_{\alpha}$ at which the application of the incrementation algorithm
necessitates a carry, which produce the \zzz-signature of the metallic code of the node.
\end{thm}

\subsection{Mid-assignments in the white metallic tree}\label{smassmid}

   Before turning to the connections between the assignments on $\cal W$ and the
\nzm-codes, we deal with a particular fixed assignment which, in some sense,
synthesizes the properties we observed on the leftmost, the penultimate and the rightmost
assignments. We say that an assignment is {\bf fixed} if the black nodes are always 
applied the same rule and if it is the same for the white nodes. 
In this sub section,we consider what we call a {\bf mid-assignment}. A mid-assignment
is defined by a constant~$k$ with \hbox{$k\in\{2..p$$-$4$\}$} which defines the 
position of the black son among the sons of a node, avoiding the positions we already 
studied. Denote by ${\cal W}_{\mu,k}$ the white metallic tree equipped with such 
an assignment. It is illustrated by Figure~\ref{fmetalblancmid} in the case when 
\hbox{$p=7$} and \hbox{$k=4$}.

\vskip 5pt
\vskip 5pt
\vtop{
\ligne{\hfill
\includegraphics[scale=0.35]{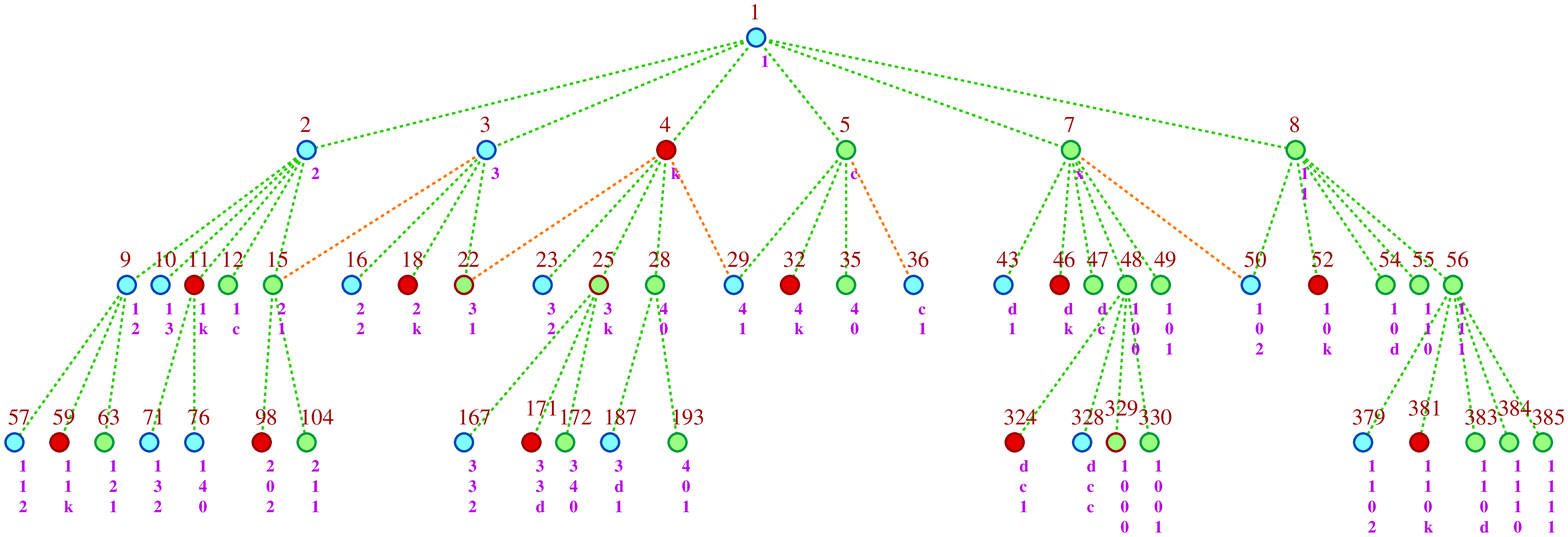}
\hfill}
\vspace{-10pt}
\ligne{\hfill
\vtop{\leftskip 0pt\parindent 0pt\hsize=300pt
\begin{fig}\label{fmetalblancmid}
\leurre
The white metallic tree. Partial representation of the first three levels of the tree
when $p=9$
with the conventions mentioned in the text.
\end{fig}
}
\hfill}
}
\vskip 5pt

The rules for the nodes are given by~$(26)$ and the sons of a node by~$(27)$.
We note that the root obeys the rule \www\aa{} of~$(26)$. From the metallic code of the 
leftmost son and from Lemma~\ref{lzzdist}, we can see that the rule \www\aa{} is applied
until we meet the sons of the first black node on the father level. We also can
check that lines~\fnb 3 to~\fnb 5 of Table~$(27)$ are observed. 
\vskip 10pt
\ligne{\hfill
$\vcenter{\vtop{\leftskip 0pt\hsize=250pt
\vskip 5pt
\ligne{\hfill\bbb\kkk{} $\rightarrow$ 
\www\numd$..$\www\kkk$^{\copy120}..$\bbb\kkk.\www\kkk$^{\copy110}..$\www\ddd.\www\zzz$,$
\hfill}
\ligne{\hfill
\www\aa{} $\rightarrow$ 
\www\numd$..$\www\kkk$^{\copy120}$\bbb\kkk.\www\kkk$^{\copy110}..$\www\ddd.%
\www\zzz.\www\uu$,$ {\rm with \zzz{} $<$ \aa{} $<$ \kkk,}\hfill}
\ligne{\hfill
\www\bbb{} $\rightarrow$ 
\www\uu$..$\www\kkk$^{\copy120}$\bbb\kkk.\www\kkk$^{\copy110}..$\www\ddd.%
\www\zzz$,$ {\rm with \kkk{} $<$ \bbb{} $\leq$ \ddd,}\hfill}
\ligne{\hfill
\www\zzz{} $\rightarrow$ 
\www\uu$..$\www\kkk$^{\copy120}$\bbb\kkk.\www\kkk$^{\copy110}..$\www\ccc. \www\zzz.%
\www\uu.\hfill}
\vskip 5pt
}}$
\hfill$(26)$\hskip 10pt}
\vskip 10pt
Then,
it is easy to see that the rule \bbb\kkk{} is applied and that the corresponding 
lines~\fnb 1 and~\fnb 2 of~$(27)$ are observed too by the metallic codes of the sons
of the node. As a black node has $p$$-$3 nodes
and not $p$$-$2 as a white one, the signature of its rightmost son is \zzz, so that
the signature of the leftmost son of the following white node on the father level
is \uu{} and not \numd{} as required by the rule~\www\aa. Accordingly the rule
\www\bbb{} is applied until the \zzz-node is met on the father level and we can see that 
the metallic sons of the corresponding nodes of the father level obey lines~\fnb 6
and~\fnb 7 of~$(27)$. Then, we apply the
rule~\www\zzz: the signature of the leftmost node is \uu{} as the signature of the 
rightmost node on the father level was \zzz. Lines~\fnb 8 is applied to get the metallic 
codes of the sons. Now, as the father node~$\nu$ is a \zzz-node, the signature of the
previous node on the level is either~\ddd{} or~\ccc{} but, in that latter case,
the metallic code of~$\nu$$-$1 has the suffix \ddd\ccc$^\ast$. It is the reason why
the metallic code of the rightmost son of~$\nu$ is $[\nu]$\zzz.
\vskip 10pt
\ligne{\hfill
$\vcenter{\vtop{\leftskip 0pt\hsize=260pt
\llaligne {$\nu$} {range} {son} {\nzm-code} {ref.}
\vskip 3pt
\hrule height 0.3pt depth 0.3pt width \hsize 
\vskip 3pt
\llaligne {\bbb\kkk} {$1..p$$-$4} {$j$} {\aa$_h$$..$\aa$_1$\aa$_0^{\copy120}$%
\jjj$^{\copy110}$} {1}
\llaligne {} {$p$$-$$3$} {} {\aa$_k$$..$\aa$_1$\aa$_0$\zzz} {2}
\llaligne {\www\aa} {$1..p$$-$4} {$j$} {\aa$_h$$..$\aa$_1$\aa$_0^{\copy120}$%
\jjj$^{\copy110}$} {3}
\llaligne {} {$p$$-$$3$} {} {\aa$_k$$..$\aa$_1$\aa$_0$\zzz} {4}
\llaligne {} {$p$$-$$2$} {} {\aa$_k$$..$\aa$_1$\aa$_0$\uu} {5}
\llaligne {\www\bbb} {$1..p$$-$3} {$j$} {\aa$_h$$..$\aa$_1$\aa$_0^{\copy120}$%
\jjj} {6}
\llaligne {} {$p$$-$$2$} {} {\aa$_k$$..$\aa$_1$\aa$_0$\zzz} {7}
\llaligne {\www\zzz} {$1..p$$-$4} {$j$} {\aa$_h$$..$\aa$_1$\aa$_0^{\copy120}$%
\jjj} {8}
\llaligne {} {$p$$-$$3$} {} {\aa$_k$$..$\aa$_1$\aa$_0$\zzz} {9}
\llaligne {} {$p$$-$$2$} {} {\aa$_k$$..$\aa$_1$\aa$_0$\uu} {10}
\vskip 5pt
\hrule height 0.3pt depth 0.3pt width \hsize 
}}$
\hfill$(27)$\hskip 10pt}
\vskip 10pt

   It is interesting to note that the rules corresponding to the leftmost, the penultimate
and the rightmost assignments can be derived from the rules~$(26)$. As an example,
consider the leftmost assignment: the rule~\bbb\uu,\bbb\numd{} of~$(22)$ comes from
the rule \www\aa{} of~$(26)$ noticing that the leftmost signature \www\numd{} becomes
\bbb\numd{} because of the leftmost position of the black nodes in ${\cal W}_{\lambda}$.
Due to the leftmost position of the black nodes, the rule~\www\aa{} of~$(22)$ comes
from the rule~\www\bbb{} of~$(26)$ up to the change in the positions of the black node.
The rule \www\zzz{} of~$(22)$ is the same of that of~$(26)$ up to the change in the 
black nodes and the rule~\www\uu{} of~$(22)$ comes from the rule \www\aa{} of
$(26)$ up to the change due to the position of the black node.

\subsection{Assignments in $\cal  W$ and the \nzm-codes}\label{snzmassign}

   As the \zzz-signature has no more any meaning in \nzm-codes, the property of
the preferred son can be reformulated as follows: is there a value \aa{} such that
each node~$\nu$ has among its $\alpha$-sons exactly one of them whose \nzm-code
is {\bf [$\nu$]\aa} for at least one assignment $\alpha$?

   Before addressing that issue, note that we can easily characterise in \nzm-terms the
nodes whose signature is \zzz{} in the metallic code. As far as the nodes do not change
but their sons according to the assignment~$\alpha$ we set on $\cal W$, let us still
call those nodes \zzz-nodes even in that context. We have:

\begin{lemm}\label{lnzmzzznodes}
Let $\nu$ be a \zzz-node and let {\bf [$\nu$] $=$ [$\omega$]\zzz$^k$} be its 
metallic code, where \hbox{$k>0$} and the signature of {\bf [$\omega$]} is not \zzz. 
Then we have:
\vskip 5pt
\ligne{\hfill
{\bf [$\omega$]\zzz $=$ [$\omega$$-$$1$]\xxx} {\rm and, when $k\geq2,$
[$\omega$]\zzz$^k$ $=$ [$\omega$$-$$1$]\ddd\ccc$^{k-2}$\ddd} 
\hfill $(28)$\hskip 10pt}
\end{lemm}

\noindent
The lemma is an immediate application of $(12)$.

The lemma tells us that the suffixes \xxx{} and \ddd\ccc$^\ast$\ddd{} cannot be
used for replacing the notion of preferred son in the context of the metallic codes:
the \nzm-code of the \zzz-son of $\nu$ contains \hbox{[$\omega$$-$1]} and not
\hbox{[$\omega$]}. 

   Now, it is not difficult to see that \uu\uu{} occurs among the sons of the root~\uu,
whatever the assignment. Also, the first nodes on the level ${\cal L}_2$ are
\uu\numd, ..., \uu\xxx, \numd\uu{} and \numd\uu{} is the $p$$-$2$^{\rm th}$ node.
Accordingly, if \numd{} is $\alpha$-white, \numd\uu{} occurs as its rightmost
$\alpha$-son. Let us call {\bf \uu-nodes} the nodes of~$\cal W$ whose signature of
their \nzm-code is \uu. We may wonder what is the distribution of the \uu-nodes in
$\cal W$? In fact, what we already said is a valuable hint to the solution: from
Lemma~\ref{lnotuniquenzmcode} and its Corollary~\ref{cnzmcodeforbid},
we know that \hbox{\bf [$\omega$$-$$1$]\xxx$\oplus$\uu{} = [$\omega$]\uu}
and that \hbox{\bf [$\omega$$-$$1$]\xxx\ddd$^{k+1}$ = [$\omega$]\uu$^{k+2}$}.
This allows us to prove:

\begin{lemm}\label{lnzmuudist}
Let $\mu$ and $\nu$ be two consecutive \uu-nodes of the level ${\cal L}_n$ with
\hbox{$\mu < \nu$}. Then \hbox{$\nu-\mu = p$$-$$3$} if and only if \uu\uu{} is a 
suffix of~$\nu$ and, when it is not the case, \hbox{$\nu-\mu = p$$-$$2$}.
\end{lemm}

\noindent
Proof. When \uu\uu{} is a suffix of~$\nu$, we can write 
\hbox{\bf [$\nu$]$_{nz}$ $=$ [$\omega$]$_{nz}$\uu$^{k+2}$}, with $k$ a natural integer.
Using Algorithm~\ref{anzmdecr}, we get that 
\hbox{\bf [$\nu$$-$$1$]$_{nz}$ $=$ [$\omega$$-$$1$]$_{nz}$\xxx\ddd$^{k+1}$}, so that
\hbox{\bf [$\mu$]$_{nz}$ $=$ [$\omega$$-$$1$]$_{nz}$\xxx\ddd$^k$\uu} whose distance 
to~$\nu$$-$1 is $p$$-$4 so that \hbox{$\nu-\nu=p$$-$3}, as announced. Assume that
\hbox{\bf [$\nu$]$_{nz}$ $=$ [$\omega$]$_{nz}$\uu}, with the signature of 
\hbox{\bf [$\omega$]$_{nz}$} being greater than \uu. Then, 
\hbox{\bf [$\mu$] $=$ [$\omega$$-$$1$]$_{nz}\uu$}, which can be checked by iterated 
applications of Algorithm~\ref{anzmdecr}. Accordingly,
\hbox{$\nu-\mu = p$$-$2}, which completes the proof of Lemma~\ref{lnzmuudist}.
\hfill $\Box$

For any node~$\nu$, call {\bf successor} of~$\nu$, denoted by {\it succ}$(\nu)$,
the node whose \nzm-code is \hbox{\bf [$\nu$]$_{nz}$\uu}.
Lemma~\ref{lnzmuudist} and our study of the penultimate assignment on $\cal W$ with 
respect to the metallic codes suggests to state:

\begin{thm}\label{tprefnzm}
Let $\cal W$ equipped with the rightmost assignment~$\rho$ and consider the \nzm-codes of
its nodes. Then, for any node~$\nu$, its successor occurs among its $\rho$-sons and no
other $\rho$-sons of~$\nu$ is a \uu-node,
so that we can call \hbox{\bf [$\nu$]$_{nz}$\uu} the \nzm-preferred son
of~$\nu$. Moreover, $\rho$ is the unique assignment~$\alpha$ such that for any node, 
its successor occurs among its $\alpha$-sons.
\end{thm}

\noindent
Proof. By induction, we prove that ${\cal W}_{\rho}$ can also be the
defined by application of the rules~$(29)$ and that the \nzm-codes of the sons of~$\nu$
are defined by Table~$(30)$.
\vskip 10pt
\ligne{\hfill
$\vcenter{\vtop{\leftskip 0pt\hsize=250pt
\vskip 5pt
\ligne{\hfill\bbb\uu{} $\rightarrow$ \www\numd.\www\numt$..$\www\ddd.\bbb\uu$,$
\www\aa{} $\rightarrow$ \www\numd$..$\www\ddd.\www\xxx.\bbb\uu$,$ \hfill}
\vskip 5pt
}}$
\hfill$(29)$\hskip 10pt}
\vskip 10pt
\ligne{\hfill
$\vcenter{\vtop{\leftskip 0pt\hsize=260pt
\llaligne {$\nu$} {range} {son} {\nzm-code} {ref.}
\vskip 3pt
\hrule height 0.3pt depth 0.3pt width \hsize 
\vskip 3pt
\llaligne {\bbb\uu} {$1..p$$-$4} {$h$} {\aa$_k$$..$\aa$_1$\aa$_0^{\copy120}$%
\hhh$^{\copy110}$} {1}
\llaligne {} {$p$$-$$3$} {} {\aa$_k$$..$\aa$_1$\aa$_0$\uu} {2}
\llaligne {\www\aa} {$1..p$$-$3} {$h$} {\aa$_k$$..$\aa$_1$\aa$_0^{\copy120}$%
\hhh$^{\copy110}$} {3}
\llaligne {} {$p$$-$$2$} {} {\aa$_k$$..$\aa$_1$\aa$_0$\uu} {4}
\vskip 5pt
\hrule height 0.3pt depth 0.3pt width \hsize 
}}$
\hfill$(30)$\hskip 10pt}
\vskip 10pt
We can see that the root is applied the rule~\www\aa{} with the exceptional value
\hbox{\aa{} = \uu} which is used for the root only. We already seen, that the rule
\www\numd{} applies to \numd, the leftmost node of ${\cal L}_1$. Then, by induction
and applying $p$$-$3 times Algorithm~\ref{anzmdecr}, we can see that the rules
\www\aa{} apply up to the node \ddd. Now, the next node on~${\cal L}_1$ after~\ddd{}
is \xxx. From~$(30)$ and Algorithm~\ref{anzmdecr}, we can see that the leftmost
$\rho$-node of~\xxx{} has \ddd\numd{} as \nzm-code. The $p$$-$4$^{\rm th}$ son of~$\nu$
is \ddd\ddd, which is a valid \nzm-code, so that the next son is \ddd\xxx{} and the
last one is \xxx\uu{} as required by the rule \www\xxx. Accordingly,
the leftmost node of \uu\uu{} is \xxx\numd. Iteratively applying Algorithm~\ref{anzmincr},
we get that the $p$$-$4$^{\rm th}$ node of~\uu\uu{} is \xxx\ddd, so that by
Corollary~\ref{cnzmcodeforbid}, the rightmost node is \uu\uu\uu. Consequently,
the rule~\bbb\uu{} applies to~\uu\uu. By the way, we can check that the $\rho$-sons
of~\uu\uu{} satisfy the lines~\fnb 3 and~\fnb 4 of $(30)$.

We can repeat this progression on the trace of each sub-tree rooted at a node of 
the level~${\cal L}_1$ on the level~${\cal L}_2$ starting from the nodes of the
level~${\cal L}_2$ in order to transport~$(29)$ and~$(30)$ on ${\cal L}_3$.
We note that when the root is a black node, we have just one missing white node 
which changes nothing in the application of the rules~\www\aa, so that the rule~\bbb\uu{}
applies as we have seen: the rule \bbb\uu{} applies when the distance from an \uu-node
to the previous one is $p$$-$3. The occurrence of the pattern \xxx\ddd$^\ast$ allows us
to spare one node, avoiding the signature \xxx, so that the following \uu-node
is applied the rule~\bbb\uu.

Presently, consider an assignment~$\alpha$. Let $\nu$ be the first node of~$\cal W$ such
that \statutdiff {\nu}. If $\nu$ is the rightmost node $\rho_n$ of the level~$n$, 
it is $\alpha$ white, so that it contains the rightmost son of~$\rho$$-$1 which is 
a \uu\uu-node. And so, as an $\alpha$-node, $\nu$ contains two \uu-sons. If
$\nu$ is not the rightmost node of a level, it is necessarily $\alpha$-black, so
that it does not contains the successor of~$\nu$ which is the rightmost $\rho$-son
of the node which is $\rho$-white. But this discrepancy induces a shift of the leftmost
$\alpha$-son with respect with the $\rho$-one: for each node~$\mu$ after~$\nu$ on the
same level, the \uu-son is the successor of $\nu$$-$1. Accordingly, the proof of 
Theorem~\ref{tprefnzm} is completed. \hfill $\Box$

\vskip 10pt
\vtop{
\ligne{\hfill
\includegraphics[scale=0.35]{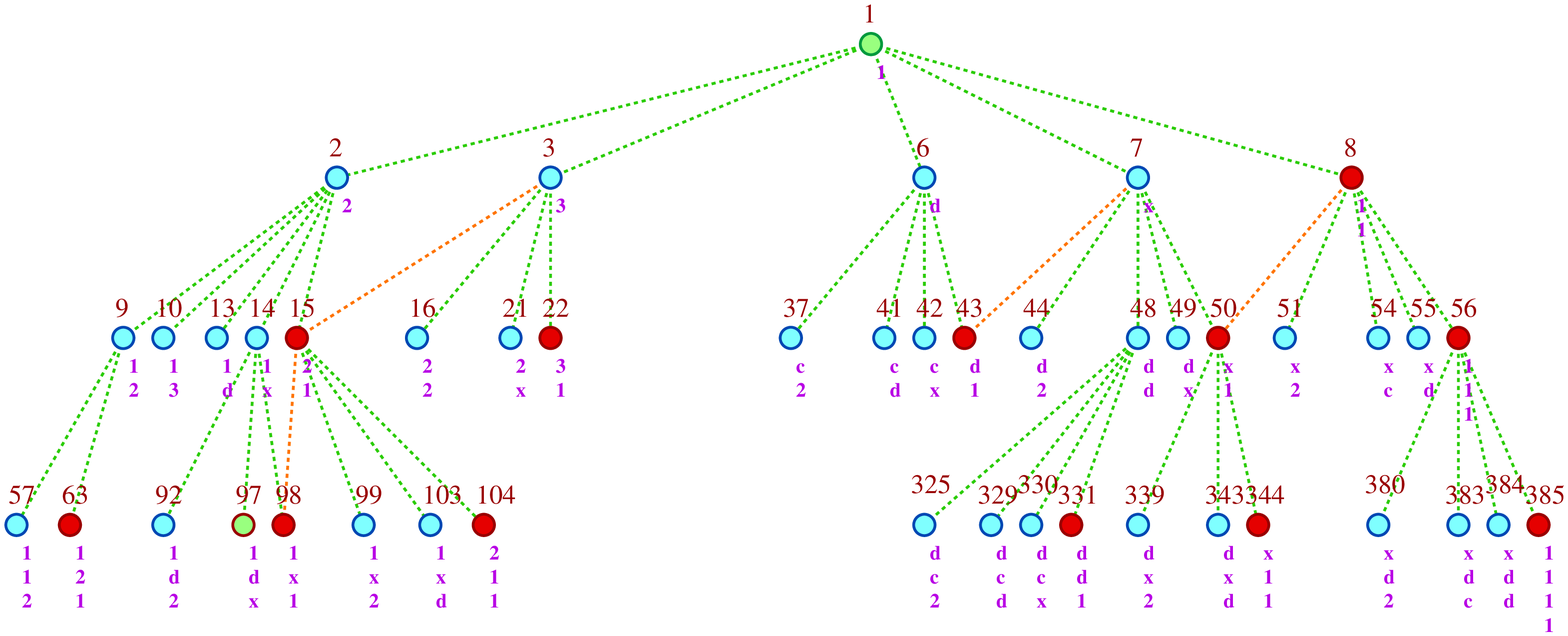}
\hfill}
\vspace{-10pt}
\ligne{\hfill
\vtop{\leftskip 0pt\parindent 0pt\hsize=300pt
\begin{fig}\label{fnzmblancder}
\leurre
The white metallic tree and the rightmost assignment under the \nzm-codes for the nodes. 
Partial representation of the first three levels of the tree
when $p=9$ with the conventions mentioned in the text.
\end{fig}
}
\hfill}
}
\vskip 10pt
Figure~\ref{fnzmblancprem} illustrates ${\cal W}_{\lambda}$. The colours are again those
of~Figures~\ref{fmetalblanc} and~\ref{fmetalblancassder}. As in those latter figures, the
blue colour indicates an application of the rules~\www\aa.
\vskip 10pt
\vtop{
\ligne{\hfill
\includegraphics[scale=0.35]{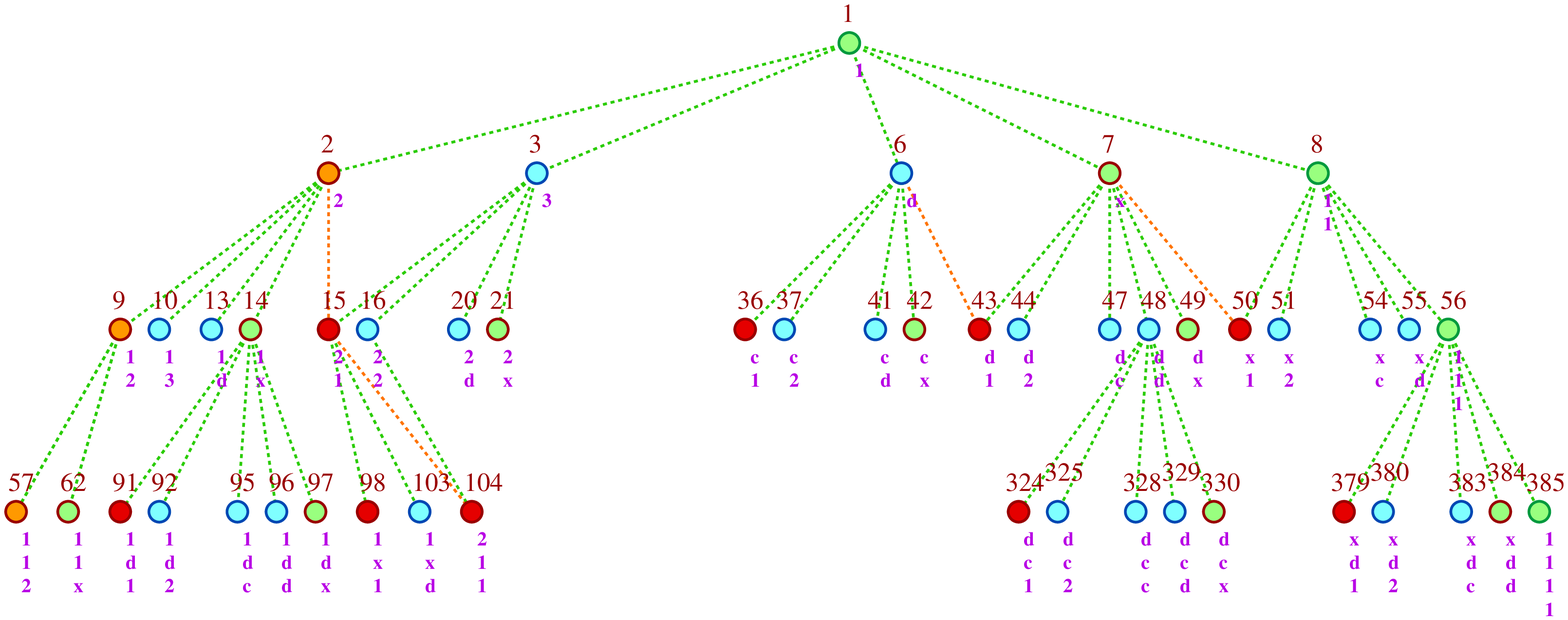}
\hfill}
\vspace{-10pt}
\ligne{\hfill
\vtop{\leftskip 0pt\parindent 0pt\hsize=300pt
\begin{fig}\label{fnzmblancprem}
\leurre
The white metallic tree and the leftmost assignment under the \nzm-codes for the nodes. 
Partial representation of the first three levels of the tree
when $p=9$ with the conventions mentioned in the text.
\end{fig}
}
\hfill}
}
\vskip 10pt
Figure~\ref{fnzmblancder} illustrates ${\cal W}_{\rho}$. The convention on the colours
are the same as for Figures~\ref{fmetalblanc} and~\ref{fmetalblancassder}. We can see on
the figure that two colours only occur: blue and red. This corresponds to the fact
that $(29)$ mentions two rules only, the rule~\bbb\uu{} and the rules~\www\aa{}
with \hbox{\aa $\in$ $\{$\numd$..$\xxx$\}$}. We can check on the figure that the
\nzm-codes which are displayed in the figure in the same way as for the metallic codes
in Figures~\ref{fmetalblanc} and~\ref{fmetalblancassder} satisfy Table~$(30)$.
The green colour with a red circle indicates the nodes which are before the \uu-nodes
on the same level and the colour green with a green circle indicate the \uu-nodes of
the rightmost branch of the tree. The black nodes lie on the leftmost branch of the
tree and also in the \uu-nodes which are not on that branch. On 
Figure~\ref{fnzmblancprem}, we
can see that the $\lambda$-assignment does not possess the \nzm-preferred son property.
For all nodes~$\nu$ except those which lie on the rightmost branch of the tree,
the successor of~$\nu$ is the leftmost $\lambda$-son of~$\nu$+1.

It is possible to transport the property stated in Theorem~\ref{tpenzzisom} and that
of Theorem~\ref{tzzpaths} to ${\cal W}_{\rho}$. We define the {\bf \uu-branch} as the
sequence of nodes \hbox{$\{\omega_n\}_{n\in\mathbb N}$}, where $\omega_0$ is the
root and $\omega_{n+1}$ is the successor of~$\omega_n$. The \uu-branch is the
analog in the \nzm-codes context of the \zzz-branch in the context of the metallic
codes. Similarly, we define the {\bf \uu-paths issued from a node~$\nu$} in 
${\cal W}_{\rho}$. Note the difference with the previous situation: a \zzz-path is
a path whose terms except the first one are sons of the previous term, whatever the
assignment given to $\cal W$. In the context of the \nzm-codes, a \uu-path
is a true path in ${\cal W}_{\rho}$ and it is not a path in any other ${\cal W}_{\alpha}$
as established in the proof of Theorem~\ref{tprefnzm}. We can state:

\begin{thm}\label{tnzmrightmostisom}
Let ${\cal V}$ be the set of \uu-nodes of ${\cal W}_{\rho}$, 
equipped with the rightmost assignment~$\rho$, \uu{} being excepted.
Define the mapping $\varphi$ from~$\cal V$ onto~${\cal W}_{\rho}$ 
by 
\vskip 5pt
\ligne{\hfill$\varphi(([\nu]_{nz}\uu)) \rightleftharpoons$ $([\nu]_{nz})$.\hfill} 
\vskip 5pt
Define the sons of~($[\nu]_{nz}$\uu) as the \uu-sons of the $\rho$-sons 
of~($[\nu]_{nz}$). Then $\varphi$ defines
an isomorphism between $\cal V$ equipped with its natural numbering and 
${\cal W}_{\rho}$ and $\varphi^{-1}$ transports the $\rho$-assignment onto
$\cal V$.
\end{thm}

\section{Properties of the black metallic tree}\label{black_metal}

As defined in Subsection~\ref{smrules}, the black metallic tree $\cal B$ is defined
by the same rules as the white one, the difference being that the root of $\cal B$
is a black node. We know that the number of nodes on the level~$n$ of~$\cal B$ is
$b_n$ which satisfies $(3)$. We also know that \hbox{$B_n=m_n$}. Accordingly,
the nodes of the rightmost branch of $\cal B$ are numbered by $m_n$, as known from 
Theorem~\ref{theadlevelwb}, and
we get from~$(19)$ that their \nzm-code is \hbox{\ddd\ccc$^{n-2}$\ddd}.

   In~\cite{mmarXiv2}, we proved the properties of the sons signatures of the nodes in 
the black metallic tree under the leftmost assignment and when the nodes are fitted with
their metallic code. The properties are different from those we have noted in the 
white one in the similar context. In Sub section~\ref{sblackmetal} we consider the
properties of~$\cal B$ when its nodes are fitted with the metallic codes.
Sub subsection~\ref{smblackprems} studies the case of ${\cal B}_{\lambda}$ when $\cal B$ is
constructed under the $\lambda$-assignment.
Figure~\ref{fmetalnoir} illustrates the black metallic 
tree in that context for \hbox{$p=9$} as in the case of Figure~\ref{fmetalblanc} to 
which the reader is referred for a comparison
between $\cal W$ and $\cal B$. We recall the results in Sub-subsection~\ref{smblackprems}.
A more detailed study of that comparison can be found
in~\cite{mmarXiv2}. In the present section, we shall stress on the the comparison with
the situation of the black metallic tree under the rightmost assignment and also,
in both the leftmost and the rightmost assignments when the nodes are equipped with 
their \nzm-codes. We shall write ${\cal B}_{\lambda}$, ${\cal B}_{\rho}$ for
$\cal B$ equipped with the $\lambda$-, $\rho$-assignments respectively.
The study of ${\cal B}_{\lambda}$ with the \nzm-codes is dealt with in 
Sub-subsection~\ref{snzmblackprems}, while the similar
study with ${\cal B}_{\rho}$ is the goal of Sub-subsection~\ref{snzmblackders}.

\subsection{The black metallic tree and the metallic codes}\label{sblackmetal}

We now turn to the black metallic tree $\cal B$ and we look at properties, similar to 
those which hold for the white metallic tree, which are still valid in that tree and
we try to see the reason why for those which are not valid. 
Sub subsection~\ref{smblackprems} looks at the situation for ${\cal B}_{\lambda}$,
the tree $\cal B$ when it is fitted with the leftmost assignment $\lambda$. 
Sub subsection~\ref{smblackders} deals with the situation for ${\cal B}_{\rho}$,
the tree $\cal B$ when it is fitted with the rightmost assignment $\rho$. We recall 
the reader that in this subsection, we consider the metallic codes for the 
representations of the numbers attached to the nodes.

\subsubsection{The black metallic tree under the leftmost assignment and the metallic 
codes}
\label{smblackprems}

Figure~\ref{fmetalnoir} shows us that the preferred is not true in ${\cal B}_{\lambda}$. 
The leftmost son
of a level, a black node, has no son whose signature is~\zzz. All other nodes
have a son whose signature is~\zzz, and among them, the last node of a level has two
sons whose signature is~\zzz, so that the leftmost assignment is not even a 
\zzz-assignment for the leftmost one. Now, for a node~$\nu$ which has a unique son 
whose signature is~\zzz, the metallic code of that node is not {\bf [$\nu$]0} but it 
is \hbox{\bf [$\nu$$-$1]0}. 

\vskip 10pt
\vtop{
\ligne{\hfill
\includegraphics[scale=0.35]{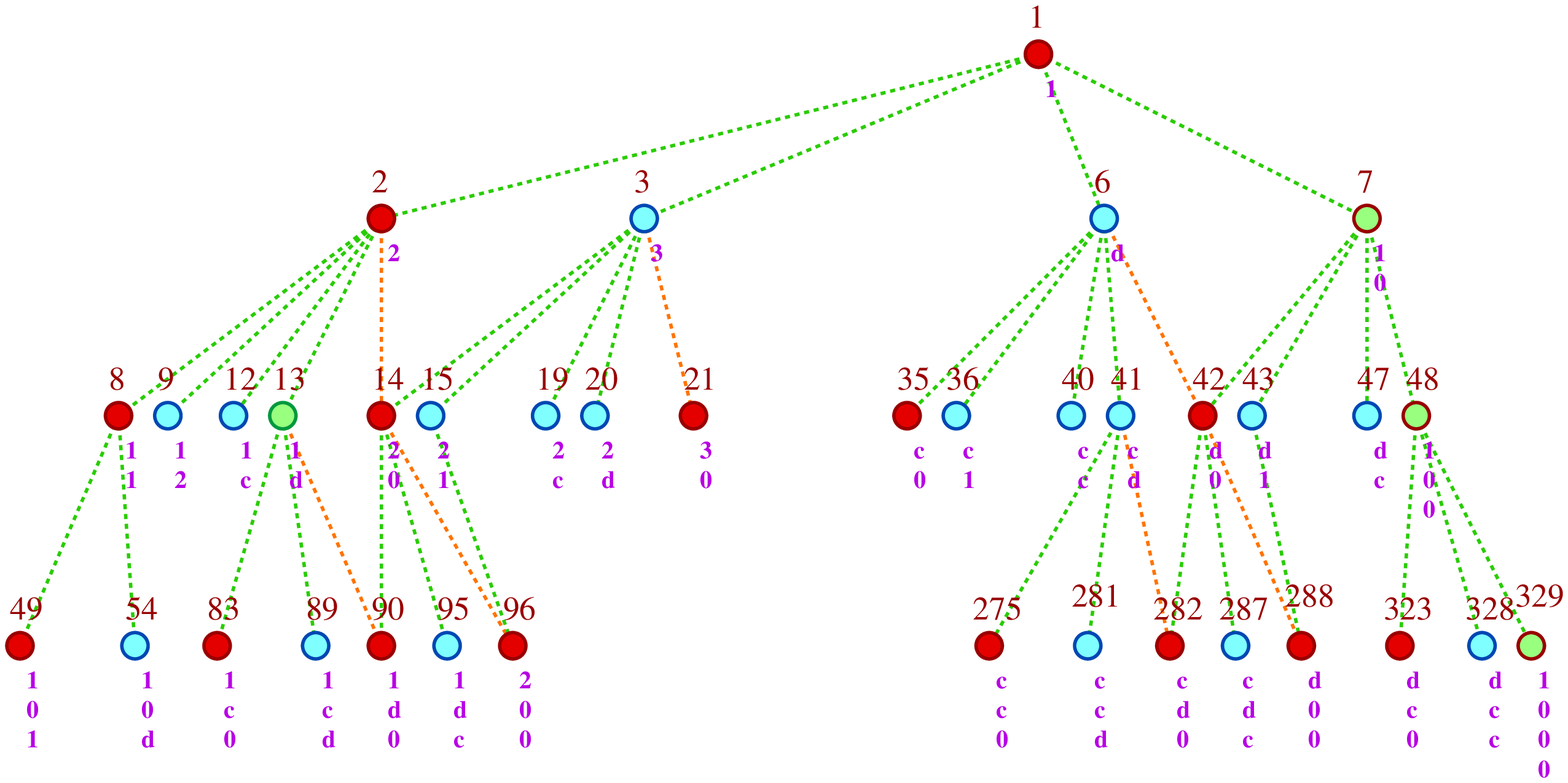}
\hfill}
\vspace{-10pt}
\ligne{\hfill
\vtop{\leftskip 0pt\parindent 0pt\hsize=300pt
\begin{fig}\label{fmetalnoir}
\leurre
The black metallic tree with the leftmost assignment and with the metallic code of the
nodes. The same convention about colours of the nodes and of the edges
between nodes as in Figure~{\rm\ref{fmetalblanc}} is used. 
We can see that the preferred son property is not true in the present setting.
\end{fig}
}
\hfill}
}
\vskip 5pt
Here too, call {\bf successor} of the node~$\nu$, the node whose metallic code 
is {\bf [$\nu$]0}. We can state:

\begin{thm}\label{tmblacksuccleft}~{\rm see \cite{mmarXiv2}}. In ${\cal B}_{\lambda}$,
the nodes are applied the rules of~$(31)$ and the metallic codes of the
$\lambda$-sons of a node~$\nu$ are given by Table~$(32)$, the root being excepted. 
The root is applied the rule 
\hbox{\bbb\uu{} $\rightarrow$ \bbb\numd\www\numt$..$\www\ddd\www\zzz}. For the other
nodes there are two kinds of black nodes, the nodes \bbb\uu{} and the
nodes \bbb\zzz which follow the rules for black nodes in~$(31)$. The rule~\bbb\uu{} is 
also followed by the node~\numd{} whose signature is~\numd. The nodes \bbb\uu{} are 
present 
on the leftmost branch of $\cal B$, the node \numd{} being excepted, and only on those 
places. The other black nodes, the leftmost one of the $\lambda$-sons of a node
are \bbb\zzz{} nodes. There are two types of white nodes, \www\zzz, and \www\aa{} 
with \hbox{\aa{} $>$ \zzz}. The metallic codes of the sons of a node are given in 
Table~$(32)$ in terms of \hbox{\bbb$_k$$..$\bbb$_0$ $\rightleftharpoons$ 
$[\nu]$$-$$1$}. The \www\zzz-nodes are exactly the  nodes of the rightmost branch of the 
tree, the root being excepted.
\vskip 0pt
The nodes of the rightmost branch of the tree being excepted, the successor of~$\nu$
is the leftmost $\lambda$-son of~$\nu$. The tree ${\cal B}_{\lambda}$ does not
observe the preferred son property. The nodes on the extremal branches of the tree
being excepted but the root being included, any other node has a \zzz-node among its
$\lambda$-sons which is not its successor. The root is the single node of the tree
which has a preferred son.
\vskip 5pt
\ligne{\hfill
$\vcenter{\vtop{\leftskip 0pt\hsize=300pt\bf
\ligne{\hfill \bbzz{} $\rightarrow$ \bbb\zzz$,$\www\uu$,..,$\www\ccc\hskip 20pt
\bbuu{} $\rightarrow$ \bbuu$,$\www\numd$,..,$\www\ddd\hfill}
\ligne{\hfill \wwzz{} $\rightarrow$ \bbb\zzz$,$\www\uu$,..,$\www\ccc$,$\www\zzz
\hskip 20pt \www\aa{} $\rightarrow$ \bbb\zzz$,$\www\uu$,..,$\www\ddd$.$\hfill}
}}$
\hfill $(31)$\hskip 10pt}
\vskip 5pt
For any node~$\nu$ which is not the rightmost one on a level,
the successor of~$\nu$ is the leftmost node of~$\nu$$+$$1$. For the rightmost node
on the level~$n$, its successor is the rightmost node on the level~$n$$+$$1$.
We also have that the type~\bbuu{} occurs for the leftmost node of a level
only and that the type~\wwzz{} occurs for the rightmost node of a level only.
Table~$(32)$ gives the metallic code of a node~$\nu$ in terms
of \hbox{\bf [$\nu$]}.
\vskip 5pt
\ligne{\hfill
$\vcenter{\vtop{\leftskip 0pt\hsize=260pt
\llaligne {$\nu$} {range} {son} {metallic code} {ref.}
\vskip 5pt
\hrule height 0.3pt depth 0.3pt width \hsize
\vskip 5pt
\llaligne {\bbb\zzz} {$1$$..$$p$$-$$3$} {$h$} 
{$[$\bbb$_k$..\bbb$_0$]\hhh$^{\copy120}$} {1}
\llaligne {\bbb\uu} {$1$$..$$p$$-$$3$} {$h$} {\uu\zzz$^{k-1}$\hhh} {2}
\llaligne {\www\aa} {$1$$..$$p$$-$$2$} {$h$} 
{$[($\bbb$_k$..\bbb$_0$)$-$$1]$\hhh$^{\copy120}$} {3}
\llaligne {\www\zzz} {$1$$..$$p$$-$$3$} {$h$} {\ddd\ccc$^{k-1}$\hhh$^{\copy120}$} {4}
\llaligne {} {$p$$-$$2$} {} {\uu\zzz$^{k+1}$} {5}
\vskip 5pt
\hrule height 0.3pt depth 0.3pt width \hsize
}}$
\hfill $(32)$\hskip 10pt}
\vskip 5pt
\end{thm}

\noindent
Proof. As usual, we proceed by induction on the level~$n$ and, on each level, by
induction on~$\nu$ from the leftmost node of the level to its rightmost one. 
The leftmost node $\lambda_n$ on the level~$n$ is $m_{n-1}$+1 whose metallic code is 
\numd{} when \hbox{$n=1$} and it is \hbox{\uu\zzz$^{n-1}$\uu} when \hbox{$n>1$}. 
Accordingly, as the leftmost son of~$\lambda_n$ is
$\lambda_{n+1}$ whose metallic code is \hbox{\uu\zzz$^n$\uu}, so that we easily
obtain the line~{\fnb 2} of Table~$(32)$. Accordingly, that line is
proved which also proves the rule \bbb\uu. Starting from~$\lambda_n$+1, we
have that the distance between two consecutive black nodes on a level which have the same
grand-father~$\varphi$ is $p$$-$2. This shows us that the rules of~$(31)$ apply to the
$\lambda$-sons of~$\varphi$. We can note that the property is true whether $\varphi$ is
black or white. The distance between the rightmost node of a level, which is a \zzz-node
and the first \zzz-node of the next level is $p$$-$2. Now, as far as the signature of the
leftmost son of a level is consequently \uu{} and as far as that node is black 
under~$\lambda$, we get that the first \zzz-node of a level is the second black node on 
the level. From that situation, we have that the case when the distance between 
two \zzz-nodes is $p$$-$3 occurs within the \www\zzz-nodes which have two \zzz-nodes 
among their $\lambda$-sons. The fact that the rightmost son of the level~$n$ is $m_n$,
as proved in Theorem~\ref{theadlevelwb}, explains the line~\fnb 3 of Table~$(32)$. 
Presently, consider the rightmost
son~$\rho_n$ of the level~$n$. We know that its number is $m_n$ so that its metallic code
is \hbox{\uu\zzz$^n$}. Applying the Algorithm for decreminting a metallic code,
see~\cite{mmarXiv2}, we get that \hbox{[$m_n$$-$1] $=$ \ddd\ccc$^{n-1}$}. Applying again
that algorithm which here consists in decrementing the last digit only, we get the
lines~{\fnb 4} and~{\fnb 5} of Table~$(32)$. Consequently, Theorem~\ref{tmblacksuccleft}
is proved. \hfill $\Box$

\def\gaa{\hbox{\bf C}}
\newdimen\llarge\llarge=25pt
\def\demiligne #1 #2 #3 #4 #5 #6{\footnotesize
\hbox to \llarge {\hfill#1\hfill}
\hbox to \llarge {#2\hfill}
\hbox to \llarge {#3\hfill}
\hbox to \llarge {#4\hfill}
\hbox to \llarge {#5\hfill}
\hbox to \llarge {#6\hfill}
}
\vskip 5pt

Note that, {\it par abus de langage}, we can also say for the black metallic tree equipped
with the leftmost assignment that $m_{k+1}$ is the preferred son of~$m_k$.

\def\ddemili #1 #2 #3 {%
\ligne{
\hbox to 20pt{\footnotesize\bf #1\hfill}
\hbox to 33pt{#2\hfill}
\hbox to 80pt{#3\hfill}
\hfill}
}

\subsubsection{The black metallic tree under the rightmost assignment and the metallic codes}
\label{smblackders}

Figure~\ref{fmetalnoirder} illustrates ${\cal B}_{\rho}$. The figure can be compared 
with Figure~\ref{fmetalblancassder}. The colours indicates that the rules in the
case of~${\cal B}_{\rho}$ seems to be simpler than the rules for~${\cal W}_{\rho}$,
see~$(24)$. If we compare Figure~\ref{fmetalnoirder} with Figure~\ref{fmetalnoir},
we can see that the preferred son property which is not observed in ${\cal B}_{\lambda}$
as stated in Theorem~\ref{tmblacksuccleft} seems to be satisfied in ${\cal B}_{\rho}$.

Indeed, we have a stronger property tightly connected with Lemma~\ref{lzzdist} and
it reminds what we noted in the case of the \nzm-codes:

\vskip 10pt
\vtop{
\ligne{\hfill
\includegraphics[scale=0.35]{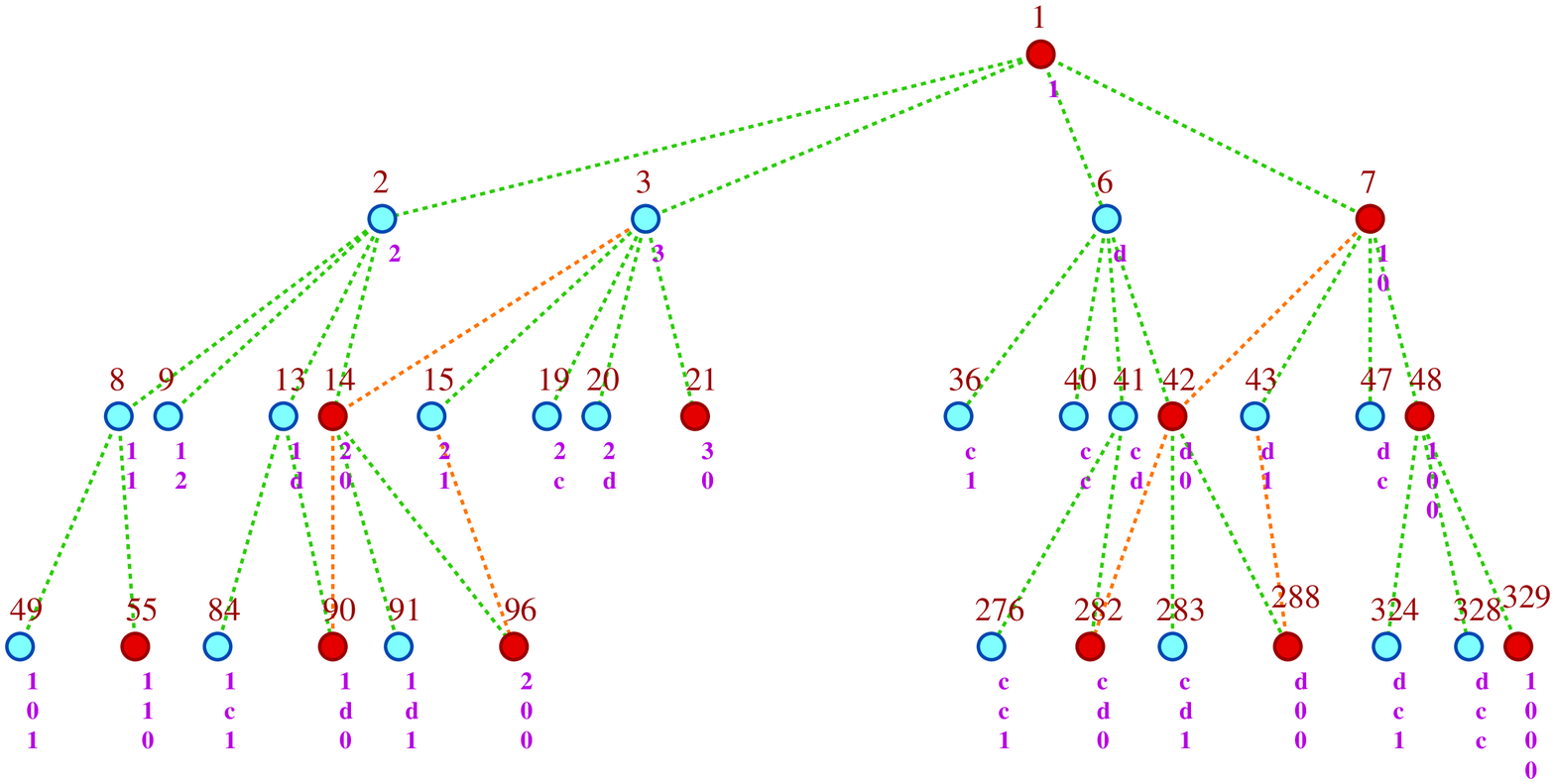}
\hfill}
\vspace{-10pt}
\ligne{\hfill
\vtop{\leftskip 0pt\parindent 0pt\hsize=300pt
\begin{fig}\label{fmetalnoirder}
\leurre
The black metallic tree with the rightmost assignment and with the metallic code of the
nodes. The same convention about colours of the nodes and of the edges
between nodes as in Figure~{\rm\ref{fmetalblancassder}} is used. 
We can see that the preferred son property is true in the present setting.
\end{fig}
}
\hfill}
}
\vskip 5pt

\begin{thm}\label{tblackprefder}
Let ${\cal B}_{\rho}$ be the black metallic tree equipped with the
rightmost assignment. Consider the metallic representations of its nodes. The rules
which may be used for constructing the tree are given by~$(33)$ and the
metallic codes of the $\rho$-sons of a node~$\nu$ are given by Table~$(34)$
in terms of $[\nu]$ and of $[\nu]$$-$$1$. Equipped with the rightmost assignment,
$\cal B$ possesses the preferred son property. But the tree does not possess
that property if it is fitted with another assignment.
\end{thm}

\noindent
Proof. Our first remark is that Lemma~\ref{lzzdist} is also true for $\cal B$.
The reason is that the property given on the lemma follows from the order on the 
numbers themselves and on the properties of the incrementation and not on the fact
that we may use the numbers to identify the nodes of an infinite finitely generated
tree.

   From that remark, we note that the first \zzz-node on level~2{} in $\cal B$
is the $p$$-$2$^{\rm th}$ of the level so that it is the rightmost $\rho$-son of
the node~2 of $\cal B$ which is the leftmost one on level~1. Accordingly, all the
$\rho$-white nodes on level~1 have the successor of their metallic code as
the metallic code of their rightmost $\rho$-son. Accordingly, the argument performed
for the analysis of the rightmost assignment on $\cal W$ can be repeated for those
white nodes. Now, the rightmost $\rho$-son of the penultimate node~\ddd{} on level~1
is \ddd\zzz, so that the leftmost $\rho$-son of the penultimate node on level~1 is
\ddd\uu{} and the penultimate node on level~2 is then \ddd\ccc, so that the last node
is \uu\zzz\zzz. Lemma~\ref{lzzdist} tells that from \ddd\uu{} up to \uu\zzz\zzz{} there
are $p$$-$3 nodes which exactly fits with the requirement of \ddd\zzz{} to be
a $\rho$-black node. Next, the argument goes on as in the proof of the rules~$(24)$.
We can see that the rules are exactly those of~$(33)$. This also proves the
codes given in Table~$(34)$ which also proves that the
preferred son property is true in ${\cal B}_{\rho}$ with respect to the metallic code.
Note that in~$(34)$, \hbox{\aa$_k$..\aa$_0$ $\rightleftharpoons$ $[\nu]$} and
that \hbox{\bbb$_k$..\bbb$_0$ $\rightleftharpoons$ $[\nu$$-$$1]$}.
\vskip 5pt
\ligne{\hfill
\bbb\zzz{} $\rightarrow$ \www\uu..\www\ccc.\bbb\zzz,\hskip 20pt
\www\aa{} $\rightarrow$ \www\uu..\www\ddd.\bbb\zzz, \hskip 10pt with \aa $>$ \zzz.
\hfill$(33)$\hskip 10pt}
\vskip 5pt
\ligne{\hfill
$\vcenter{\vtop{\leftskip 0pt\hsize=260pt
\llaligne {$\nu$} {range} {son} {metallic code} {ref.}
\vskip 5pt
\hrule height 0.3pt depth 0.3pt width \hsize
\vskip 5pt
\llaligne {\bbb\zzz} {$1$$..$$p$$-$$4$} {$h$} 
{[\bbb$_k$..\bbb$_0$]\hhh} {1}
\llaligne {} {$p$$-$$3$} {} {\aa$_k$..\aa$_0$\zzz} {2}
\llaligne {\www\aa} {$1$$..$$p$$-$$3$} {$h$} 
{[\bbb$_k$..\bbb$_0$]\hhh} {3}
\llaligne {} {$p$$-$$2$} {} {\aa$_k$..\aa$_0$\zzz} {4}
\vskip 5pt
\hrule height 0.3pt depth 0.3pt width \hsize
}}$
\hfill $(34)$\hskip 10pt}
\vskip 10pt
Now, we can repeat the argument of Theorem~\ref{tprefnzm} as the \zzz-$\rho$-son is
the rightmost son whatever the node. Indeed, considering another assignment $\alpha$, 
take the first node~$\nu$ whose status is not the same under~$\alpha$ and under $\rho$.
If $\nu$ is a $\rho$ black node, as its $\alpha$-leftmost son is the rightmost 
$\rho$-son of $\nu$$-$1, it contains two \zzz-nodes. If $\nu$ is a $\rho$ white node,
as it is the first node where the statuses are different, the rightmost $\alpha$-son
of $\nu$ is a node whose signature is \ddd, so that no $\alpha$-son of~$\nu$ is a
\zzz-node. 

The proof of Theorem~\ref{tblackprefder} is now completed. \hfill $\Box$

We may repeat the proof of Theorem~\ref{tnzmrightmostisom} and prove the following result:

\begin{thm}\label{tblackrightmostisom}
Let ${\cal V}$ be the set of \zzz-nodes of ${\cal B}_{\rho}$, 
equipped with the rightmost assignment~$\rho$.
Define the mapping $\varphi$ from~$\cal V$ onto~${\cal B}_{\rho}$ 
by \hbox{$\varphi(([\nu]\zzz)) \rightleftharpoons$ $([\nu])$}. 
Define the sons of~($[\nu]$\zzz) as the \zzz-sons of the $\rho$-sons 
of~($[\nu]$). Then $\varphi$ defines
an isomorphism between $\cal V$ equipped with its natural numbering and 
${\cal B}_{\rho}$ and $\varphi^{-1}$ transports the $\rho$-assignment onto
$\cal V$.
\end{thm}

\noindent
The proof combines the argument of the proof of Theorem~\ref{tpenzzisom} and that
of Theorem~\ref{tnzmrightmostisom}. It is left to the reader as an exercise.

\subsection{The black metallic trees and the \nzm-codes}

   We now turn to the study of $\cal B$ when the numbers of its nodes are written
as \nzm-codes. In Sub subsection~\ref{snzmblackprems} we investigate the properties
for ${\cal B}_{\lambda}$ while Sub subsection~\ref{snzmblackders} is devoted to those
of~${\cal B}_{\rho}$.

\subsubsection{The black metallic tree under the leftmost assignment and the \nzm-codes}
\label{snzmblackprems}

Figure~\ref{fnzmetalnoirprem} illustrates ${\cal B}_{\lambda}$ when the nodes are fitted
with their \nzm-codes. At first glance, whatever the digit \aa, no $\lambda$-son
of a node $\nu$ has the \nzm-code [$\nu$]$_{nz}$\aa. Accordingly, the preferred
son cannot be defined for ${\cal B}_{\lambda}$, a situation which reminds us that of the
same tree when we consider the metallic codes of the nodes.

\vskip 10pt
\vtop{
\ligne{\hfill
\includegraphics[scale=0.35]{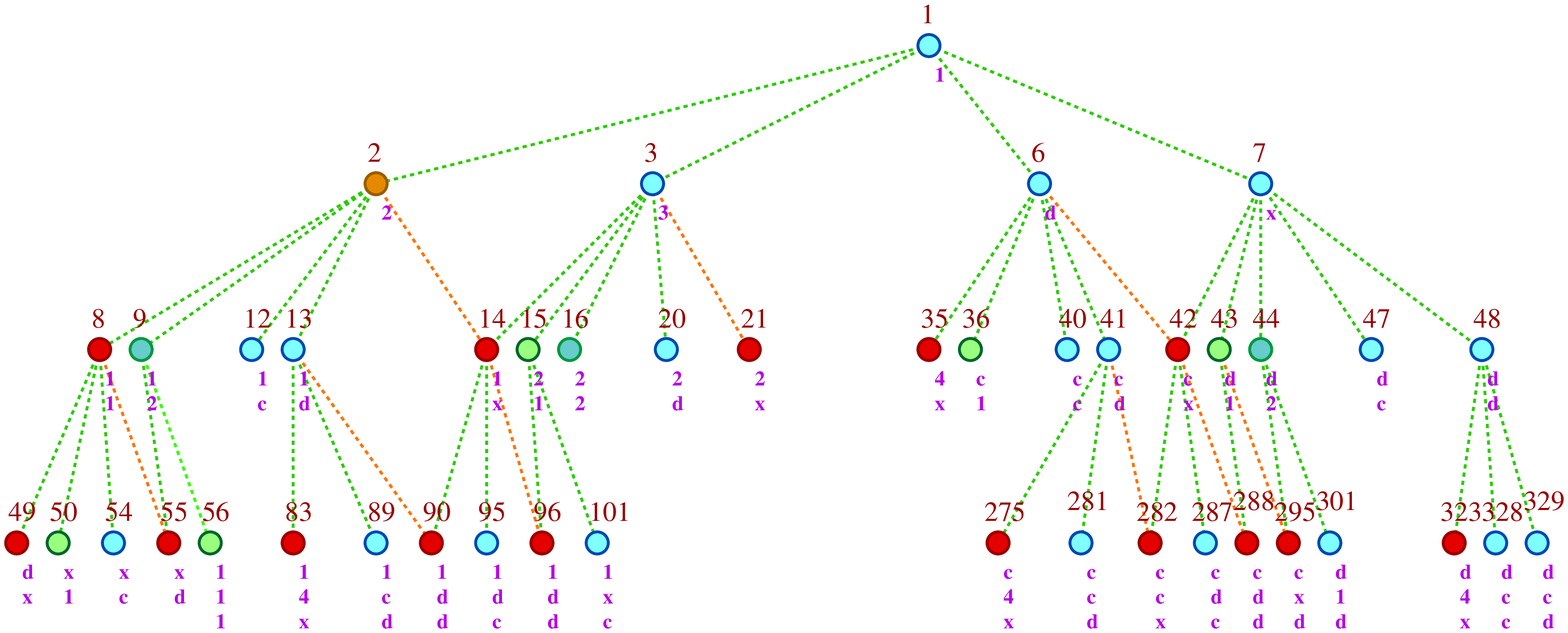}
\hfill}
\vspace{-10pt}
\ligne{\hfill
\vtop{\leftskip 0pt\parindent 0pt\hsize=300pt
\begin{fig}\label{fnzmetalnoirprem}
\leurre
The black metallic tree still with the leftmost assignment but with the \nzm-codes
of the nodes. This time it seems that we have five types of rules for the nodes
in order to define the sons signature. We can see that the preferred son property is 
not true in the present setting.
\end{fig}
}
\hfill}
}
\vskip 5pt

In this situation we can state:

\begin{thm}\label{tnzmblackprem}
In ${\cal B}_{\lambda}$, the black metallic tree dotted with the leftmost assignment,
the rules giving the status and the \nzm-signatures of the sons of a node are given
in~$(35)$, the root being excepted.and the \nzm-codes of the $\lambda$-sons of 
a node~$\nu$ are given by Table~$(36)$. The tree ${\cal B}_{\lambda}$ under the 
leftmost assignment has no preferred son property in term of the \nzm-codes of its nodes.
\end{thm}

\noindent 
Proof.
Indeed, it is immediate to see that the root of~${\cal B}_{\rho}$ is applied the
rule~\rrr\uu{} of~$(35)$ and that its leftmost son, \numd, is applied the 
rule~\bbb\numd{} of~$(35)$. The \nzm-codes of the respective sons are given in 
Table~$(34)$, on lines~\fnb 1 and~\fnb 2 for \rrr\uu, online~\fnb 3 for \bbb\numd. 
Lemma~\ref{lnzmuudist} and Algorithm~\ref{anzmincr} show us that the rule~\www\aa{}
applies to the white sons of a node, provided that the rule~\bbb{} holds for the black 
nodes. The rightmost node on level~1 is \xxx{} and it is \ddd\ccc$^{k-2}$\ddd{}
on the level~$k$, starting from \hbox{$k=2$}. Consequently,
The leftmost node of the level~2 is \uu\uu{} and the leftmost node of the
further levels is \ddd\ccc$^{k-2}$\xxx{} which explains the rules \bbb\uu{} and
\bbb\xxx. Those latter rules are contained in the rule~\bbb{} of~$(35)$
which stands for any black node, \uu{} and \numd{} being applied the rules \rrr\uu{}
and \bbb\numd{} of~$(35)$ as already noticed.
\vskip 5pt
\ligne{\hfill
$\vcenter{\vtop{\leftskip 0pt\hsize=260pt
\ligne{\hfill
\rrr\uu{} $\rightarrow$ \bbb\numd$,$\www\numt$,..,$\www\xxx$,$\hskip 10pt
\bbb\numd{} $\rightarrow$ \bbb\uu$,$\www\numd$,..,$\www\ddd$,$ \hskip 10pt
\bbb{} $\rightarrow$ \bbb\xxx$,$\www\uu$,..,$\www\ccc$.$
\hfill}
\ligne{\hfill
\www\uu{} $\rightarrow$ \bbb\ddd$,$\www\xxx$,..,$\www\ccc$,$ \hskip 10pt 
\www\numd{} $\rightarrow$ \bbb\ddd$,$\www\uu$,..,$\www\ddd$,$
\hfill}
\ligne{\hfill
\www\aa{} $\rightarrow$ \bbb\xxx$,$\www\uu$,..,$\www\ddd$.$ \hskip 10pt 
with \aa $>$ \numd$.$
\hfill}
}}$
\hfill$(35)$\hskip 10pt}
\vskip 10pt
The particular forms of the \nzm-codes of the leftmost nodes of a 
level together with Algorithm~\ref{anzmincr} explain lines~\fnb 4 and~\fnb 5 of 
Table~$(36)$. The \nzm-codes of the rightmost node of a level and Algorithm~\ref{anzmdecr}
explain lines~\fnb {11} and~\fnb {12} of the table.
\vskip 10pt
In Table~$(36)$, we need to consider 
\hbox{\bbb$_k..$\bbb$_0$ $\rightleftharpoons$ $[\nu$$-$$1]$} together with
\hbox{\fff$_k..$\fff$_0$ $\rightleftharpoons$ $[\nu$$-$$2]$}.
\vskip 5pt
\ligne{\hfill
$\vcenter{\vtop{\leftskip 0pt\hsize=260pt
\llaligne {$\nu$} {range} {son} {metallic code} {ref.}
\vskip 5pt
\hrule height 0.3pt depth 0.3pt width \hsize
\vskip 5pt
\llaligne {\rrr\uu} {$1$$..$$p$$-$$4$} {$h$} {\hhh$^{\copy110}$} {1}
\llaligne {} {$p$$-$$3$} {} {\xxx} {2}
\llaligne {\bbb\numd} {$1$$..$$p$$-$$3$} {$h$} {\uu\hhh} {3}
\llaligne {\bbb} {$1$} {} {[\fff$_k$..\fff$_0$]\xxx} {4}
\llaligne {} {$2$$..$$p$$-$$3$} {$h$} {$[$\bbb$_k$$..$\bbb$_0]$\hhh$^{\copy120}$} {5}
\llaligne {\www\uu} {$1$} {} {[\fff$_k$..\fff$_0$]\ddd} {6}
\llaligne {} {$2$} {} {[\fff$_k$..\fff$_0$]\xxx} {7}
\llaligne {} {$3$$..$$p$$-$$2$} {$h$} {[\bbb$_k$..\bbb$_0$]\hhh$^{\copy120\copy120}$} {8}
\llaligne {\www\numd} {$1$} {} {[\fff$_k$..\fff$_0$]\ddd} {9}
\llaligne {} {$2$$..$$p$$-$$2$} {$h$} {[\bbb$_k$..\bbb$_0$]\hhh$^{\copy120}$} {10}
\llaligne {\www\aa} {$1$} {} {[\fff$_k$..\fff$_0$]\xxx} {11}
\llaligne {} {$2$$..$$p$$-$$2$} {$h$} {[\bbb$_k$..\bbb$_0$]\hhh$^{\copy120}$} {12}
\vskip 5pt
\hrule height 0.3pt depth 0.3pt width \hsize
}}$
\hfill $(36)$\hskip 10pt}
\vskip 10pt
Let us see that the table is relevant for the other nodes. Starting from the leftmost
node, we can see that Algorithm~\ref{anzmincr} and Lemma\ref{lnzmuudist} show
us that for a white sons of a node~$\nu$, the first one has the signature \xxx{}
and the \nzm-code is based on~[$\nu$$-$2]: it can be seen on the leftmost $\lambda$-son
of the second node on a level. It explains the lines~\fnb {11} and~\fnb {12} as long as
a black son is not met. The black son interrupts the sequence \hbox{\xxx\uu...\ddd}
of the \nzm-signatures on \ccc. So that for \www\uu, the signature of its leftmost 
$\lambda$-son is \ddd. Accordingly, the sequence of \nzm-signatures becomes
\hbox{\ddd\uu..\ddd}, so that the sequence of \nzm-signatures defined for the rule
\www\aa{} may again apply. This corresponds with the occurrence of a pattern
\xxx\ddd$^\ast$ which is a reason why after \ddd{} we have \uu{} in the signatures,
but it comes from the \nzm-signature of~$\nu$ which is then \uu: but necessarily,
the \nzm-signature of~$nu$$-$1 was \ddd. This also explains the \nzm-codes
given in Table~$(36)$. The lack of preferred son property is also a consequence of
the table. Accordingly, the proof of Theorem~\ref{tnzmblackprem} is completed.
\hfill $\Box$

\subsubsection{The black metallic tree under the rightmost assignment and the \nzm-codes}
\label{snzmblackders}

   Figure~\ref{fnzmetalnoirder} illustrates ${\cal B}_{\rho}$. The conventions for
the representation are the same as for Figure~\ref{fnzmetalnoirprem}. At first glance,
the structure seems to be more regular than in the case of the leftmost assignment.
However, it also seems to do not observe the preferred son property, whatever the digit
\aa{} chosen in \hbox{\uu,..,\ddd,\xxx}. The rules for the nodes are given in~$(35)$
and the \nzm-codes for the $\rho$-sons of a node are given by Table~$(36)$.

We can see that the rule is applied the rule~\rrr\uu{} of~$(37)$ and that
\numd{} is applied the rule~\www\aa{} of~$(37)$. Lemma~\ref{lnzmuudist} explains
that the \nzm-signatures \hbox{\uu..\ddd\xxx} appearing in the rule~\www\aa{} is
repeated as long as we are in the white $\rho$-sons of a node. When a black son
occurs, the sequence stops at \ddd{} as indicated in the rule~\bbb{} of~$(37)$.
Let $\nu$ be the $\rho$-black node for which the sons \nzm-signature is thus
\hbox{\uu..\ddd}, we have that the signature of the leftmost $\rho$-son of $\nu$+1 
is~\xxx. Accordingly, the sons \nzm-signature for~$\nu$+1 is \hbox{\xxx\uu..\ddd}
which appears in the rule~\www\uu{} of~$(37)$. In order to see why the \nzm-signature
of the rightmost $\rho$-son of~$\nu$+1, namely \ddd, is followed by the \nzm-signature
\uu, we have to look at Table~$(38)$.
\vskip 10pt
\vtop{
\ligne{\hfill
\includegraphics[scale=0.35]{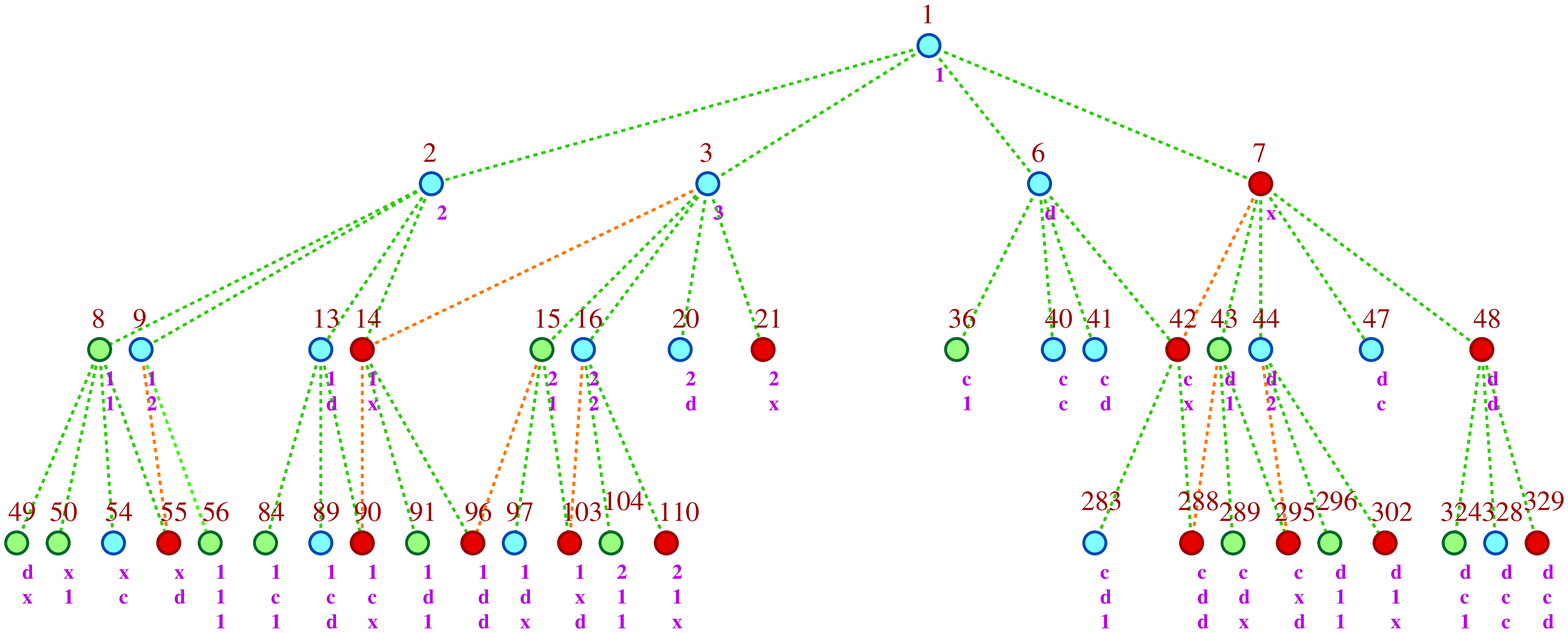}
\hfill}
\vspace{-10pt}
\ligne{\hfill
\vtop{\leftskip 0pt\parindent 0pt\hsize=300pt
\begin{fig}\label{fnzmetalnoirder}
\leurre
The black metallic tree still with the rightmost assignment but with the \nzm-codes
of the nodes. This time it seems that we have three types of rules for the nodes
in order to define the sons signature. We can see that the preferred son property is 
not true in the present setting.
\end{fig}
}
\hfill}
}
\vskip 5pt
\vskip 5pt
\ligne{\hfill
$\vcenter{\vtop{\leftskip 0pt\hsize=260pt
\ligne{\hfill
\rrr\uu{} $\rightarrow$ \www\numd$,..,$\www\ddd$,$\bbb\xxx$,$\hskip 10pt
\bbb{} $\rightarrow$ \www\uu$,..,$\www\ccc$,$\bbb\ddd$.$
\hfill}
\ligne{\hfill
\www\aa{} $\rightarrow$ \www\uu$,..,$\www\ddd$,$\bbb\xxx$,$ \hskip 10pt 
with \uu{} $<$ \aa $<$ \xxx$,$\hskip 10pt
\www\uu$,$\www\xxx{} $\rightarrow$ \www\xxx$,$\www\uu$,..,$\www\ccc$,$\bbb\ddd$.$
\hfill}
}}$
\hfill$(37)$\hskip 10pt}
\vskip 10pt
The particular forms of the \nzm-codes of the leftmost nodes of a 
level together with Algorithm~\ref{anzmincr} explain lines~\fnb 4 and~\fnb 5 of 
Table~$(38)$.
\vskip 5pt
\ligne{\hfill
$\vcenter{\vtop{\leftskip 0pt\hsize=260pt
\llaligne {$\nu$} {range} {son} {metallic code} {ref.}
\vskip 5pt
\hrule height 0.3pt depth 0.3pt width \hsize
\vskip 5pt
\llaligne {\rrr\uu} {$1$$..$$p$$-$$4$} {$h$} {\hhh$^{\copy110}$} {1}
\llaligne {} {$p$$-$$3$} {} {\xxx} {2}
\llaligne {\bbb} {$1$$..$$p$$-$$3$} {$h$} {$[$\bbb$_k$$..$\bbb$_0]$\hhh} {3}
\llaligne {\www\uu,\www\xxx} {$1$} {} {[\fff$_k$..\fff$_0$]\xxx} {4}
\llaligne {} {$2$$..$$p$$-$$2$} {$h$} {[\bbb$_k$..\bbb$_0$]\hhh$^{\copy120}$} {5}
\llaligne {\www\aa} {$1$$..$$p$$-$$3$} {$h$} {[\bbb$_k$..\bbb$_0$]\hhh} {6}
\llaligne {} {$p$$-$$2$} {} {[\bbb$_k$..\bbb$_0$]\xxx} {7}
\vskip 5pt
\hrule height 0.3pt depth 0.3pt width \hsize
}}$
\hfill $(38)$\hskip 10pt}
\vskip 10pt
Let us see that the table is relevant for the other nodes. Starting from the leftmost
node on a level, we can see that lines~\fnb 4 and~\fnb 5 apply: from our study of
Sub-subsection~\ref{snzmblackprems}, the leftmost node on level~$n$+2 with 
\hbox{$n\in\mathbb N$} is \ddd\ccc$^n$\xxx. This proves lines~\fnb 4 and~\fnb 5 for 
the leftmost node $\lambda_{n+2}$ of the level~$n$+2. Now, as the \nzm-signatures 
of~$\lambda_{n+2}$ are \hbox{\xxx\uu..\numd}, the \nzm-code of the leftmost $\rho$-node 
of $\lambda_{n+2}$+1 is \ddd\ccc$^{n-1}$\ddd\xxx{}, so that line~\fnb 4 and~\fnb 5
again apply but this time, the \nzm-code of the rightmost $\rho$-son of~$\lambda_{n+2}$+1
is \ddd\ccc$^{n-1}$\xxx\ddd, so that the leftmost $\rho$-son of $\lambda_{n+2}$+2 is
\ddd\ccc$^{n-2}$\ddd\uu\uu: accordingly, the rule~\www\aa{} apply to $\lambda_{n+2}$+2
and the \nzm-codes of its $\rho$-sons are those which are indicated by lines~\fnb 6
and~\fnb 7 of Table~$(38)$. As in the case of Table~$(36)$, the fact that a white node
\www\uu{} whose rightmost son has the signature \ddd{} is followed by a white node
\www\numd{} whose leftmost son has the signature \uu{} comes from two features.
The first one is that, according to our study, the successor of~$\nu$ occurs in as the 
leftmost or the second $\rho$-son of~$\nu$+1. Note that the \nzm-signature \xxx{}
is that of the leftmost $\rho$-son of a \www\uu- or \www\xxx-node or that of the 
rightmost $\rho$-son of a \www\aa-node with \hbox{\uu{} $<$ \aa{} $<$ \xxx}. Let $\nu$
be such a node. Accordingly, the suffix \xxx\ddd{} occurs in the \nzm-code of
$\nu$+1 as its leftmost $\rho$-son if $\nu$ is a \www\aa-node, so that the \uu-node
is the next son of~$\nu$+1. If $\nu$ is a \www\uu-node, the suffix \xxx\ddd$^\ast$
is the rightmost $\rho$-son of~$\nu$, so that the the \nzm-code of the leftmost
$\rho$-son of~$\nu$+1 is the successor of~$\nu$. If $\nu$ is a \www\xxx-node,
the suffix \xxx\ddd$^\ast$ is the rightmost $\rho$-son of~$\nu$+1 so that the leftmost
$\rho$-son of~$\nu$+2 is the successor of~$\nu$+1. This can be checked by induction on
the rules of Table~$(38)$. This allows us to state:

\begin{thm}\label{tnzmblackder}
Let ${\cal B}_{\rho}$ be $\cal B$, the black metallic tree, equipped with the rightmost 
assignment $\rho$. The rules which allow us to construct the tree under that assignment
are given in~$(37)$ and the \nzm-codes of the $\rho$-sons of a node~$\nu$ are 
given in $(38)$ in terms of the \nzm-codes of $\nu$$-$$1$ and of $\nu$$-$$2$.
Under that assignment, the tree does not observe the preferred son property, whatever the
digit chosen for that purpose. The successor of the node~$\nu$ is a $\rho$-son of 
$\nu$$+$$1$: its leftmost $\rho$-son or the next $\rho$-son of $\nu$$+$$1$. No assignment
allows to establish any preferred son property on $\cal B$. 
\end{thm}

\noindent
Proof. The largest part of the proof is given with the proof of~$(37)$ and of 
Table~$(38)$. For what is the assignment and a preferred son property, the fact that 
the successor of~$\nu$ is a son of~$\nu$+1 prevents the definition of another assignment
which would contain [$\nu$]$_{nz}$\aa, whatever the digit \aa{} as far as 
\hbox{\uu $\leq$ \aa}: the rightmost assignment allows us to grasp as far as possible
a complete set of \nzm-signatures, some circular permutation on
\hbox{$\{$\uu,..,\ddd,\xxx$\}$}.
If we change the
the $\rho$-status of the black son of the node~$\nu$ to an $\alpha$-white one, we must
change the $\rho$-status of a white son of~$\nu$ to an $\alpha$-black one, which reduces
the range grasped by the $\alpha$-white nodes which lies after the $\alpha$-black son.
Accordingly, no preferred son property can be observed in~${\cal B}_{\alpha}$ for all
the nodes of the tree. The proof of Theorem~\ref{tnzmblackder} is now completed.
\hfill $\Box$.

\section{Connection of the metallic trees with the tilings $\{p,4\}$
and $\{p$+$2,3\}$ of the hyperbolic plane}~\label{tilings}

   With the previous sections, we established the properties of the metallic trees.
As already mentioned in~\cite{mmarXiv2}, the metallic trees are connected with
two families of tilings of the hyperbolic plane: the tilings $\{p,4\}$ and the
tilings $\{p$+$2,3\}$. The first tiling is generated by the regular convex polygon
with $p$~sides and the right angle as interior angle at each vertex by reflections
in its sides and, recursively, by the reflections of the images in their sides. The 
second one is generated in the same way from the regular convex polygon with $p$+2
sides and with the angle $\displaystyle{{2\pi}\over3}$ as interior angle at each vertex.
Those angles indicate that four tiles share the same vertex in $\{p,4\}$ and
that three of them do the same in $\{p$+$2,3\}$. There is another way to generate 
those tilings which rely on the metallic trees. Figure~\ref{f7_9til} illustrates the
considered tilings in the case when \hbox{$p=7$} and Figure~\ref{fspan7_9til} illustrates
the role of the metallic trees in the same tilings.

\vskip 10pt
\vtop{
\ligne{\hfill
\includegraphics[scale=0.6]{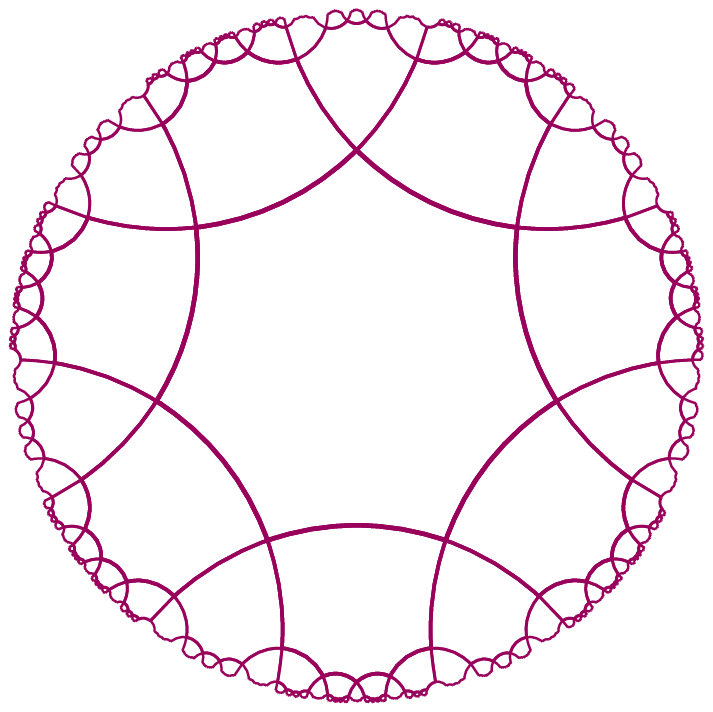}
\includegraphics[scale=0.6]{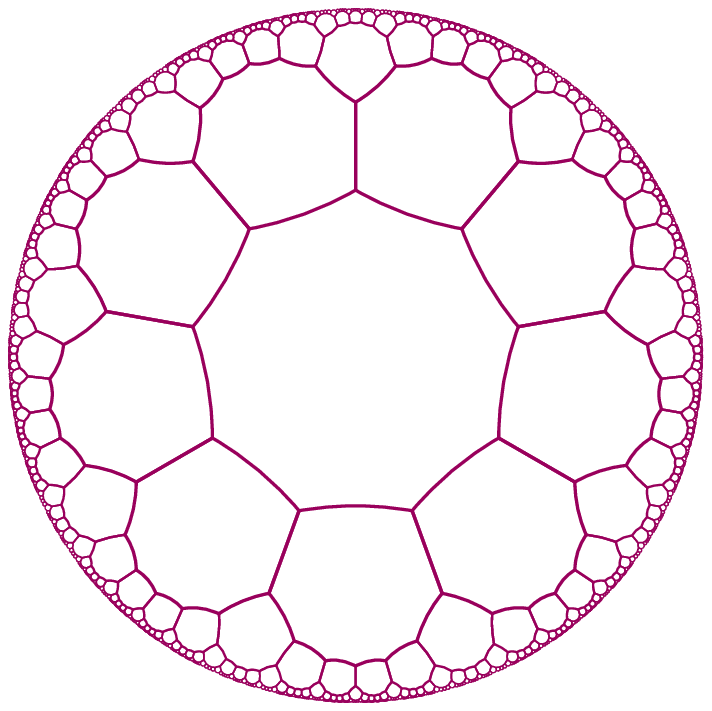}
\hfill}
\vspace{-10pt}
\ligne{\hfill
\vtop{\leftskip 0pt\parindent 0pt\hsize=300pt
\begin{fig}\label{f7_9til}
\leurre
The tilings generated by the white metallic tree with $p=7$.
To left, the the tiling $\{7,4\}$ to right, the tiling $\{9,3\}$
\end{fig}
}
\hfill}
}
\vskip 10pt
In Sub section~\ref{stilsctrstrp}, we define the regions of the tilings which are
associated with the metallic trees and in Sub section~\ref{sconnectsctrstrp}, we 
explain the correspondence between the trees and the regions. In Sub 
section~\ref{smwhitetil} we look carefully at the case of the white metallic tree
and in Sub section~\ref{smblacktil}, we study the case of the black one.

\subsection{Sectors and strips in the tilings $\{p,4\}$ and $\{p$+$2,3\}$ of the
hyperbolic plane}\label{stilsctrstrp}

The metallic trees are associated with two kinds of regions of the considered tilings.
The sub section is devoted to the definition of those regions.

The regions addressed by the white metallic tree is called a {\bf sector}.
In $\{p,4\}$ a sector of the tiling is defined by two rays~$u$ and~$v$ issued from a 
vertex~$V$ of a tile~$T$, $u$ and~$v$ being supported by the sides of~$T$ which meet 
at~$V$. The sector defined by~$u$ and~$v$ is the set of tiles whose center is contained
in the right angle defined by those rays.
The left hand-side picture of Figure~\ref{fsect7_9til} illustrates how $p$~sectors 
can be displayed around a once and for all fixed tile which we call 
the {\bf central tile}, say $T_0$. The sectors and the central tile cover the hyperbolic 
plane with no hole and their interiors do not intersect.

The right hand-side picture of the figure illustrates the same display of sectors around
$T_0$ in the tiling $\{p$+$2,3\}$ with, this time, $p$+2 sectors around
the central tile. However, in $\{p$+$2,3\}$, the definition of a sector is more 
complicate. It is again defined by two rays~$u$ and~$v$. Consider a tile~$T$, a 
vertex~$V$ of~$T$. Two sides of~$T$ meet at~$V$, say \aa{} and~\bbb, and a third 
side~\ccc, belonging to the other tiles sharing~$V$ with~$T$, also meets~$V$. Then
$u$ and~$v$ are issued from the midpoint of~\ccc, $u$ and~$v$ passing through the
midpoints of~\aa{} and~\bbb{} respectively. The sector defined by~$u$ and~$v$ is the
set of tiles whose center lies in the acute angle defined by~$u$ and~$v$. The $p$+2
sectors around~$T_0$ in $\{p$+$2,3\}$ and $T_0$ cover the hyperbolic plane with no hole
and their interiors do not intersect.

\vskip 10pt
\vtop{
\ligne{\hfill
\includegraphics[scale=0.6]{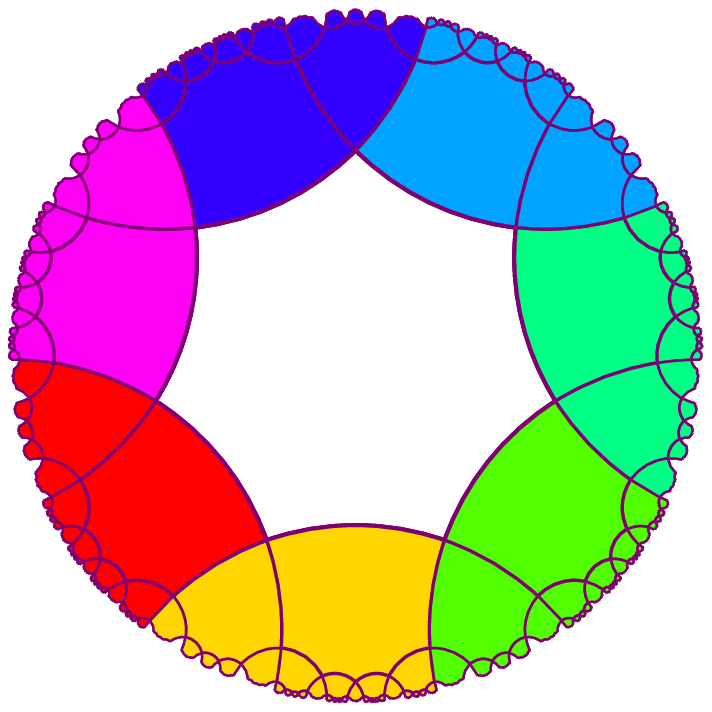}
\includegraphics[scale=0.6]{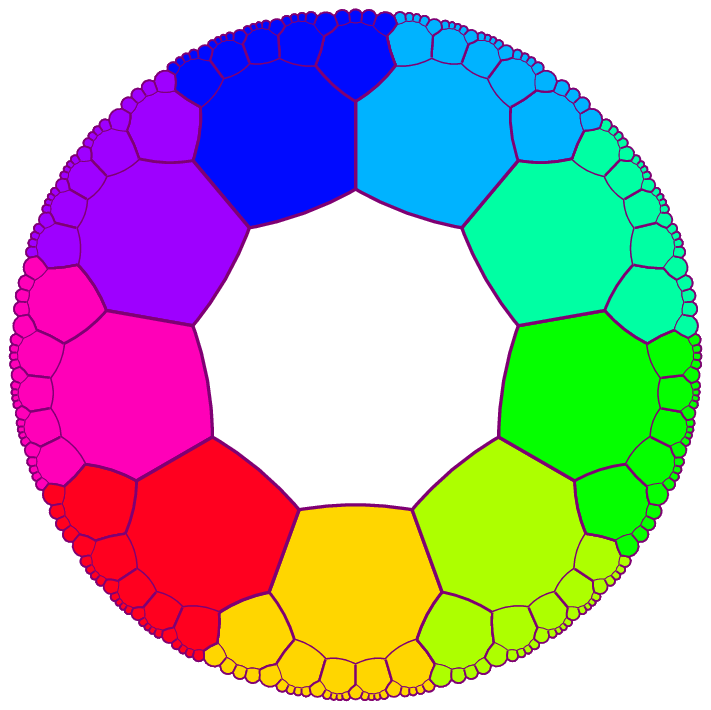}
\hfill}
\vspace{-10pt}
\ligne{\hfill
\vtop{\leftskip 0pt\parindent 0pt\hsize=300pt
\begin{fig}\label{fsect7_9til}
\leurre
The sectors around the central tile fixed once and for all. 
\end{fig}
}
\hfill}
}
\vskip 10pt
In both tilings, the tile~$T$ we above considered to define a sector is called the
{\bf head} of the sector or, also, its {\bf leading tile}.
\vskip 5pt
Presently, let us define the {\bf strips} in those tilings. 

\vskip 10pt
\vtop{
\ligne{\hfill
\includegraphics[scale=0.6]{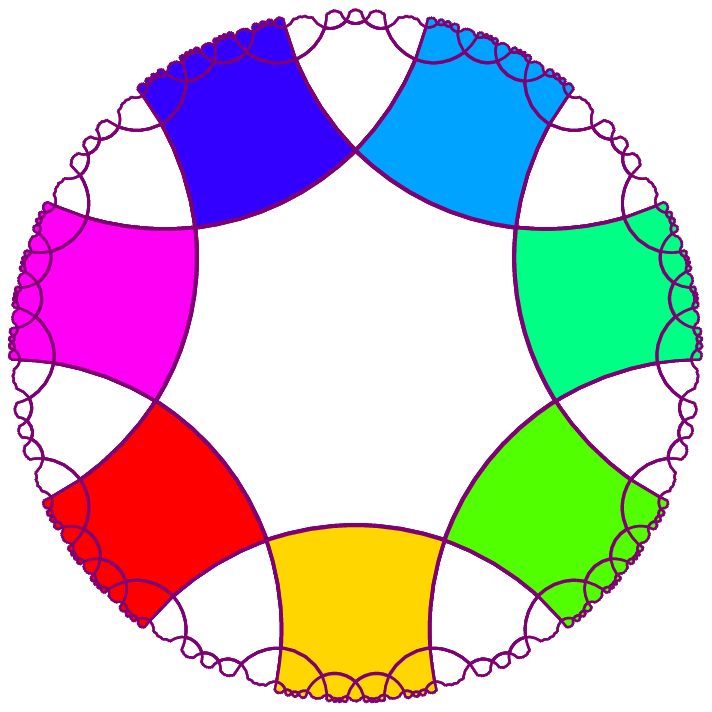}
\includegraphics[scale=0.6]{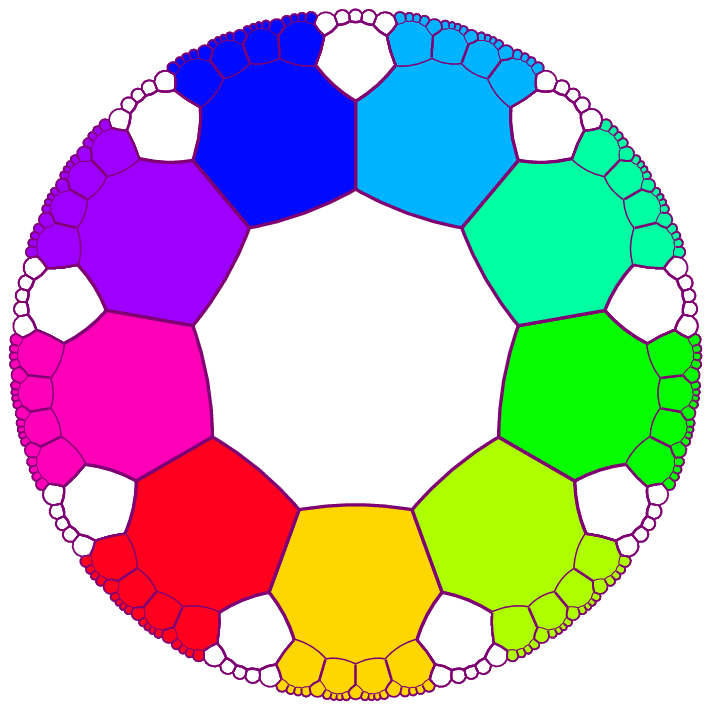}
\hfill}
\vspace{-10pt}
\ligne{\hfill
\vtop{\leftskip 0pt\parindent 0pt\hsize=300pt
\begin{fig}\label{fstrip7_9til}
\leurre
The strips around the central tile fixed once and for all. 
\end{fig}
}
\hfill}
}
\vskip 5pt
In $\{p,4\}$, a strip is defined by two rays~$u$ and~$v$ together with a side~\aa{}
of a tile~$T$, $u$ and~$v$ being issued from the ends of~\aa{} and being supported by 
the sides of~$T$ which meet~\aa. The left hand-side of Figure~\ref{fstrip7_9til} 
illustrates the strip in $\{p,4\}$. The strip, in the tiling, is the set of tiles whose
centre lies in the intersection of the three closed half-planes defined by~$u$, $v$ 
and~\aa{} which contain \aa{} and the rays. We can see in the figure that a strip is, in 
some sense, smaller than a sector. As the figure points at that, the $p$~strips
displayed around the central tile and~$T_0$ itself do not cover the hyperbolic plane.
As can be seen on the figure, in between two strips associated by two consecutive sides 
of~$T_0$, there is a sector.

In $\{p$+$2,3\}$, a strip is also defined by two rays~$u$ and~$v$ together with a 
side~\aa{} of a tile~$T$. Let \bbb{} and~\ccc{} be the sides of~$T$ which share a vertex
with~\aa. Then, $u$, $v$ is the ray issued from the foot of the perpendicular to~\aa{}
issued from the midpoint of~\bbb, \ccc{} respectively which pass through the midpoint 
by which it is defined, also see the right hand-side picture of Figure~\ref{fbandes}. 
In the tiling, the strip is the set of tiles whose centre lies
in the intersection of the three closed half-planes defined by~$u$, $v$ and the line 
supporting~\aa{} which contains that side and the rays. On the right-hand side picture
of Figure~\ref{fstrip7_9til}, we can see that the strips around~$T_0$ together with that
tile do not cover the hyperbolic plane. Applying the definition of a sector in that
context, we can see that in between the strips defined by two consecutive sides of~$T_0$,
there is a sector.

Here too, in both tilings, the tile~$T$ we considered for defining the strip is called
the {\bf head} of the strip or also, its {\bf leading tile}.

\subsection{Connections between sectors and strips as connections between
white and black metallic trees}\label{sconnectsctrstrp}

   It is the time to precisely describe the connection between the metallic trees
and the regions defined in Sub section~\ref{stilsctrstrp}. Figures~\ref{fspan7_9til}
and~\ref{fbandes} illustrate these connections. As shown in~\cite{mmgsJUCS,mmbook1},
there is a bijection between the white metallic tree and a sector of both
$\{p,4\}$ and $\{p$+$2,3\}$ for the same value of~$p$ used for defining the tree.

From now on, if $T$ is a tile of the tiling, we number its side starting from~\fnb 1
up to~$h$ with \hbox{$h = p$} or \hbox{$h = p$+2}, depending on whether $T$ belongs 
to $\{p,4\}$ or to $\{p$+$2,3\}$ respectively. Once side~\fnb 1 is fixed, the other
sides are increasingly numbered from~1 while counterclockwise turning around the tile
starting from side~\fnb 1. Denote by $(T)_i$, with \hbox{$i\in\{1..h\}$} the tile which 
shares the side~\fnb i of $T$ with that latter tile. In a tiling, a tile which shares 
a side with~$T$ is called a {\bf neighbour} of~$T$. 

The comparison between Figure~\ref{fsect7_9til} and~\ref{fspan7_9til} allows us to better
see the tree structure in a sector. The idea is to associate white nodes to the head of
a sector and black nodes to the head of a strip. 

First, consider the case of the tiling $\{p,4\}$. The root of the white metallic tree
is associated with the head~$T$ of a sector $\cal S$. Let $u$ and~$v$
be the rays defining $\cal S$. We fix number~\fnb 1{} in such a way that side~\fnb~1
is supported by~$u$, so that side~\fnb p is supported by~$v$, exchanging the names 
of~$u$ and~$v$ if necessary for the numbering of the sides of~$T$. From that numbering 
and the definition of a sector, $(T)_1$ and $(T)_p$ are outside $\cal S$. 
From \cite{mmgsJUCS,mmbook1}, we know that the neighbours $(T)_i$ of $T$ with 
\hbox{$i\in \{2..p$$-$$1\}$} are in $\cal S$. We precisely 
associate the $\lambda$-sons of the root in the order of their numbers
to the $(T)_i$'s inside $\cal S$ in the order of their numbers too.
Next, consider a tile~$\tau$ already associated with a node~$\nu$ of $\cal W$. We 
number the sides of~$\tau$ as already mentioned, the number~\fnb 1 being given to the 
side shared with the tile associated to the father of~$\nu$. If $\nu$ is white, we 
associate its $\lambda$-sons in the order of their numbers to the neighbours $(\tau)_i$ 
of~$\tau$ with $i$ in \hbox{$\{2..p$$-$$1\}$} in that order.
If $\nu$ is black, we associate its $\lambda$-sons in the order of their numbers
to the neighbours $(\tau)_j$ of~$\tau$ with $j$ in \hbox{$\{3..p$$-$$1\}$} in that order
too. From~\cite{mmgsJUCS,mmbook1}, it is known that this process establishes a bijection
between the nodes of $\cal W$ and the tiles of~$\cal S$. 

Similarly, consider a strip $\mathfrak S$
defined by the rays $u$, $v$ and the side~\aa{} of $T$, its leading tile. Fix \aa{} as
side~\fnb 1 of~$T$ and let side~\fnb 2 be supported by~$u$ and side~\fnb p be supported
by~$v$, exchanging the names of~$u$ and~$v$ if needed by the numbering of the sides
of~$T$. Then $(T)_1$, $(T)_2$ and $(T)_p$ are outside $\mathfrak S$ while the 
neighbours $(T)_j$ of~$T$ with $j$ in \hbox{$\{3..p$$-$$1\}$} are in the strip, 
see \cite{mmgsJUCS,mmbook1}. We can repeat the above process, considering the
head of $\mathfrak S$ as associated to the root of $\cal B$ as there are exactly
$p$$-$3 neighbours of~$T$ inside $\mathfrak S$.
It is not difficult to prove from that that the same process as for $\cal S$
starting from the head $T$ of $\mathfrak S$ establishes a bijection between the nodes 
of $\cal B$ and the tiles of~$\mathfrak S$. The reason is that $\cal B$ can be obtained 
from ${\cal W}_{\lambda}$ by removing the sub tree rooted at the rightmost son of the 
root of $\cal W$, and that subtree is isomorphic to $\cal W$. Now, it is proved 
in~\cite{mmgsJUCS,mmbook1}, that a strip $\mathfrak R$ can be obtained from a
sector $\cal S$ with head $T$ by removing the image of the sector defined by the 
sides~\fnb 1 and~\fnb p of $(T)_{p-1}$, the last neighbour of~$T$ in $\cal S$,
see also Figure~\ref{fbandes}, and the head of $\mathfrak R$ is $T$ too.

\vskip 10pt
\vtop{
\ligne{\hfill
\includegraphics[scale=0.6]{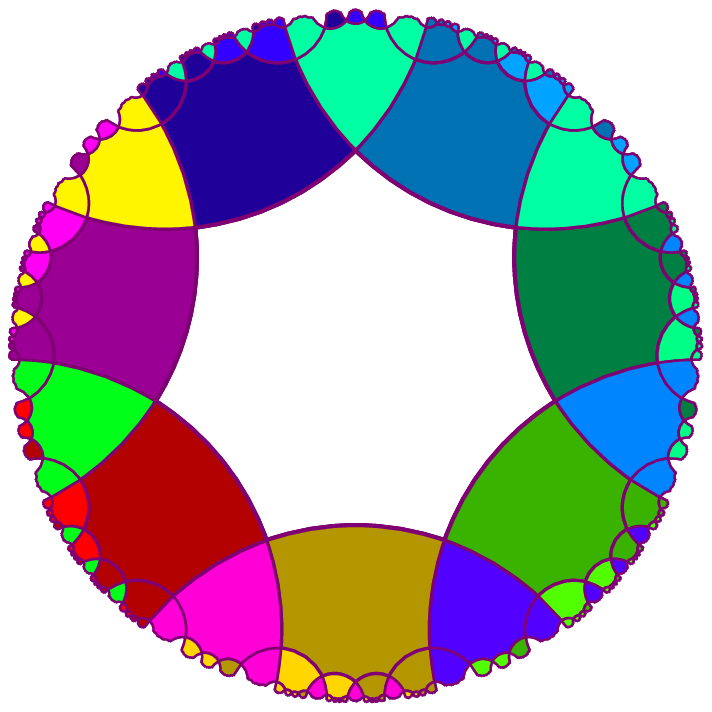}
\includegraphics[scale=0.6]{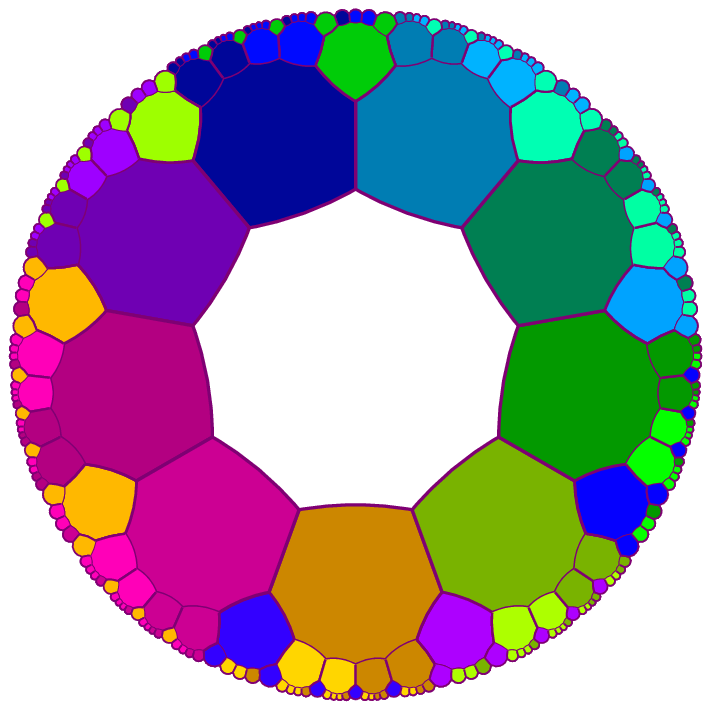}
\hfill}
\vspace{-10pt}
\ligne{\hfill
\vtop{\leftskip 0pt\parindent 0pt\hsize=300pt
\begin{fig}\label{fspan7_9til}
\leurre
How the white metallic tree generates the tilings $\{7,4\}$ and $\{9,3\}$:
the sectors are delimited by colours, each sector being associated with three colours
which are attached to the status of the nodes. Each sector in the above figures is 
spanned by the white metallic tree.
\end{fig}
}
\hfill}
}
\vskip 10pt
Secondly, consider the case of the tiling $\{p$+$2,3\}$. Again, we associate the
root of $\cal W$ with the head~$T$ of a sector~$\cal S$. Let $u$ and $v$ be the
rays defining $\cal S$ and let \aa{} be the side of another tile which meets~$T$ at 
the vertex belonging to the consecutive sides of~$T$ met by~$u$ and~$v$ at their 
midpoints. Let the side~\fnb 1 met by~$u$ while the side~\fnb {p+2} is met by~$v$,
exchanging the names of~$u$ and~$v$ if needed in order to be coherent with the numbering
of the sides of~$T$. We can see that the tiles $(T)_1$, $(T)_2$ and~$(T)_{p+2}$ have
their centre outside $\cal S$. It is proved in~\cite{mmbook1} that the neighbours
$(T)_i$ of~$T$ with \fnb i in \hbox{$\{3..p\}$} have their centre in $\cal S$.
We apply the same process as in the case of the tiling $\{p,4\}$ with this difference 
that to the $\lambda$-sons of a node~$\nu$ associated to the tile~$\tau$,
we associate in the order of the numbers of the sons the neighbours $(\tau)_i$
with $i$ in \hbox{$\{3..p\}$} in this order if $\nu$ is white and if $\nu$ is black,
we associate the the neighbours $(\tau)_j$ with $j$ in \hbox{$\{4..p\}$}. It is proved
in \cite{mmbook1} that the just described process establishes a bijection between
$\cal S$ and the tiles of a sector in the tiling $\{p$+$2,3\}$. The right hand-side of 
Figure~\ref{fbandes} illustrates the structure of the tree in $\cal S$. It also 
illustrates the fact that the same process establishes a bijection between $\cal B$
and the tiles of a strip $\mathfrak S$ in the tiling $\{p$+$2,3\}$.

\vskip 10pt
\vtop{
\ligne{\hfill
\includegraphics[scale=0.6]{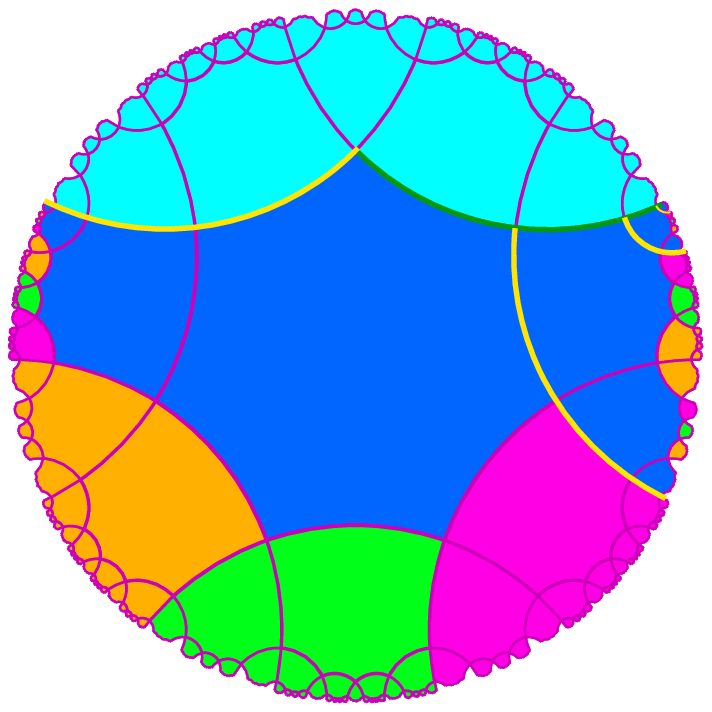}
\includegraphics[scale=0.6]{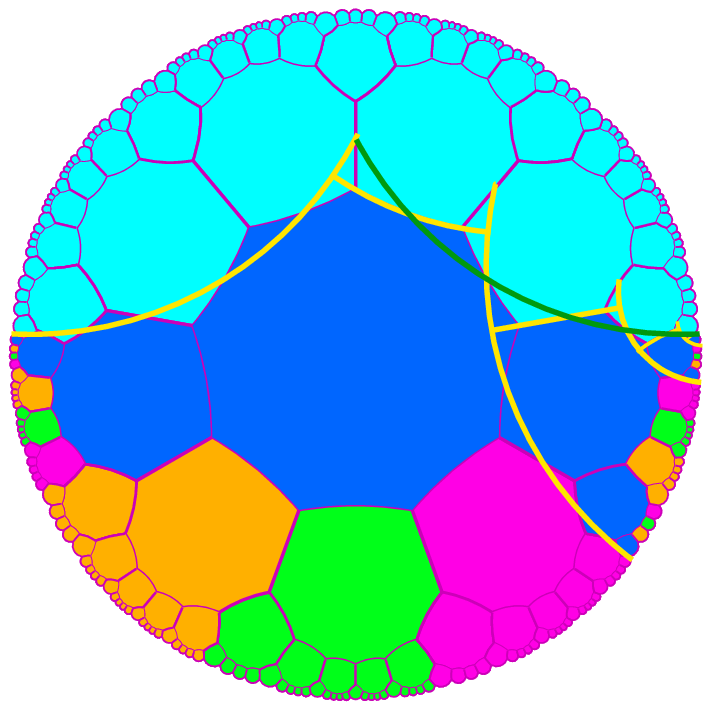}
\hfill}
\vspace{-10pt}
\ligne{\hfill
\vtop{\leftskip 0pt\parindent 0pt\hsize=300pt
\begin{fig}\label{fbandes}
\leurre
The decomposition of a sector spanned by the white metallic tree into a tile, then
two copies of the same sector and a strip spanned by the black metallic tree.
To left: the decomposition in the tiling $\{p,4\}$; to right, the 
decomposition in the tiling $\{p$$+$$2,3\}$.
In both cases, the dark blue colour indicates the black nodes while the white ones
are indicated in dark yellow, in green and in purple.
\end{fig}
}
\hfill}
}
\vskip 5pt

We note that the same tree is in bijection of a sector both in $\{p,4\}$ and
$\{p$+$2,3\}$. The difference of two sides for the regular convex polygons generating
those tilings lies in the fact that as three tiles meet at a vertex instead of four of
them, the number of neighbours of the head which are outside the sector is bigger
in $\{p$+$2,3\}$ than in $\{p,4\}$. It is also the same situation for a strip.
Let $u$, $v$ be the rays and \aa{} be the side of its leading tile~$T$ which define a
strip $\mathfrak S$. Take the side~\fnb 1 of~$T$ as \aa{} and 
Number the other sides as already indicated, exchanging the names of~$u$ and~$v$ 
if needed for side~\fnb {p+2} to be identified with the side which is crossed by~$v$
and which shares a vertex with \aa. Then, it is not difficult
to see that the centres of the neighbours $(T)_1$, $(T)_2$, $(T)_3$ and
$(T)_{p+2}$ are outside $\mathfrak S$. The other neighbours have their centres inside
$\mathfrak S$ and there are $p$$-$3 of them which explains the bijection with
$\cal B$.

   We close this sub section by reminding something we already mentioned in 
\cite{mmarXiv2}. Indeed, we indicated there a property mentioned too in \cite{mmbook2}:
a sector $\cal S$ can be spit into a sequence $\{{\mathfrak S}_{n}\}_{n\in\mathbb N}$.
The first term of the sequence is the strip ${\mathfrak S}_0$ whose head is the head too
of~$\cal S$. Note that according to our conventions, the side~\fnb 1 of~$T$ as the
head of ${\mathfrak S}_0$ is the side~\fnb p of~$T$ as the head of~$\cal S$. 
The ray $u_0$ defining ${\mathfrak S}_0$ is the ray~$u$ defining $\cal S$. The ray~$v_0$
defining ${\mathfrak S}_0$ passes through the midpoint of the side~\fnb p of $T$
as head of ${\mathfrak S}_0$. We take 
this occasion to note that the same side of a tile may receive different numbers 
depending on the context which defines the choice of the side~\fnb 1 which may differ
from one situation to another one. The head of ${\mathfrak S}_{n+1}$ is the
neighbour $(\tau_n)_p$ of the head $\tau_n$ of ${\mathfrak S}_n$. The ray~$u_{n+1}$
which defines ${\mathfrak S}_{n+1}$ is the ray~$v_n$ which defines ${\mathfrak S}_n$,
and the ray~$v_{n+1}$ defining ${\mathfrak S}_{n+1}$ passes through the midpoint of 
the side~\fnb {p+2} of $(\tau_n)_p$, the side~\fnb 1 of that neighbour being the
side it shares with $\tau_n$. The construction of the first elements of that
sequence is illustrated by Figure~\ref{fbandes}, for the tiling $\{p,4\}$ by the
left hand-side picture and by the right hand-side one for the tiling $\{p$+$2,3\}$.
  
   The following sub sections, \ref{smwhitetil} and~\ref{smblacktil} study the
applications of the numbering and their representations to location problems of the
tiles in a sector and in a strip. Such problems are at the basis of an implementation
of cellular automata in the settings of those hyperbolic tilings.

\subsection{The case of the white metallic tree}\label{smwhitetil}

   As explained in Sub section~\ref{sconnectsctrstrp}, the white metallic tree is 
connected with the tilings $\{p,4\}$ and $\{p$+$2,3\}$ of the hyperbolic plane with 
\hbox{$p\geq5$}. Recall that Figure~\ref{f7_9til} illustrates the tiling $\{7,4\}$, left 
hand side, and the tiling $\{9,3\}$, right hand side, associated to $p=7$. In the present
sub section, we take use of the studies of Sections~\ref{assign} and~\ref{black_metal}
in order to solve two location problems of the tiles in a sector of
those tilings. The first problem which we address is to find an algorithm computing
the path from a tile to the head of a sector. The problem is addressed by Sub 
subsection~\ref{smwalgopaths}. The second problem is to compute the codes of the
neighbours of a tile, which is solved in Sub subsection~\ref{smwneighcodes}.

\subsubsection{Algorithm for the path from a tile to the head of a sector}
\label{smwalgopaths}

   In~\cite{mmarXiv2}, we provided an algorithm to compute the path from a tile~$\tau$
of a sector $\cal S$ to the head of~$\cal S$ which was based on the metallic code
of~$\tau$, and $\cal W$ was supposed to be fitted with the leftmost assignment.
Here, we revisit the algorithm, assuming that $\cal W$ is fitted with the
rightmost assignment. In~\cite{mmarXiv2} two algorithms were provided, the first one
reading the digits of $[\tau]$ from the lowest to the highest and the second one
performs the same in the reverse order. In that second algorithm two paths are 
constructed, one to right, the second to left and, eventually the expected path
is the to left one. The second algorithm of~\cite{mmarXiv2} has a decisive advantage:
its complexity is linear in the size of the metallic code of~$\tau$. Accordingly,
we provide a similar algorithm based on the metallic codes as codes for the nodes
of ${\cal W}_{\rho}$.

   To that purpose, let us have a look on Figure~\ref{fmetalblancassder} which we
reproduce as Figure~\ref{fmetalblancassderbis} for the convenience of the reader. Note 
that for a node~$\nu$ of the level~$n$
such that \hbox{$\nu<m_n$}, the metallic code has $n$ digits and when, on the same level,
\hbox{$\nu\geq m_n$}, the metallic code has $n$+1 digits. In that latter case, the
highest digit is \uu. If we look at the highest digit of the metallic codes of the
nodes on level~2, we note it is~\aa{} for the last two sons of the node 
\aa and for all sons of the node \hbox{$(\aa)$+1}, its last two sons being excepted.
At this level, there is an exception when \aa{} is \uu: the last two sons of~\uu\zzz,
the sons of~\uu\uu{} and those of~$\numd$, its last two sons being excepted. And so,
in that case, three sons of three nodes are concerned with \uu{} as the highest digit.
If we look at the nodes of level~3, the second highest digit of their metallic code
is connected with the last digit in the metallic code of nodes of level~2.

More generally. Assume that \hbox{$\nu$ = \aa$_k..$\aa$_1$\aa$_0$}. Table~$(25)$
that for most nodes, the metallic code of their sons but the last two ones
are based on [$\nu$$-$1]. In particular, the second lowest digit
is the last one of [$\nu$$-$1]. Accordingly, if we know the path from the head of a 
sector down to~$\nu$, we know that the node whose metallic code is
[$\nu$]\bbb$_h$..\bbb$_0$ with \bbb$_i$ in \hbox{$\{\zzz,\uu..\ddd\}$} is either
in the sub tree issued from [$\nu$\bbb$_h$] or [([$\nu$\bbb$_h$])+1] at a time
when we know [$\nu$]\bbb$_h$ without knowing the digits \bbb$_i$ with
\hbox{$i<h$}. Algorithm~\ref{apathwmder} answers allows us to compute the path
from the head of~$\cal S$. The path is given as a table whose length is that of
the metallic code of~$\nu$. Each entry of the table contains the indication of
the son~$\sigma$ of a node~$\nu$ as the rank of $\sigma$ among the sons of~$\nu$,
the leftmost son being given rank~1, and it also contains the status of~$\nu$. 
What just mentioned and table~$(25)$ allows us to devise the algorithm.
\vskip 5pt
\vtop{
\ligne{\hfill
\includegraphics[scale=0.35]{metalblancassder.ps}
\hfill}
\vspace{-10pt}
\ligne{\hfill
\vtop{\leftskip 0pt\parindent 0pt\hsize=300pt
\begin{fig}\label{fmetalblancassderbis}
\leurre
The white metallic tree under the rightmost assignment. We are here interested in
the metallic codes.
\end{fig}
}
\hfill}
}
\vskip 5pt

Let us know see the details of that computation.
according to this general principle. Nevertheless,
it is needed to lightly tune the computation. Indeed, if the highest digit is \uu,
we need to know the next one: if the next digit is \zzz, then the left hand-side path
goes through the preferred son and the right hand-side path goes through the rightmost 
son. If the digit is~\uu{} or greater than \uu, the left hand-side path goes through the 
rightmost son and the right hand-side path goes through the leftmost son. Then,
the \ffor-loop deals with the other digits from high ones to low ones.

   In the working of the algorithm, it is assumed that when we examine the current digit
\aa, the left hand-side list$_\ell$ path goes through a node~$\nu$ and the right 
hand-side path list$_r$ goes through~$\nu$+1. The last registered digit \bbb{} occurs 
in the signature of~$\nu$. Let \aa{} be the digit we examine: it is the signature of a 
son of~$\nu$ or of~$\nu$+1. If $\nu$ is white and if \hbox{\bbb{} = \zzz}, then if 
\hbox{\aa{} = \numd}, list$_\ell$ goes on the rightmost branch of the tree rooted 
at~$\nu$ and list$_r$ goes through the leftmost branch of the 
\ligne{\hfill}
\ligne{\hfill
\vtop{
\begin{algo}\label{apathwmder}
The path from the root to the node~$\nu$ in a sector~$\cal S$ in bijection with 
${\cal W}_{\rho}$. We set \hbox{\aa$_k$$..$\aa$_0$ $\rightleftharpoons$ $[\nu]$}. 
The lists register the status of the
current node and its rank among the sons of its father, the leftmost son having
rank~$1$.
\end{algo}
\vskip-10pt
\ligne{\hfill
\vtop{\leftskip 0pt\hsize=310pt
\hrule depth 0.6pt height 0pt width \hsize
\vskip 5pt
\ligne{\hskip 23pt\pproc{} update (side; from; upto) \iis\hfill}
\ligne{\hskip 23pt\bbegin{} \iff{} side = \fnb {left}\hfill}
\ligne{\hskip 33pt\hskip 23pt \tthen{} \ffor{} $j$ \iin{} $\{$from..upto$\}$ 
\lloop{} list$_r$($j$) := list$_\ell$($j$); \endloop;\hfill}
\ligne{\hskip 33pt\hskip 23pt \eelse{} \ffor{} $j$ \iin{} $\{$from..upto$\}$
\lloop{} list$_\ell$($j$) := list$_r$($j$); \endloop;\hfill}
\ligne{\hskip 33pt\hskip 18pt \endif;
\hskip 10pt restart := upto+1; handside := side;\hfill}
\ligne{\hskip 23pt\endproc;\hfill}
\ligne{restart := 0; handside := \fnb {left};\hfill}
\ligne{\iff{} (\aa$_k$ = \uu) \eet{} (\aa$_{k-1}$ = \zzz)\hfill}
\ligne{\hskip 10pt \tthen{} list$_\ell$(0) := \www.$p$$-$3; 
list$_r$(0) := \bbb.$p$$-$2;\hfill}
\ligne{\hskip 10pt \eelse{} \iff{} (\aa$_k$ = \uu) 
\tthen{} list$_\ell$(0) := \bbb.$p$$-$2; list$_r$(0) := \www.1;\hfill}
\ligne{\hskip 33pt \eelse{} list$_\ell$ := \www.(\aa$_i$)$-$1; 
list$_r$ := \www.(\aa$_i$); \endif;\hfill}
\ligne{\endif{};\hfill}
\ligne{\ffor{} $i$ \iin{} \hbox{$\{0..k$$-$1$\}$} \inreverse\hfill}
\ligne{\lloop{} 
\iff{} (status(list$_\ell$($k$$-$$i$$-$1)) = \www)
\etaussi{} (\aa$_{i+1}$ = \zzz)\hfill}
\ligne{\hskip 23pt\hskip 5pt \tthen{} \ccase{} \aa$_i$ \iis\hfill}
\ligne{\hskip 33pt\hskip 25pt \wwhen{} \numd{} $\Rightarrow$
list$_\ell$($k$$-$$i$) = \bbb.$p$$-$2; list$_r$($k$$-$$i$) = \www.1;\hfill}
\ligne{\hskip 33pt\hskip 25pt \wwhen{} \zzz{} $\vert$ \uu{} $\Rightarrow$ \hfill}
\ligne{\hskip 33pt\hskip 23pt\hskip 13pt
update (side: \fnb {left}, from: restart, upto: $k$$-$$i$$-$1);\hfill}
\ligne{\hskip 33pt\hskip 23pt\hskip 13pt
list$_\ell$($k$$-$$i$) = \www.$p$$-$4+(\aa$_i$); 
list$_r$($k$$-$$i$) = \www.$p$$-$3+(\aa$_i$);\hfill}
\ligne{\hskip 33pt\hskip 25pt \wwhen{} \oothers{} $\Rightarrow$ \hfill}
\ligne{\hskip 33pt\hskip 23pt\hskip 13pt 
update (side: \fnb {right}, from: restart, upto: $k$$-$$i$$-$1);\hfill}
\ligne{\hskip 33pt\hskip 23pt\hskip 13pt
list$_\ell$($k$$-$$i$) = \www.(\aa$_i$)$-$2; 
list$_r$($k$$-$$i$) = \www.(\aa$_i$)$-$1;\hfill}
\ligne{\hskip 33pt\hskip 18pt\endcase;\hfill}
\ligne{\hskip 23pt\hskip 5pt \eelse{} last := $p$$-$2; place := (\aa$_i$)$-$1;\hfill}
\ligne{\hskip 23pt\hskip 28pt\iff{} status(list$_\ell$($k$$-$$i$$-$$1$)) = \bbb
\hfill}
\ligne{\hskip 23pt\hskip 33pt \tthen{} last := last$-$1; place := place$-$1;
\endif;\hfill}
\ligne{\hskip 33pt\hskip 33pt \ccase{} \aa$_i$ \iis\hfill}
\ligne{\hskip 33pt\hskip 38pt \wwhen{} \uu{} $\Rightarrow$
list$_\ell$($k$$-$$i$) = \bbb.last; list$_r$($k$$-$$i$) = \www.1;\hfill}
\ligne{\hskip 33pt\hskip 38pt \wwhen{} \zzz{} $\Rightarrow$
update (side: \fnb {left}, from: restart, upto: $k$$-$$i$$-$1);\hfill}
\ligne{\hskip 33pt\hskip 23pt\hskip 24pt
list$_\ell$($k$$-$$i$) = \www.last$-$1; 
list$_r$($k$$-$$i$) = \bbb.last;\hfill}
\ligne{\hskip 33pt\hskip 38pt \wwhen{} \oothers{} $\Rightarrow$ \hfill}
\ligne{\hskip 33pt\hskip 23pt\hskip 24pt 
update (side: \fnb {right}, from: restart, upto: $k$$-$$i$$-$1);\hfill}
\ligne{\hskip 33pt\hskip 23pt\hskip 24pt 
list$_\ell$($k$$-$$i$) = \www.place; list$_r$($k$$-$$i$) = \www.place+1;\hfill}
\ligne{\hskip 33pt\hskip 23pt\endcase;\hfill}
\ligne{\hskip 23pt\endif;\hfill}
\ligne{\endloop;\hfill}
\ligne{\iff{} handside = \fnb {right}\hfill}
\ligne{\hskip 5pt\tthen{} update(side: \fnb {right}, from: restart, upto: $k$);\hfill} 
\ligne{\endif;\hfill}
\vskip 5pt
\hrule depth 0pt height 0.6pt width \hsize
}
\hfill}
}
\hfill}
\vskip 5pt
\noindent
tree rooted at $\nu$+1
so that if $\sigma$ is the new end of list$_\ell$, $\sigma$+1 is that of list$_r$.
The same 
situation occurs for~$\nu$ if its signature is not~\zzz{} and if 
\hbox{\aa{} = \uu}. In the other cases, the number to be remembered
is (\aa)$-$1 for the edge going from the father of~$\nu$ to~$\nu$ for the
left hand-side path and it is (\aa) for the edge to $\nu$+1 for $\nu$+1 stored in
the right hand-side path. Now, if the node to be remembered is the son of the last node
of the left hand-side path, the right hand-side path is identified with 
the
left hand-side one by taking its values. If the node to be remembered is the son of the 
last node stored in the right hand-side path, the same is performed by exchanging 
the roles of the paths. That task is performed by the procedure 'update'. 
The procedure remembers the position of the current digit which will be the 
starting point of the next updating. The procedure also remembers the side to which
the path goes during the updating. It allows the algorithm to perform the possible last 
updating after the \ffor-loop in such a way that the result is the left hand-side path.

\ligne{\hfill
\vtop{
\begin{algo}\label{anzmpathw}
Algorithm for constructing the path from a tile~$\nu$ to the leading tile of its sector
from \hbox{$[\nu]_{nz}$ $=$ \aa$_k$..\aa$_1$\aa$_0$}. In the first \iff, {\fnb 0}
	means that the path remains on the root. The result is in {\rm list$_r$}.
\end{algo}
\vskip -5pt
\ligne{\hfill
$\vcenter{\vtop{\leftskip 0pt\hsize=300pt
\hrule height 0.3pt depth 0.3pt width \hsize
\vskip 5pt
\ligne{$j$ := $k$; update := 0;\hfill}
\ligne{\iff{} \aa$_k$ = \uu\hfill}
\ligne{\hskip 10pt \tthen{} list$_{\ell}$(0) := \rrr.{\fnb 0}; 
list$_r$(0) := \www.{\fnb 1};\hfill}
\ligne{\hskip 10pt \eelse{} 
list$_{\ell}$(0) := \www.{\fnb {(\aa$_k$)$-$1}}; list$_r$(0) := \www.{\fnb {(\aa$_k$)}}; 
\hfill}
\ligne{\endif;\hfill}
\ligne{\ffor{} $j$ \iin{} [0..$k$$-$1] \inreverse\hfill}
\ligne{\lloop{}\hfill}
\ligne{\hskip 23pt \iff{} \aa$_j$ = \uu\hfill}
\ligne{\hskip 23pt\hskip 10pt \tthen{} 
\iff{} status($\pi_\ell$($j$+1)) \iin{} $\{$\www,\rrr$\}$\hfill}
\ligne{\hskip 23pt\hskip 10pt\hskip 23pt\hskip 10pt \tthen{}  
list$_\ell$($k$$-$$j$) := \bbb.{\fnb {p$-$2}};\hfill}
\ligne{\hskip 23pt\hskip 10pt\hskip 23pt\hskip 10pt \eelse{}
list$_\ell$($k$$-$$j$) := \bbb.{\fnb {p$-$3}}; \hfill}
\ligne{\hskip 23pt\hskip 10pt \hskip 23pt\endif;\hfill}
\ligne{\hskip 23pt\hskip 10pt \hskip 23pt list$_r$($k$$-$$j$) := \www.{\fnb 1};\hfill}
\ligne{\hskip 23pt\hskip 10pt \eelse{} 
\ffor{} $i$ \iin{} [update+1..$k$$-$$j$$-$1] \lloop{} 
list$_\ell$($i$) := list$_r$($i$); \endloop;\hfill}
\ligne{\hskip 23pt\hskip 10pt\hskip 23pt update := $k$$-$$j$$-$1;\hfill}
\ligne{\hskip 23pt\hskip 10pt\hskip 23pt 
list$_\ell$($k$$-$$j$) := \www.{\fnb {(\aa$_j$)$-$1}}; 
list$_r$($k$$-$$j$) :=  \www.{\fnb {(\aa$_j$)}}; 
\hfill} 
\ligne{\hskip 23pt\endif;\hfill}
\ligne{\endloop;\hfill}
\ligne{\iff{} \aa$_0$ = \uu\hfill}
\ligne{\hskip 10pt \tthen{} \ffor{} $i$ \iin{} [update+1..$k$] 
\lloop{} list$_r$($i$) := list$_\ell$($i$); \endloop;\hfill}
\ligne{\endif;\hfill}
\vskip 5pt
\hrule height 0.3pt depth 0.3pt width \hsize
}
}$
\hfill}
}
\hfill}
\vskip 10pt

   Algorithm~\ref{anzmpathw} does a similar computation when the \nzm-codes are used
as coordinates of the nodes of ${\cal W}_\rho$. As there are only two rules for the 
rightmost assignment, two rules which are very similar due to the fact that a black node
occurs when the signature~\ddd{} is followed by~\uu{} in the \nzm-code of the next
value of the number, the algorithm is simpler. The path goes to left when
the current digit is~\uu, in case the next digit will be again~\uu. Otherwise, the
path goes to right. Accordingly, the updating occurs only for following the right 
hand-side path. As mentioned in the caption of the algorithm, the result is in the
right hand-side path.

\subsubsection{The codes for the neighbours of a tile in a white metallic tree}
\label{smwneighcodes}

   In the present sub subsection, we turn to another problem. Knowing the coordinate
\hbox{$[\nu]$} or \hbox{$[\nu]_{nz}$} of a tile, how to get the coordinates of the same
type for its neighbours? The answer is given by Table~\ref{tbmetalneighw} for the
metallic codes and by Table~\ref{tbnzmneighw} for the \nzm-ones. Both tables 
consider ${\cal W}_\rho$, {\it i.e.} the metallic tree equipped with the rightmost
assignment. The tables give the neighbours both in $\{p,4\}$ and $\{p$+$2,3\}$.

Table~\ref{tbnzmneighw} is shorter as there are only two rules for ${\cal W}_\rho$
with the \nzm-codes.

\ligne{\hfill
\vtop{
\begin{tab}\label{tbnzmneighw}
Table of the neighbours of~$\nu$: to left, in the tiling $\{p,4\}$, to right,
in the tiling $\{p$$+$$2,3\}$. In the table, the black son of $(\omega)$
is denoted by $(\omega)_b$.
\end{tab}
\ligne{\hfill
\vtop{\leftskip 0pt\hsize=300pt
\ligne{\hfill in $\{p,4\}$\hfill\hfill in $\{p$+$2,3\}$\hfill}
\ligne{\hfill \www-node\hfill}
\vskip 7.5pt
\ligne{\hfill
$\vcenter{
\vtop{\leftskip 0pt\hsize=145pt
\hrule height 0.3pt depth 0.3pt width \hsize
\vskip 5pt
\ddemili {rep.} {tile} {\hfill \nzm-code}
\ddemili {1} {$(\nu)_1$} {[(\aa$_k$..\aa$_1$)$-$1]}
\ddemili {2} {$(\nu$$-$$1)_b$} {[(\aa$_k$..\aa$_1$\aa$_0$)$-$1]\uu}
\ddemili {j} {\www-sons} {[(\aa$_k$..\aa$_1$\aa$_0$)$-$1]\iii}
\ddemili {} {} {$i\in\{2..p$-$2\}$, $j=i$+1}
\ddemili {p} {$(\nu)_b$} {\aa$_k$..\aa$_1$\aa$_0$\uu}
\vskip 5pt
\hrule height 0.3pt depth 0.3pt width \hsize
}}$
\hfill
$\vcenter{
\vtop{\leftskip 0pt\hsize=145pt
\hrule height 0.3pt depth 0.3pt width \hsize
\vskip 5pt
\ddemili {rep.} {tile} {\hfill \nzm-code}
\ddemili {1} {$(\nu)_1$} {[(\aa$_k$..\aa$_1$)$-$1]}
\ddemili {2} {$\nu$$-$1} {[(\aa$_k$..\aa$_1$\aa$_0$)$-$1]}
\ddemili {3} {$(\nu$$-$$1)_b$} {[(\aa$_k$..\aa$_1$\aa$_0$)$-$1]\uu}
\ddemili {j} {\www-sons} {[(\aa$_k$..\aa$_1$\aa$_0$)$-$1]\iii}
\ddemili {} {} {$i\in\{2..p$-$2\}$, $j=i$+2}
\ddemili {p+1} {$(\nu)_b$} {\aa$_k$..\aa$_1$\aa$_0$\uu}
\ddemili {p+2} {$\nu$+1} {[(\aa$_k$..\aa$_1$\aa$_0$)+1]}
\vskip 5pt
\hrule height 0.3pt depth 0.3pt width \hsize
}}$
\hfill}
\vskip 7.5pt
\ligne{\hfill \bbb-node\hfill}
\vskip 7.5pt
\ligne{\hfill
$\vcenter{
\vtop{\leftskip 0pt\hsize=145pt
\hrule height 0.3pt depth 0.3pt width \hsize
\vskip 5pt
\ddemili {rep.} {tile} {\hfill \nzm-code}
\ddemili {1} {$(\nu)_1$} {\aa$_k$..\aa$_1$}
\ddemili {2} {$(\nu$$-$$1)_b$} {[(\aa$_k$..\aa$_1$\aa$_0$)$-$1]\uu}
\ddemili {j} {\www-sons} {[(\aa$_k$..\aa$_1$\aa$_0$)$-$1]\iii}
\ddemili {} {} {$i\in\{2..p$-$3\}$, $j=i$+1}
\ddemili {p$-$1} {$(\nu)_b$} {\aa$_k$..\aa$_1$\aa$_0$\uu}
\ddemili {p} {$(\nu)_1$+1} {[(\aa$_k$..\aa$_1$)+1]}
\vskip 5pt
\hrule height 0.3pt depth 0.3pt width \hsize
}}$
\hfill
$\vcenter{
\vtop{\leftskip 0pt\hsize=145pt
\hrule height 0.3pt depth 0.3pt width \hsize
\vskip 5pt
\ddemili {rep.} {tile} {\hfill \nzm-code}
\ddemili {1} {$(\nu)_1$} {\aa$_k$..\aa$_1$$\ominus$\uu}
\ddemili {2} {$\nu$$-$1} {[(\aa$_k$..\aa$_1$\aa$_0$)$-$1]}
\ddemili {3} {$(\nu$$-$$1)_b$} {[(\aa$_k$..\aa$_1$\aa$_0$)$-$1]\uu}
\ddemili {j} {\www-sons} {[(\aa$_k$..\aa$_1$\aa$_0$)$-$1]\iii}
\ddemili {} {} {$i\in\{2..p$-$3\}$, $j=i$+2}
\ddemili {p} {$(\nu)_b$} {\aa$_k$..\aa$_1$\aa$_0$\uu}
\ddemili {p+1} {$\nu$+1} {[(\aa$_k$..\aa$_1$\aa$_0$)+1]}
\ddemili {p+2} {$(\nu)_1$+1} {[(\aa$_k$..\aa$_1$)+1]}
\vskip 5pt
\hrule height 0.3pt depth 0.3pt width \hsize
}}$
\hfill}
}
\hfill}
}
\hfill}
\vskip 10pt
Its construction is easy from Table~$(30)$ which gives the neighbours $(\nu)_j$
for \hbox{$j\in\{3..p\}$} for a white node and \hbox{$j\in\{3..p$$-$$1\}$} for a
black one. In both cases, $(\nu)_1$ is the father and $(\nu)_2$ is the
rightmost son of $\nu$$-$1 whose \nzm-codes can easily be derived from Table~$(30)$.
In the case of the black node, $(\nu)_p$ is $(\nu)_1$+1, the node which lies just after
the father of~$\nu$ on the level of~$(\nu)_1$. The part of the tables devoted
to $\{p$+$2,3\}$ involves two specific neighbours: $\nu$$-$1 and $\nu$+1. For a white
node they are $(\nu)_2$ and $(\nu)_{p+2}$ respectively. For a black node, they are
$(\nu)_2$ and $(\nu)_{p+1}$ respectively as far as in that case $(\nu)_{p+2}$ is
$(\nu)_1$+1.

\ligne{\hfill
\vtop{
\begin{tab}\label{tbmetalneighw}
Table of the neighbours of~$\nu$: to left, in the tiling $\{p,4\}$, to right,
in the tiling $\{p$$+$$2,3\}$. In the table, the black son of $(\omega)$
is denoted by $(\omega)_b$, its preferred son by $(\omega)_\pi$.
\end{tab}
\ligne{\hfill
\vtop{\leftskip 0pt\hsize=300pt
\ligne{\hfill in $\{p,4\}$\hfill\hfill in $\{p$+$2,3\}$\hfill}
\ligne{\hfill \www\aa-node\hfill}
\vskip 7.5pt
\ligne{\hfill
$\vcenter{
\vtop{\leftskip 0pt\hsize=145pt
\hrule height 0.3pt depth 0.3pt width \hsize
\vskip 5pt
\ddemili {rep.} {tile} {\hfill metalic code}
\ddemili {1} {$(\nu)_1$} {[(\aa$_k$..\aa$_1$)$-$1]}
\ddemili {2} {$(\nu$$-$$1)_b$} {[(\aa$_k$..\aa$_1$\aa$_0$)$-$1]\uu}
\ddemili {j} {\www-sons} {[(\aa$_k$..\aa$_1$\aa$_0$)$-$1]\iii}
\ddemili {} {} {$i\in\{2..p$-$3\}$, $j=i$+1}
\ddemili {p$-$1} {$(\nu)_\pi$} {\aa$_k$..\aa$_1$\aa$_0$\zzz}
\ddemili {p} {$(\nu)_b$} {\aa$_k$..\aa$_1$\aa$_0$\uu}
\vskip 5pt
\hrule height 0.3pt depth 0.3pt width \hsize
}}$
\hfill
$\vcenter{
\vtop{\leftskip 0pt\hsize=145pt
\hrule height 0.3pt depth 0.3pt width \hsize
\vskip 5pt
\ddemili {rep.} {tile} {\hfill metalic code}
\ddemili {1} {$(\nu)_1$} {[(\aa$_k$..\aa$_1$)$-$1]}
\ddemili {2} {$\nu$$-$1} {[(\aa$_k$..\aa$_1$\aa$_0$)$-$1]}
\ddemili {3} {$(\nu$$-$$1)_b$} {[(\aa$_k$..\aa$_1$\aa$_0$)$-$1]\uu}
\ddemili {j} {\www-sons} {[(\aa$_k$..\aa$_1$\aa$_0$)$-$1]\iii}
\ddemili {} {} {$i\in\{2..p$-$3\}$, $j=i$+2}
\ddemili {p} {$(\nu)_\pi$} {\aa$_k$..\aa$_1$\aa$_0$\zzz}
\ddemili {p+1} {$(\nu)_b$} {\aa$_k$..\aa$_1$\aa$_0$\uu}
\ddemili {p+2} {$\nu$+1} {[(\aa$_k$..\aa$_1$\aa$_0)+1]$}
\vskip 5pt
\hrule height 0.3pt depth 0.3pt width \hsize
}}$
\hfill}
\vskip 7.5pt
\ligne{\hfill \www\zzz-node\hfill}
\vskip 7.5pt
\ligne{\hfill
$\vcenter{
\vtop{\leftskip 0pt\hsize=145pt
\hrule height 0.3pt depth 0.3pt width \hsize
\vskip 5pt
\ddemili {rep.} {tile} {\hfill metallic code}
\ddemili {p$-$2} {$(\nu)_\pi$} {\aa$_k$..\aa$_1$\aa$_0$\zzz}
\ddemili {p$-$1} {$(\nu)_{p-1}$} {\aa$_k$..\aa$_1$\aa$_0$\uu}
\ddemili {p} {$(\nu)_b$} {\aa$_k$..\aa$_1$\aa$_0$\numd}
\vskip 5pt
\hrule height 0.3pt depth 0.3pt width \hsize
}}$
\hfill
$\vcenter{
\vtop{\leftskip 0pt\hsize=145pt
\hrule height 0.3pt depth 0.3pt width \hsize
\vskip 5pt
\ddemili {rep.} {tile} {\hfill metallic code}
\ddemili {p$-$1} {$(\nu)_\pi$} {\aa$_k$..\aa$_1$\aa$_0$\zzz}
\ddemili {p} {$(\nu)_{p-1}$} {\aa$_k$..\aa$_1$\aa$_0$\uu}
\ddemili {p+1} {$(\nu)_b$} {\aa$_k$..\aa$_1$\aa$_0$\numd}
\vskip 5pt
\hrule height 0.3pt depth 0.3pt width \hsize
}}$
\hfill}
\vskip 7.5pt
\ligne{\hfill \www\uu-node\hfill}
\vskip 7.5pt
\ligne{\hfill
$\vcenter{
\vtop{\leftskip 0pt\hsize=145pt
\hrule height 0.3pt depth 0.3pt width \hsize
\vskip 5pt
\ddemili {rep.} {tile} {\hfill metallic code}
\ddemili {1} {$(\nu)_1$} {[(\aa$_k$..\aa$_1$)$-$1]}
\ddemili {2} {$(\nu$$-$$1)_b$} {[(\aa$_k$..\aa$_1$\aa$_0$)$-$1]\numd}
\ddemili {j} {\www-sons} {[(\aa$_k$..\aa$_1$\aa$_0$)$-$1]\iii}
\ddemili {} {} {$i\in\{3..p$-$3\}$, $j=i$}
\ddemili {p$-$2} {$(\nu)_\pi$} {\aa$_k$..\aa$_1$\aa$_0$\zzz}
\ddemili {p$-$1} {$(\nu)_{p-1}$} {\aa$_k$..\aa$_1$\aa$_0$\uu}
\ddemili {p} {$(\nu)_b$} {\aa$_k$..\aa$_1$\aa$_0$\numd}
\vskip 5pt
\hrule height 0.3pt depth 0.3pt width \hsize
}}$
\hfill
$\vcenter{
\vtop{\leftskip 0pt\hsize=145pt
\hrule height 0.3pt depth 0.3pt width \hsize
\vskip 5pt
\ddemili {rep.} {tile} {\hfill metallic code}
\ddemili {1} {$(\nu)_1$} {[(\aa$_k$..\aa$_1$)$-$1]}
\ddemili {2} {$\nu$$-$1} {[(\aa$_k$..\aa$_1$\aa$_0$)$-$1]}
\ddemili {3} {$(\nu$$-$$1)_b$} {[(\aa$_k$..\aa$_1$\aa$_0$)$-$1]\numd}
\ddemili {j} {\www-sons} {[(\aa$_k$..\aa$_1$\aa$_0$)$-$1]\iii}
\ddemili {} {} {$i\in\{3..p$-$3\}$, $j=i$+1}
\ddemili {p$-$1} {$(\nu)_\pi$} {\aa$_k$..\aa$_1$\aa$_0$\zzz}
\ddemili {p} {$(\nu)_p$} {\aa$_k$..\aa$_1$\aa$_0$\uu}
\ddemili {p+1} {$(\nu)_b$} {\aa$_k$..\aa$_1$\aa$_0$\numd}
\ddemili {p+2} {$\nu$+1} {[(\aa$_k$..\aa$_1$\aa$_0)+1]$}
\vskip 5pt
\hrule height 0.3pt depth 0.3pt width \hsize
}}$
\hfill}
\vskip 7.5pt
\ligne{\hfill \bbb-node\hfill}
\vskip 7.5pt
\ligne{\hfill
$\vcenter{
\vtop{\leftskip 0pt\hsize=145pt
\hrule height 0.3pt depth 0.3pt width \hsize
\vskip 5pt
\ddemili {p} {$(\nu)_1$+1} {\aa$_k$..\aa$_1$}
\vskip 5pt
\hrule height 0.3pt depth 0.3pt width \hsize
}}$
\hfill
$\vcenter{
\vtop{\leftskip 0pt\hsize=145pt
\hrule height 0.3pt depth 0.3pt width \hsize
\vskip 5pt
\ddemili {p+1} {$\nu$+1} {[(\aa$_k$..\aa$_1$\aa$_0$)+1]}
\ddemili {p+2} {$(\nu)_1$+1} {\aa$_k$..\aa$_1$}
\vskip 5pt
\hrule height 0.3pt depth 0.3pt width \hsize
}}$
\hfill}
}
\hfill}
}
\hfill}
\vskip 10pt
Table~\ref{tbmetalneighw} displays more cases as far as there are more rules for 
${\cal W}_\rho$ when the metallic codes are used. In order to reduce the number of 
repetitions, the lines for \www\zzz-nodes again takes the lines for \www\aa-nodes
except for the lines~\fnb {p-2}, \fnb {p-1} and~\fnb p in the case of $\{p,4\}$
and the lines \fnb {p-1}, \fnb p and \fnb {p+1} in the case of $\{p$+$2,3\}$. The
same thing was done in the case of the \bbb-nodes which takes the lines of 
the \www\uu-nodes except the line \fnb p for $\{p,4\}$ and the lines 
\fnb {p+1} and~\fnb {p+2} for $\{p$+$2,3\}$. The table rewrites the exceptional
lines accordingly. The reason of the changes is clear. The codes for the first sons
of a \www\aa-node and for a \www\zzz-one are built one the same way from the metallic
code of the node. For a \bbb-node, a similar remark is relevant.

\subsection{The case of the black metallic tree}\label{smblacktil}

In this section, we consider the same problems for the black metallic tree.
Sub-subsection~\ref{smbalgopaths} deals with the path from a node to the root
of ${\cal B}_\rho$, while
Sub-subsection~\ref{smbneighcodes} computes the codes of the neighbours of a tile
in a strip.

\subsubsection{Algorithm for the path from a tile to the head of a strip}
\label{smbalgopaths}

Algorithm~\ref{apathbmder} constructs the path from a node~$\nu$ in ${\cal B}_\rho$ 
to the root of the tree. It makes use of the same procedure 'update' as in
Algorithm~\ref{apathwmder}. 
\vskip -5pt
\ligne{\hfill
\vtop{
\begin{algo}\label{apathbmder}
The path from the root to the node~$\nu$ in ${\cal B}_{\rho}$. Here, 
\hbox{\aa$_k$$..$\aa$_0$ $\rightleftharpoons$ $[\nu]$}. The input is $[\nu]$.
	The result is in {\rm list$_\ell$}.
\end{algo}
\vskip-10pt
\ligne{\hfill
\vtop{\leftskip 0pt\hsize=310pt
\hrule depth 0.6pt height 0pt width \hsize
\vskip 5pt
\ifnum 1=0 {
\ligne{\hskip 23pt\pproc{} update (side; from; upto) \iis\hfill}
\ligne{\hskip 23pt\bbegin{} \iff{} side = \fnb {left}\hfill}
\ligne{\hskip 33pt\hskip 23pt \tthen{} \ffor{} $j$ \iin{} $\{$from..upto$\}$ 
\lloop{} list$_r$($j$) := list$_\ell$($j$); \endloop;\hfill}
\ligne{\hskip 33pt\hskip 23pt \eelse{} \ffor{} $j$ \iin{} $\{$from..upto$\}$
\lloop{} list$_\ell$($j$) := list$_r$($j$); \endloop;\hfill}
\ligne{\hskip 33pt\hskip 18pt \endif;
\hskip 10pt restart := upto+1; handside := side;\hfill}
\ligne{\hskip 23pt\endproc;\hfill}
} \fi
\ligne{restart := 0; handside := \fnb {left};\hfill}
\ligne{\iff{} \aa$_k$ = \zzz \hfill}
\ligne{\hskip 10pt\tthen{} list$_\ell$(0) := \rrr.0;
list$_r$(0) := \www.1;\hfill}
\ligne{\hskip 10pt\eelse{} \iff{} \aa$_k$ = \uu{}
\tthen{} list$_\ell$(0) := \bbb.$p$$-$3; list$_r$(0) := \www.1;\hfill}
\ligne{\hskip 33pt \eelse{} list$_\ell$ := \www.(\aa$_i$); 
list$_r$ := \www.(\aa$_i$)+1; \endif;\hfill}
\ligne{\endif{};\hfill}
\ligne{\ffor{} $i$ \iin{} \hbox{$\{0..k$$-$1$\}$} \inreverse\hfill}
\ligne{\lloop{} 
\iff{} status(list$_\ell$($k$$-$$i$$-$1)) = \www\hfill}
\ligne{\hskip 33pt \tthen{} last := $p$$-$2;\hfill}
\ligne{\hskip 33pt \eelse{} last := $p$$-$3;\hfill}
\ligne{\hskip 23pt \endif;\hfill}
\ligne{\hskip 23pt \iff{} \aa$_i$ = \zzz\hfill}
\ligne{\hskip 23pt\hskip 10pt \tthen{} \iff{} handside = \fnb {right}\hfill}
\ligne{\hskip 23pt\hskip 33pt\hskip 10pt
\tthen{} update (side: \fnb {left}, from: restart, upto: $k$$-$$i$$-$1);\hfill}
\ligne{\hskip 23pt\hskip 33pt\endif;\hfill}
\ligne{\hskip 23pt\hskip 33pt
list$_\ell$($k$$-$$i$) := \bbb.last; list$_r$($k$$-$$i$) := \www.1;\hfill}
\ligne{\hskip 23pt\hskip 10pt \eelse{} 
update (side: \fnb {right}, from: restart, upto: $k$$-$$i$$-$1);\hfill}
\ligne{\hskip 23pt\hskip 33pt
list$_\ell$($k$$-$$i$) := \www.\aa$_i$; list$_r$($k$$-$$i$) := \www.(\aa$_i$)+1;\hfill}
\ligne{\hskip 23pt \endif;\hfill}
\ligne{\endloop;\hfill}
\vskip 5pt
\hrule depth 0pt height 0.6pt width \hsize
}
\hfill}
}
\hfill}
\vskip 5pt
The algorithm makes use of the metallic code of~$\nu$. The lists register the status 
of the current node and its rank among the sons of its father, the leftmost son 
having rank~$1$.  

The justification of Algorithm~\ref{apathwbder} is straightforward. It is similar 
to the case
of Algorithm~\ref{anzmpathw}. It is based on the fact that outside the case when 
the current digit \aa{} is \zzz, the path necessarily goes within the tree rooted
at the node whose signature is \aa+1 and it is the \aa$^{\rm th}$ node among the sons
of its father. When \hbox{\aa{} = \zzz}, except the case when the path has to move
to right, it is needed to go on on both paths at a distance~1 from each other as at that
moment, the next digit is not known.

\vskip -5pt
\ligne{\hfill
\vtop{
\begin{algo}\label{anzmpathbder}
The path from the root to the node~$\nu$ in a strip~$\mathfrak S$ in bijection with 
${\cal B}_{\rho}$. We set \hbox{\aa$_k$$..$\aa$_0$ $\rightleftharpoons$ $[\nu]$}. 
The lists register the status of the
current node and its rank among the sons of its father, the leftmost son having
rank~$1$.
\end{algo}
\vskip-10pt
\ligne{\hfill
\vtop{\leftskip 0pt\hsize=310pt
\hrule depth 0.6pt height 0pt width \hsize
\vskip 5pt
\ifnum 1=0 {
\ligne{\hskip 23pt\pproc{} update (side; from; upto) \iis\hfill}
\ligne{\hskip 23pt\bbegin{} \iff{} side = \fnb {left}\hfill}
\ligne{\hskip 33pt\hskip 23pt \tthen{} \ffor{} $j$ \iin{} $\{$from..upto$\}$ 
\lloop{} list$_r$($j$) := list$_\ell$($j$); \endloop;\hfill}
\ligne{\hskip 33pt\hskip 23pt \eelse{} \ffor{} $j$ \iin{} $\{$from..upto$\}$
\lloop{} list$_\ell$($j$) := list$_r$($j$); \endloop;\hfill}
\ligne{\hskip 33pt\hskip 18pt \endif;
\hskip 10pt restart := upto+1; handside := side;\hfill}
\ligne{\hskip 23pt\endproc;\hfill}
} \fi
\ligne{restart := 0; handside := \fnb {left}; last := $k$$-$1;\hfill}
\ligne{\iff{} \aa$_k$ = \uu \hfill}
\ligne{\hskip 10pt\tthen{} list$_\ell$(0) := \bbb.$p$$-$3;
list$_r$(0) := \www.\numd; \hfill}
\ligne{\hskip 10pt\eelse{} \iff{} \aa$_k$ = \xxx{}\hfill}
\ligne{\hskip 10pt\hskip 33pt \tthen{} list$_\ell$(0) := \www.1; 
list$_r$(0) := \www.1;\hfill}
\ligne{\hskip 10pt\hskip 46pt\hskip 10pt
list$_\ell$(1) := \www.1; list$_r$(1) := \www.2; last := last$-$1;\hfill}
\ligne{\hskip 33pt\hskip 10pt \eelse{} \iff{} \aa$_i$ $<$ \ddd\hfill}
\ligne{\hskip 33pt\hskip 33pt\hskip 10pt
\tthen{} list$_\ell$(0) := \www.(\aa$_i$)+1; list$_r$(0) := \www.(\aa$_i$)+2; \hfill}
\ligne{\hskip 33pt\hskip 33pt\hskip 10pt 
\eelse{} list$_\ell$(0) := \rrr.0; list$_r$(0) := \www.\numd;\hfill}
\ligne{\hskip 33pt\hskip 33pt\hskip 31pt 
list$_\ell$(1) := \bbb.$p$$-$3; list$_r$(0) := \www.\uu; last := last$-$1;\hfill}
\ligne{\hskip 66pt \endif;\hfill}
\ligne{\hskip 33pt \endif{};\hfill}
\ligne{\endif{};\hfill}
\ligne{\ffor{} $i$ \iin{} \hbox{$\{0..{\rm final}\}$} \inreverse\hfill}
\ligne{\lloop{} last := $p$$-$2;\hfill}
\ligne{\hskip 23pt
\iff{} status(list$_\ell$($k$$-$$i$$-$1)) = \bbb{} \tthen{} last := $p$$-$3; 
\endif;\hfill}
\ligne{\hskip 23pt \ccase{} \aa$_i$ \iis\hfill}
\ligne{\hskip 23pt\hskip 10pt \wwhen{} \xxx{} $\Rightarrow$
update (side: \fnb {right}, from: restart, upto: $k$$-$$i$$-$1);\hfill}
\ligne{\hskip 23pt\hskip 33pt
list$_\ell$($k$$-$$i$) = \www.1; list$_r$($k$$-$$i$) = \www.2;\hfill}
\ligne{\hskip 23pt\hskip 10pt \wwhen{} \ddd{} $\vert$ \ccc{} $\Rightarrow$
update (side: \fnb {right}, from: restart, upto: $k$$-$$i$$-$1);\hfill}
\ligne{\hskip 23pt\hskip 33pt
list$_\ell$($k$$-$$i$) = \bbb.last; list$_r$($k$$-$$i$) = \www.1;\hfill}
\ligne{\hskip 23pt\hskip 10pt \wwhen{} \oothers{} $\Rightarrow$ \hfill}
\ligne{\hskip 23pt \hskip 33pt \iff{} \aa$_i$ \iin{} $\{$\ccc,\ddd$\}$\hfill}
\ligne{\hskip 23pt\hskip 33pt\hskip 10pt
\tthen{}
update (side: \fnb {right}, from: restart, upto: $k$$-$$i$$-$1);\hfill}
\ligne{\hskip 23pt\hskip 33pt\hskip 10pt
\eelse{}
update (side: \fnb {left}, from: restart, upto: $k$$-$$i$$-$1);\hfill}
\ligne{\hskip 23pt\hskip 33pt \endif;\hfill}
\ligne{\hskip 23pt\hskip 33pt
list$_\ell$($k$$-$$i$) = \www.(\aa$_i$)+1; list$_r$($k$$-$$i$) = \www.(\aa$_i$)+2;\hfill}
\ligne{\hskip 23pt \endcase;\hfill}
\ligne{\endloop;\hfill}
\vskip 5pt
\hrule depth 0pt height 0.6pt width \hsize
}
\hfill}
}
\hfill}
\vskip 10pt
Algorithm~\ref{anzmpathbder} addresses the same issue when the coordinates of the nodes
are given through their \nzm-code. We can see that its structure is more complex
than that of Algorithm~\ref{apathbmder}.

The reason is not only the fact this time we have three rules for the nodes instead of
two for Algorithm~\ref{apathbmder}, it is also due to the fact that the occurrence
of the pattern {\bf dc$^\ast$d} entails that when the pattern is followed by
a digit \aa{} with \hbox{\aa{} $<$ \ccc}, the appropriate node is in the node pointed
at by the right hand-side path: it follows from Table~$(38)$.

\subsubsection{The codes for the neighbours of a tile in a  black metallic tree}
\label{smbneighcodes}

We now turn to the computation of the coordinates of the neighbours of a tile~$\nu$
which lies in a strip~$\mathfrak S$ in bijection with ${\cal B}_\rho$. we first study
that computation when it is based on $[\nu]$. The computation should be easier
as there are two rules only for the sons of a node.
\vskip -5pt
\vtop{
\begin{tab}\label{tbneighbmder}
Table of the metallic codes of the neighbours of a tile~$\nu$ in both tilings
$\{p,4\}$ and $\{p$+$2,3\}$ in the black metallic tree under the rightmost assignment.
We assume that \hbox{\aa$_k$$..$\aa$_1$\aa$_0$ $\rightleftharpoons$ {\bf [$\nu$]}}
and $(\nu)_i$ indicates the neighbour~$i$. 
\end{tab}
\vskip -5pt
\ligne{\hfill
\vtop{\leftskip 0pt\hsize=300pt
\ligne{\hfill in $\{p,4\}$\hfill\hfill in $\{p$+$2,3\}$\hfill}
\ligne{\hfill \www\aa-node\hfill}
\vskip 7.5pt
\ligne{\hfill
$\vcenter{
\vtop{\leftskip 0pt\hsize=145pt
\ddemili {rep.} {tile} {\hfill metallic code}
\vskip 5pt
\hrule height 0.3pt depth 0.3pt width \hsize
\vskip 5pt
\ddemili {1} {$(\nu)_1$} {[(\aa$_k$..\aa$_1$)+1]}
\ddemili {2} {$(\nu$$-$$1)_p$} {[$\nu$$-$1]\zzz}
\ddemili {\jjj} {\www-sons} {[$\nu$$-$1]\iii}
\ddemili {} {} {$i\in\{1..p$$-$$3\}$, $j = i$+2}
\ddemili {p} {\bbb-son} {[$\nu$]\zzz}
\vskip 5pt
\hrule height 0.3pt depth 0.3pt width \hsize
}}$
\hfill
$\vcenter{
\vtop{\leftskip 0pt\hsize=145pt
\ddemili {rep.} {tile} {\hfill metallic code}
\vskip 5pt
\hrule height 0.3pt depth 0.3pt width \hsize
\vskip 5pt
\ddemili {1} {$(\nu)_1$} {[(\aa$_k$..\aa$_1$)+1]}
\ddemili {2} {$\nu$$-$1} {[$\nu$$-$1]}
\ddemili {3} {$(\nu$-$1)_{p+1}$} {[$\nu$$-$1]\zzz}
\ddemili {\jjj} {\www-sons} {[$\nu$$-$1]\iii}
\ddemili {} {} {$i\in\{1..p$$-$$3\}$, $j = i$+3}
\ddemili {p+1} {\bbb-son} {[$\nu$]\zzz}
\ddemili {p+2} {$\nu$+1} {[$\nu$+1]}
\vskip 5pt
\hrule height 0.3pt depth 0.3pt width \hsize
}}$
\hfill}
\vskip 7.5pt
\ligne{\hfill \bbb\zzz-node\hfill}
\vskip 7.5pt
\ligne{\hfill
$\vcenter{
\vtop{\leftskip 0pt\hsize=145pt
\ddemili {rep.} {tile} {\hfill metallic code}
\vskip 5pt
\hrule height 0.3pt depth 0.3pt width \hsize
\vskip 5pt
\ddemili {1} {$(\nu)_1$} {[\aa$_k$..\aa$_1$]}
\ddemili {2} {$(\nu$$-$$1)_p$} {[$\nu$$-$1]\zzz}
\ddemili {\jjj} {\www-sons} {[$\nu$$-$1]\iii}
\ddemili {} {} {$i\in\{1..p$$-$$3\}$, $j = i$+2}
\ddemili {p} {\bbb-son} {[$\nu$]\zzz}
\vskip 5pt
\hrule height 0.3pt depth 0.3pt width \hsize
}}$
\hfill
$\vcenter{
\vtop{\leftskip 0pt\hsize=145pt
\ddemili {rep.} {tile} {\hfill metallic code}
\vskip 5pt
\hrule height 0.3pt depth 0.3pt width \hsize
\vskip 5pt
\ddemili {1} {$(\nu)_1$} {[\aa$_k$..\aa$_1$]}
\ddemili {2} {$\nu$$-$1} {[$\nu$$-$1]}
\ddemili {3} {$(\nu$-$1)_{p+1}$} {[$\nu$$-$1]\zzz}
\ddemili {\jjj} {\www-sons} {[$\nu$$-$1]\iii}
\ddemili {} {} {$i\in\{1..p$$-$$4\}$, $j = i$+3}
\ddemili {p} {\bbb-son} {[$\nu$]\zzz}
\ddemili {p+1} {$\nu$+1} {[$\nu$+1]}
\ddemili {p+2} {$(\nu)_1$+1} {[(\aa$_k$..\aa$_1$)+1]}
\vskip 5pt
\hrule height 0.3pt depth 0.3pt width \hsize
}}$
\hfill}
}
\hfill}
}
\vskip 10pt
Table~\ref{tbneighbmder} follows immediately from the examination of rules~$(33)$
and Table~$(34)$. The additional neighbours of~$\nu$ in~$\{p$+$2,3\}$ are the nodes
$\nu$$-$1 and $\nu$+1 which are on the same level as~$\nu$ in ${\cal B}_\rho$. This
introduce a small change in the numbering of the sons of the node compared with their
numbering in $\{p,4\}$.

\vskip -5pt
\vtop{
\begin{tab}\label{tbnzmneighbder}
Table of the \nzm-codes of the neighbours of a tile~$\nu$ in both tilings
$\{p,4\}$ and $\{p$+$2,3\}$ in the black metallic tree under the rightmost assignment.
We assume that \hbox{\aa$_k$$..$\aa$_1$\aa$_0$ $\rightleftharpoons$ {\bf [$\nu$]$_{nz}$}}
and $(\nu)_i$ indicates the neighbour~$i$. When references of neighbours for a type of 
nodes are missing, they have to be seen in the same column at the previous type and,
again at the previous one if they are still missing.
\end{tab}
\vskip -5pt
\ligne{\hfill
\vtop{\leftskip 0pt\hsize=300pt
\ligne{\hfill in $\{p,4\}$\hfill\hfill in $\{p$+$2,3\}$\hfill}
\ligne{\hfill \www\aa-node\hfill}
\vskip 7.5pt
\ligne{\hfill
$\vcenter{
\vtop{\leftskip 0pt\hsize=145pt
\ddemili {rep.} {tile} {\hfill \nzm-code}
\vskip 5pt
\hrule height 0.3pt depth 0.3pt width \hsize
\vskip 5pt
\ddemili {1} {$(\nu)_1$} {[(\aa$_k$..\aa$_1$)$+$1]$_{nz}$}
\ddemili {2} {$(\nu$$-$1$)_p$} {[$\nu$$-$2]$_{nz}$\xxx}
\ddemili {j} {\www-sons} {[$\nu$$-$1]$_{nz}$\iii}
\ddemili {} {} {$i\in\{1..p$-$3\}$, $j=i$+2}
\ddemili {p} {\bbb-son} {[$(\nu$$-$1$)]_{nz}$\xxx}
\vskip 5pt
\hrule height 0.3pt depth 0.3pt width \hsize
}}$
\hfill
$\vcenter{
\vtop{\leftskip 0pt\hsize=145pt
\vskip 5pt
\ddemili {rep.} {tile} {\hfill \nzm-code}
\hrule height 0.3pt depth 0.3pt width \hsize
\vskip 5pt
\ddemili {1} {$(\nu)_1$} {[(\aa$_k$..\aa$_1$)$+$1]$_{nz}$}
\ddemili {2} {$\nu$$-$1} {[$\nu$$-$1]$_{nz}$}
\ddemili {3} {$(\nu$-1$)_{p+1}$} {[$\nu$$-$2]$_{nz}$\xxx}
\ddemili {j} {\www-sons} {[$\nu$$-$1]$_{nz}$\iii}
\ddemili {} {} {$i\in\{1..p$-$3\}$, $j=i$+3}
\ddemili {p+1} {\bbb-son} {[$(\nu$$-$1$)]_{nz}$\xxx}
\ddemili {p+2} {$\nu$+1} {[$\nu$+1]$_{nz}$}
\vskip 5pt
\hrule height 0.3pt depth 0.3pt width \hsize
}}$
\hfill}
\vskip 7.5pt
\ligne{\hfill \www\uu-node\hfill}
\vskip 7.5pt
\ligne{\hfill
$\vcenter{
\vtop{\leftskip 0pt\hsize=145pt
\ddemili {rep.} {tile} {\hfill \nzm-code}
\vskip 5pt
\hrule height 0.3pt depth 0.3pt width \hsize
\vskip 5pt
\ddemili {2} {$(\nu$$-$1$)_p$} {[$\nu$$-$2]$_{nz}$\ddd}
\ddemili {3} {$(\nu)_3$} {[$\nu$$-$2]$_{nz}$\xxx}
\ddemili {j} {\www-sons} {[$\nu$$-$1]$_{nz}$\iii}
\ddemili {} {} {$i\in\{1..p$-$4\}$, $j=i$+3}
\ddemili {p} {\bbb-son} {[$(\nu$$-$1$)]_{nz}$\ddd}
\vskip 5pt
\hrule height 0.3pt depth 0.3pt width \hsize
}}$
\hfill
$\vcenter{
\vtop{\leftskip 0pt\hsize=145pt
\ddemili {rep.} {tile} {\hfill \nzm-code}
\vskip 5pt
\hrule height 0.3pt depth 0.3pt width \hsize
\vskip 5pt
\ddemili {3} {$(\nu$$-$1$)_p$} {[$\nu$$-$2]$_{nz}$\ddd}
\ddemili {4} {$(\nu)_3$} {[$\nu$$-$2]$_{nz}$\xxx}
\ddemili {j} {\www-sons} {[$\nu$$-$1]$_{nz}$\iii}
\ddemili {} {} {$i\in\{1..p$-$4\}$, $j=i$+4}
\ddemili {p+1} {\bbb-son} {[$(\nu$$-$1$)]_{nz}$\ddd}
\vskip 5pt
\hrule height 0.3pt depth 0.3pt width \hsize
}}$
\hfill}
\vskip 7.5pt
\ligne{\hfill \www\xxx-node\hfill}
\vskip 7.5pt
\ligne{\hfill
$\vcenter{
\vtop{\leftskip 0pt\hsize=145pt
\ddemili {rep.} {tile} {\hfill \nzm-code}
\vskip 5pt
\hrule height 0.3pt depth 0.3pt width \hsize
\vskip 5pt
\ddemili {1} {$(\nu)_1$} {[(\aa$_k$..\aa$_1$)$+$2]$_{nz}$}
\vskip 5pt
\hrule height 0.3pt depth 0.3pt width \hsize
}}$
\hfill
$\vcenter{
\vtop{\leftskip 0pt\hsize=145pt
\ddemili {rep.} {tile} {\hfill \nzm-code}
\vskip 5pt
\hrule height 0.3pt depth 0.3pt width \hsize
\vskip 5pt
\ddemili {1} {$(\nu)_1$} {[(\aa$_k$..\aa$_1$)$+$2]$_{nz}$}
\vskip 5pt
\hrule height 0.3pt depth 0.3pt width \hsize
}}$
\hfill}
\vskip 7.5pt
\ligne{\hfill \bbb\ddd,\bbb\xxx-nodes\hfill}
\vskip 7.5pt
\ligne{\hfill
$\vcenter{
\vtop{\leftskip 0pt\hsize=145pt
\ddemili {rep.} {tile} {\hfill metallic code}
\vskip 5pt
\hrule height 0.3pt depth 0.3pt width \hsize
\vskip 5pt
\ddemili {1} {$(\nu)_1$} {[(\aa$_k$..\aa$_1$)$+$1]$_{nz}$}
\ddemili {2} {$(\nu$$-$$1)_p$} {[$\nu$$-$2]$_{nz}\xxx$}
\ddemili {j} {\www-sons} {[$\nu$$-$1]$_{nz}$\iii}
\ddemili {} {} {$i\in\{1..p$-$3\}$, $j=i$+2}
\ddemili {p} {$(\nu)_p$} {[$\nu$+1]$_{nz}$}
\vskip 5pt
\hrule height 0.3pt depth 0.3pt width \hsize
}}$
\hfill
$\vcenter{
\vtop{\leftskip 0pt\hsize=145pt
\ddemili {rep.} {tile} {\hfill metallic code}
\vskip 5pt
\hrule height 0.3pt depth 0.3pt width \hsize
\vskip 5pt
\ddemili {1} {$(\nu)_1$} {[(\aa$_k$..\aa$_1$)$+$1]$_{nz}$}
\ddemili {2} {$\nu$$-$1} {[$\nu$$-$1]$_{nz}$}
\ddemili {3} {$(\nu$$-$$1)_p$} {[$\nu$$-$2]$_{nz}\xxx$}
\ddemili {j} {\www-sons} {[$\nu$$-$1]$_{nz}$\iii}
\ddemili {} {} {$i\in\{1..p$-$3\}$, $j=i$+3}
\ddemili {{p+1}} {$(\nu)_{p+1}$} {[$\nu$+1]$_{nz}$}
\ddemili {{p+2}} {$(\nu)_1$+1} {[(\aa$_k$..\aa$_1$)+2]$_{nz}$}
\vskip 5pt
\hrule height 0.3pt depth 0.3pt width \hsize
}}$
\hfill}
}
\hfill}
}
\vskip 10pt
Table~\ref{tbnzmneighbder} shows the \nzm-codes of the neighbours $(\nu)_i$ of~$\nu$
in ${\cal B}_\rho$. The table follows from the rules~$(37)$ and~$(38)$. It should be 
remarked that the \nzm-codes of the neighbours of a \www\xxx-node are very similar to
those of a \www\uu-one. The difference is in computation of the \nzm-code of the father.
For the \www\xxx-node, the \nzm-code of the father is given by 
\hbox{[(\aa$_k$..\aa$_1$)+2]$_{nz}$} and not by \hbox{[(\aa$_k$..\aa$_1$)+1]$_{nz}$}
as it the case for a \www\uu-node.

\section*{Conclusion}\label{conclude}

    We can conclude the paper with several remarks.

    The present paper deepens the research started with \cite{mmarXiv1} and which was
continued by~\cite{mmarXiv2}.

    The extension addressed both the connection of the trees with tilings of the
hyperbolic plane and the generalization of the golden sequence to the metallic ones
performed in \cite{mmarXiv2}. A comparison was made in that latter paper between the 
white metallic tree and the black one, finding an explanation of the surprising results 
obtained in the case of the black metallic tree were the preferred son property was
no more true.

    The present paper was an occasion to revisit the previous results with two new
features: the \nzm-codes and the notion of assignment, a notion that was already
introduced in~\cite{mmJUCStools} in the case of the Fibonacci trees. A key result
here is Theorem~\ref{tpref} which partially solves a question raised 
in~\cite{mmJUCStools} as far as any assignment possesses the preferred son property
in the frame of the metallic codes applied to white metallic trees. The introduction of 
the \nzm-codes radically changed the situation even for the white metallic trees: the 
preferred son property is true for a single assignment as shown by Theorem~\ref{tprefnzm}.
A unique assignment also possesses the preferred son property for the black metallic tree
in the frame of the metallic codes. But no assignment possess that property when the
black metallic tree is fitted with the \nzm-codes.

    The present paper also computes the metallic and the \nzm-codes of the neighbours of
a tile, taking advantage of the isomorphism established between the metallic trees
and particular regions of the tilings $\{p,4\}$ and $\{p$+$2,3\}$ of the hyperbolic
plane. The paper also investigates algorithms to compute the path from a node to the
root of its tree using metallic codes and also using \nzm-ones. All these algorithms are
linear in the size of the code which is used;
    
    The paper is more conclusive than \cite{mmarXiv2} claimed for itself. Other problems
are probably still open: the present author does not pretend to have solved any
possible problems in these settings, just to have closed one or two issues.


\begin{thebibliography}{5}
\newcount\bibi\bibi=1

\bibitem{ImaIwaMorita}
C. Iwamoto, T. Andou, K. Morita, K. Imai,
Computational Complexity in the Hyperbolic Plane, 
{\it Lecture Notes in Computer Science}, {\bf 2420}, (2002), 
Proceedings of {\bf MFCS'2002}, 365-374.

\bibitem{mmJUCStools}
M. Margenstern,
New Tools for Cellular Automata of the Hyperbolic Plane,
{\it Journal of Universal Computer Science},
{\bf 6}(12), (2000), 1226--1252.

\bibitem{mmbook1}
M. Margenstern,
{\it Cellular Automata in Hyperbolic Spaces, vol. $I$, Theory},
Collection: {\it Advances in Unconventional Computing and Cellular Automata},
Editor: Andrew Adamatzky,
Old City Publishing, Philadelphia, (2007), 424p.

\bibitem{mmbook2}
M. Margenstern,
{\it Cellular Automata in Hyperbolic Spaces, vol. $II$, Implementation and
computations},
Collection: {\it Advances in Unconventional Computing and Cellular Automata},
Editor: Andrew Adamatzky,
Old City Publishing, Philadelphia, (2008), 360p.

\bibitem{mmarXiv1}
M. Margenstern,
About Fibonacci trees $-$ I $-$, {\it arXiv}:1904.12135, cs.DM., (2019),
17pp.

\bibitem{mmarXiv2}
M. Margenstern,
About Fibonacci trees $-$ II $-$ : generalized Fibonacci trees, {\it arXiv}:1907.04677, 
cs.DM., (2019), 35pp.

\bibitem{mmgsJUCS}
M. Margenstern, G. Skordev,
Fibonacci Type Coding for the Regular Rectangular Tilings of the
Hyperbolic Plane,
{\it Journal of Universal Computer Science},
{\bf 9}(5), (2003), 398-422.

\end{thebibliography}
\end{document}